\documentclass[12pt,a4paper]{article}
\usepackage[english]{babel}
\usepackage[T1]{fontenc}
\usepackage[adobe-utopia]{mathdesign}
\usepackage[intlimits]{amsmath}
\usepackage{graphicx}
\usepackage{a4wide}
\usepackage{amsbsy}
\usepackage{hyperref}
\usepackage{hyphenat}
\usepackage{physics}
\usepackage{german}
\usepackage[square, authoryear]{natbib}
\usepackage{authblk}
\usepackage{braket}
\bibpunct{[}{]}{;}{a}{}{,}
\begin{document}

\title{\textsc{\huge{The state is not abolished, it withers away: how quantum field theory became a theory of scattering\footnote{This paper was originally published in Studies in the History and Philosophy of Modern Physics, Volume 60 (2017), pp. 46-80. Due to formatting mistakes that appeared only after proof-reading, parts of that paper are garbled. An erratum was published in SHPMP that links to this version, which is the definitive version of the paper.} }}}
\author{Alexander S. Blum\thanks{ablum@mpiwg-berlin.mpg.de}}
\affil{Max Planck Institute for the History of Science, Boltzmannstra"se 22, 14195 Berlin, Germany}
\maketitle

Learning quantum field theory (QFT) for the first time, after first learning quantum mechanics (QM), one is (or maybe, rather, I was) struck by the change of emphasis: The notion of the quantum state, which plays such an essential role in QM, from the stationary states of the Bohr atom, over the Schr"odinger equation to the interpretation debates over measurement and collapse, seems to fade from view when doing QFT. Not that it's gone - as any physicist will be quick to tell you, QFT is simply a quantum theory, with all the general structure of QM taken over unchanged. But the quantum state is hardly mentioned, when dealing with Feynman diagrams, path integrals and all the other mainstays of an introductory QFT course.

This was not always so: The QFT of the late 1920s and 1930s developed as a straightforward extension and generalization of QM; consequently, writing down Schr"odinger equations and calculating the energies of stationary states were the prime concerns of the physicists working with QFT at the time.\footnote{One just needs to look at the first paper on the full theory of quantum electrodynamics \citep{heisenberg_1929_zur-quantendynamik}.} But, as is well-known, this early QFT suffered from crippling defects, most notably the divergence problem, i.e., that all calculations appeared to give nonsensical, infinite results, once one went past the first approximation. The divergence difficulties of QFT (or at least of quantum electrodynamics) were solved through the renormalization techniques developed in the late 1940s. As was frequently stressed already at the time, the success of the renormalization program meant that the conceptual foundations of QM could be taken over to field theory with only slight modifications, as opposed to what physicists had generally believed all through the 1930s and early 1940s. But even though the foundations did not change (or change only just enough so that they could stay the same, as Weinberg has characterized this development \citep[p.18]{weinberg_1977_the-search}), the formalism of renormalized QFT (finding its first definitive formulation in Freeman Dyson's systematization of Richard Feynman's modular diagrammatic approach) looked quite different from that of 1930s QFT and a lot more like what we call QFT today.

What we observe can be called a paradigm shift. The term ``paradigm shift'' can be used in two ways, to designate a change of worldview and conceptual foundation or to designate a change in the paradigmatic problem to be calculated from the theoretical foundations. Oftentimes these two changes occur simultaneously and a distinction need not be made. Our case is different: Although the theoretical basis remained almost the same, the paradigmatic problem that was to be calculated from this basis changed. QM had been all about properties of the quantum state, most importantly the associated energy levels. QFT was all about scattering. Both frameworks were of course able to address other problems, but the calculations were in general modeled after the paradigmatic calculation: So, in QM, scattering was traditionally modeled as a stationary state, assuming a continuous influx of scattered particles.\footnote{See, e.g., the standard textbook by \citep[Section IX]{dirac_1935_the-principles}.} Conversely, bound-state problems could be treated in renormalized QFT using Feynman graphs, whose design alone was a constant reminder of their origins in the calculation of scattering matrix elements.

This shift of emphasis is certainly not ignored in the Sam Schweber's major work on the re-invention of QED in the late 1940s \citep{schweber_1994_qed}, but it is merely alluded to at several points and not analyzed or identified in greater detail. It is clearly emphasized and discussed explicitly by  \cite{wuethrich_2010_the-genesis}. Both books serve as an invaluable foundation for the research presented here. But also in the latter study, the transition is looked at only very locally, both in space and time (Feynman's study of the Dirac equation in the years 1946-1949) and in ``conceptual space'' (the modes of representation and the shift from spectroscopic term schemes to Feynman diagrams).

In this paper, I attempt to give a broader reconstruction of how this paradigm shift came about. I will attempt to identify several distinct historical developments that contributed to it and to identify several different factors that were essential for the possibility and the actual occurrence of this shift. For the organization of this paper, I have opted for a focus on the former, i.e., on a more narrative, rather than a thematic, structure. To counterbalance this, I will begin by briefly presenting the major recurring themes that the reader should look out for in the following (double) narrative.

The first impetus towards this paradigm shift, which therefore shows up as a starting point in several of the narrative threads below, is the attempt to formulate quantum theory in a more explicitly relativistic manner. The great champion of these attempts is Paul Dirac, who throughout the 1930s provided various starting points for such an explicitly relativistic formulation of quantum theory. What all these formulations had in common was that in some sense they problematized the quantum mechanical notion of an instantaneous state and tended towards replacing it with a focus on overall processes. This stemmed from the relativistic need to treat space and time on the same footing and the consequent tendency of relativity towards a block universe view.

Such attempts at making quantum theory more relativistic would hardly have been necessary if there hadn't been the grave divergence difficulties of QFT. For Dirac and others the relativistic reformulations were just preliminary exercises for a successor theory to QFT, which would circumvent the divergence difficulties and provide a consistent and divergence-free quantum theory of electrodynamic (and nuclear) interactions. Two radical attempts at creating such an entirely new theory  in the 1930s and 1940s play an important role in my reconstruction: Heisenberg's S-Matrix theory and the Wheeler-Feynman theory of action-at-a-distance electrodynamics. I have accordingly structured my narrative along these two attempts, how they arose and how they influenced the formulation of renormalized QFT. For while they ultimately were generally viewed as having gone way too far - demanding a total overhaul of the foundations of the theory, when in fact small, conservative modifications were all that was needed in order to construct a workable QFT - they provided essential insights and methods to the new scattering-focused formalism of QFT.

The two approaches I will be studying were very different, but they shared one central aspect: In attempting to solve the difficulties of QFT, they got rid of the notion of a (quantum) state altogether. The framework they provided for such a theory was taken over into ``regular'' QFT in the late 1940s, and defined that theory in an essential way, by providing calculational techniques and, perhaps even more importantly, by preparing a mindset in which scattering could be thought of as the primary, paradigmatic thing to be calculated from a theory. They thus paved the way for the marginalization of the quantum state, even as it turned out that it would not entirely be abolished, as had been the original expectation.

This paper has a clear focus on theoretical and conceptual developments, but experimental developments played an important role, which will also be duly addressed. In particular, the growing importance of scattering experiments in cosmic ray physics was essential for Heisenberg's path to the S-Matrix approach. But the role played by experiments is more complex than that. In Feynman's approach, they initially hardly played a role, as his was merely an attempt to reconstruct the well-established results of electrodynamics in a theory without instantaneous states (and without localized fields). Also, the experimental developments that led up to the development of renormalized QED were not scattering experiments. In particular, the Lamb Shift was really a classical spectroscopic measurement, similar to the spectroscopic experiments that had played such a central role in the development of Bohr's theory of the atom and the consequent conceptualization of quantum theory as a theory of stationary states and transitions between them. What we will thus encounter several times in the following narrative, is the difficulty of adapting the newly emerging scattering theories to the calculation of traditional quantum theory observables, such as the energies of stationary states. The partial successes in this direction played an important role in the establishment of the scattering paradigm.

I now turn to my two central narratives: In the first half of the paper, I will describe the genesis and initial success of Heisenberg's S-Matrix theory, culminating in Stueckelberg's theory of the causal S-Matrix, which might have been the starting point for the establishment of a scattering-centric reformulation of QFT, if it hadn't been for Feynman's modular diagrammatic approach appearing at the same time. The development of this latter approach is recounted in the second half of the paper, beginning from Wheeler and Feynman's attempts to reformulate electrodynamics without fields and Feynman's attempts to quantize such a theory using path integrals. The second half concludes with Dyson's merging of Heisenberg's S-Matrix approach with Feynman's techniques, to create the new formulation of QFT and thereby conclude the paradigm shift from energy levels to scattering.

\section{The S-Matrix}

Heisenberg's theory of the S-Matrix was laid out in a series of papers published during World War II. But we will go back somewhat further and study the origins of this theory in Heisenberg's attempts at incorporating a smallest, fundamental length into quantum theory. The fact that Heisenberg's S-Matrix has its origins in his theory of the fundamental length has often been remarked. In the following, I will be discussing this development specifically with an eye to the abolishment of the quantum state.

\subsection{Heisenberg and the fundamental length}

The development of Heisenberg's work on a fundamental length is described in  \citep{kragh_1995_arthur}. The important point for our purposes is that initially the fundamental length was intended solely to remove the divergence difficulties of  QED, acting as a cutoff scale for the divergent integrals appearing in higher order calculations in perturbation theory. The fundamental length was introduced into the theory by modifying the Hamiltonian, first by replacing differentials by differences (the 1930 lattice world, discussed in detail in \citep{carazza_1995_heisenbergs}), later by smearing out the energy density at a point in space with the help of a regularizing function (the 1935 $\Delta$-formalism, discussed in \citep{miller_1994_early}). These attempts always implied absolute limits on position measurements (or the measurements of field strengths at a point in space), but did not alter the general structure of the fundamental dynamical equations, which were still based on Hamiltonians and wave functions. These attempts did not go very far, running afoul, e.g., of their lack of relativistic invariance, and neither of them was ever published.

In 1936, Heisenberg turned to Fermi's theory of $\beta$ decay, which implicitly contained a parameter with the units of a length in the form of the dimensionful coupling constant $g$. Initially, Heisenberg's interest in Fermi theory (as laid out in a letter to Pauli on 26 May 1936)\footnote{All letters from and to Pauli in the 1930s are reproduced in \citep{hermann_1985_wissenschaftlicher}. All translations are by me.} was not related to the divergence difficulties. Instead, he hoped that the dimensionful coupling constant of Fermi theory would help to explain the occurrence of cosmic ray showers. 

What exactly was there to be explained? The difficulty was that the showers seemed to emanate from a single point, implying that all of the secondary particles were created in a single event. In QED, the creation of a large number of particles (photons and electron-positron pairs) in a single, localized event was very improbable: The more particles are involved in a single electromagnetic process, the higher is the power of the fine structure constant $\alpha = e^2/\hbar c \approx 1/137$ appearing in the rate of said process, thus making the probability of a single explosive shower event incredibly small. 

Heisenberg now realized that in Fermi theory the expansion parameter in a perturbative calculation is not simply proportional to the equivalent of the fine structure constant, $g' = g/\hbar c$, which by itself is still dimensionful, but  also to a power of the total momentum $k$ involved in the process.\footnote{Which power depends on the formulation of Fermi theory one is using. Heisenberg was at the time considering the version of \cite{konopinski_1935_on-the-fermi}, which includes an additional factor of neutrino momentum, so that the third power of the momentum enters the expansion paramenter.} Heisenberg argued that for large momenta, the expansion parameter also gets large and the situation is very different than in  QED: many-particle processes become just as probable as one- or two-particle processes. As he wrote to Pauli:

\begin{quote}
It thus seems to me as though one could understand the existence of cosmic ray showers directly from Fermi's $\beta$ theory.
\end{quote}

Now, at first this seems to be totally unrelated to Heisenberg's earlier work on the fundamental length. That he was thinking of a connection already at the time of the quoted letter to Pauli, can be surmised from the fact that he chose a system of units (by introducing the necessary factors of $\hbar c$), where the coupling constant $g'$ has the units $\mathrm{cm}^3$, so that the expansion parameter is $g'/\lambda^3$, where $\lambda$ is the wavelength of the incoming particle. This implied a reading where the explosive cosmic ray showers occurred when the wavelength of the exploding particle became smaller than a critical length $\sqrt[3]{g'}$, where the expansion coefficient becomes of order one.

I have, and will be, distinguishing between the terms ``fundamental length,'' to denote a length scale below which the usual notions of space no longer apply, and ``critical length,'' to denote a length scale below which certain phenomena (e.g., explosions) occur. With the discovery of the critical length in Fermi theory, it was certainly tempting to connect this with the idea of a fundamental length. However, the critical length of Fermi theory was as yet unrelated to the problem the fundamental length had been supposed to solve: The divergence difficulties of QFT.

Pauli's reply is not extant, but he appears to have addressed precisely this point: Fermi theory diverged at higher orders of perturbation theory, just like QED did, so it was not permissible to explain cosmic ray showers through such higher order calculations. But Heisenberg was not to be deterred, and in his reply to Pauli on 30 May 1936 insisted on the differences between Fermi theory and QED: In QED the expansion parameter was small, so perturbation theory should work, but doesn't. In Fermi theory, on the other hand, the theory itself already indicated that perturbation theory should break down for high energies, so the divergence difficulties were not a defect of the theory as a whole, but only of the perturbative calculations:

\begin{quote}
The non-convergence of the self-energy for large momenta in Fermi's theory is therefore not an argument against the theory, but - as long as one does not develop new mathematical techniques to treat the domain of high energies - only an argument for the lack of mathematical understanding of the calculating physicist. I do not claim that I am already sure that all self-energy difficulties will disappear in the future. But I do want to claim that all arguments so far concerning the infinite energies in the Fermi theory were nonsense and that it is now the most important task to check, how such a theory behaves qualitatively for high energies. To check this in quantum electrodynamics, however, would be pointless, since one knows that for dimensional reasons ($e^2/\hbar c$) there is no qualitative difference between high and low energies.
\end{quote} 

The critical length thus did not function as a cutoff scale, as the fundamental length had. Instead, Heisenberg argued, for a theory including a critical length in the form of a dimensionful coupling constant, which was thereby able to qualitatively explain the cosmic ray explosions, one could hope that the divergences might simply be an artifact of perturbation theory. But this was very hard to verify, given that perturbation theory was the only known way to do calculations in QFT. Pauli and Heisenberg made some attempts at doing their calculations in discretized space, i.e., on the lattice (now just as a calculational tool, not as a model of physical space as in 1930), but their results were inconclusive. Pauli became skeptical of the whole approach early on and became convinced that one could not construct a theory that was simultaneously free of divergences and able to describe explosions (Letter to Heisenberg, 26 October 1936):

\begin{quote}
[S]o it always seems to be the case that \emph{as soon as we have an increase of shower processes for large} $k$\emph{, the eigenvalues of the Hamilton operator will also always come out infinite} [...] 
\end{quote}

By the end of the year, Heisenberg gave in and gave up, not just on the lattice, but on QFT as a whole. Already when originally proposing his idea of explosions and a fundamental length in Fermi theory, he had anticipated (Letter to Pauli, 30 May 1936):

\begin{quote}
Of course, it is also conceivable that the formalism of wave quantization will have to be modified when introducing Fermi's $g$, just like one had to modify the earlier physics when introducing the universal dimensionful constants $c$ and $h$. 
\end{quote}

And this was the position to which he now again retreated (Letter to Pauli, 7 December 1936):

\begin{quote}
I am now again totally convinced that the quantization rules are in need of reform [...]
\end{quote}

What were to be the elements of such a new theory that renounced the quantum theory of fields, a theoretical framework that had, in spite of its apparent difficulties, matured in the preceding decade? This is the theme that we will now be following up until the formulation of S-Matrix theory.

One element that certainly remained was the critical length, which was to be related to the occurrence of explosive cosmic ray showers. This was now wedded to Heisenberg's earlier notion of a fundamental length, which provided absolute limits on measurement, a connection which had not been made explicitly, while Heisenberg still hoped to show the non-perturbative finiteness of Fermi theory. The connection between critical and fundamental length could in fact be made independently of all formalism through a physical argument by Bohr. It is first found in writing in a letter from Bohr to Dirac, dating from 2 July 1936.\footnote{Bohr Scientific Correspondence, Niels Bohr Archive, Copenhagen.} Since Heisenberg had presented his work on the critical length in Copenhagen just a week earlier,\footnote{The conference had taken place from 21 to 26 June 1936, see the timeline in \citep{hermann_1985_wissenschaftlicher}. Bohr mentioned that Heisenberg had presented his work at this conference in the letter to Dirac.} it is quite probable that Bohr had communicated similar thoughts to Heisenberg himself:

\begin{quote}
[T]hese new and most promising considerations of Heisenberg appear to me to offer a most important clue to the old problem of the limitation of the very ideas of space and time imposed by the atomistic structure of all measuring instruments. You may remember that we have often discussed such questions but hitherto it seemed most difficult to find an unambiguous starting point. It now appears, however, that any measurement of such short lengths and intervals where the conjugated momenta and energy will cause all matter to split into showers will be excluded in principle.
\end{quote}

The same argument later shows up in published work of Heisenberg's, in which he also explicitly acknowledges discussions with Bohr \citep{heisenberg_1938_uber-die-in-der-theorie}. The conclusion was: The critical length of explosion theory was also a fundamental length, because when trying to measure at scales smaller than the critical length, the measuring apparatus literally blew up in your face.

With this conceptual foundation, what was the formal basis on which Heisenberg continued his work? As witnessed by his last quote, his final disillusionment with QFT was directed at a very specific point of the theory, namely, the quantization, i.e., the transition from the classical to the quantum field theory. This meant that he believed that the unquantized Fermi field theory (including, in particular, the non-linear terms with the dimensionful coupling constant) was to be considered a correct basis for a future quantum theory (a hypothesis which he backed up by some semi-classical calculations on the creation of showers in \citep{heisenberg_1936_zur-theorie}), but that the manner in which such a quantum theory was constructed from it had to be rethought. On 18 December 1936 he wrote to Pauli:

\begin{quote}
The current situation in field theory can be described in the following manner: We have a \emph{korrespondenzm"a"sige}\footnote{This German neologism is easy to translate but only with a quite cumbersome English phrase. It means ``corresponding to the actual theory in a manner analogous to the way in which classical mechanics corresponds to quantum mechanics.''} theory, which is about as good and as bad as Bohr's theory was at the time. In this approximate (``semi-classical'') theory, the universal length is already \emph{correctly} included, if one writes it as the factor multiplying a non-linear interaction term in the Hamiltonian. What is missing is the step from here to a consistent theory.
\end{quote}

At this time, the analogy with the situation before the creation of quantum mechanics becomes a recurring theme, and it will not be the last time we encounter it. That nostalgia for the revolution of their youth drove the protagonists of the creation of quantum mechanics to look for far too radical solutions to the woes of QFT is a clich\'e, but certainly has some truth to it. It even guided their search for solutions in even more specific ways: QM, according to matrix mechanics, was to be a discrete, algebraic theory. This vision was undermined by the triumph of Schr"odinger's differential equation. Devising a new quantization procedure for field theory, now offered the possibility of undoing this historical error. Outlining his vision of a new quantization procedure, Heisenberg further wrote:

\begin{quote}
[T]he particles of the future theory will not be smaller than the universal length, it will not even by possible to talk of a local interaction with a precision greater than the universal length (for example, I do not see the necessity for having a differential equation for $\psi$) [...]
\end{quote}

To which Pauli remarked in the margins: ``Lorentz Group.'' Relativity remained the bugbear of the fundamental length, as it had already been in 1930. In fact, it might well be argued that relativity already was the bugbear of matrix mechanics: Aside from a few awkward attempts by Dirac to turn time into a matrix, the union of quantum mechanics and relativity only began in earnest when the Schr"odinger equation opened up the possibility of constructing generalized, relativistic wave equations.

\subsection{The Beginnings of S-Matrix Theory}

In early 1937, Heisenberg began to formulate a program that addressed this central difficulty by reformulating QFT in such a manner that the Lorentz invariance was more clearly apparent. The usual methods, starting with Heisenberg and Pauli's initial formulation of 1929 were not explicitly covariant, and relativistic invariance needed to be proven in quite a roundabout manner. Ideally, this reformulation would move away from the usual Schr"odinger equation, which Heisenberg felt would have no place in the future theory. The reformulation would then be modified by integrating the fundamental length in such a manner that the equivalence with the regular Schr"odinger equation was lost, thereby establishing a new quantization formalism that no longer relied on continuous functions and differential equations, but still guaranteed relativistic invariance.

It was already well-known that QFT could be formulated in a (more) covariant fashion by moving to the interaction picture. This had been established in 1932 through the work of Dirac, Rosenfeld, Vladimir Fock and Boris Podolsky \citep{dirac_1932_relativistic, rosenfeld_1932_uber-eine-mogliche, dirac_1932_on-quantum}. The interaction picture by itself did not provide full relativistic invariance; that also required the introduction of Dirac's many-time formalism, which, however, worked only for a finite number of electrons and not for electron-positron theory. This defect was only cured by Tomonaga's introduction of super-many-times theory \citep{tomonaga_1946_on-a-relativistically}. In 1937, the universal-time Schr"odinger equation in the interaction picture was the most covariant formulation of QFT, as given in a letter from Heisenberg to Pauli of 16 January 1937,

\begin{equation}
i \frac{\partial \varphi}{\partial t} = \overline{H_1} \varphi = \left( \int H_1 dV \right) \varphi
\end{equation}

where $H_1$ is the Hamiltonian density in the interaction picture, i.e., just the interaction term, which is a relativistic invariant. The equation itself, however, is plainly non-covariant, since it makes explicit reference both to time and to space (in the volume integral).

It should be noted that the interaction picture, as introduced by Dirac, only had a time-dependent (and no time-independent) Schr"odinger equation. This explains Heisenberg's starting point - in his entirely non-covariant 1929 formulation with Pauli, they had taken the time-independent Schr"odinger equation as the central equation \citep{heisenberg_1929_zur-quantendynamik}. In addition, his central concern at the time was scattering processes in cosmic rays (the explosions), something which could be addressed with a time-dependent Schr"odinger equation, even if the usual treatment (see, e.g., the second, 1935, edition of Dirac's textbook \citep{dirac_1935_the-principles}) relied on the time-independent Schr"odinger equation. We see here for the first time, how demands of relativistic invariance joined with a new experimental focus (from the spectroscopic data of the first decades of the 20th century to the cosmic ray physics of the 1930s) to move the theory from the stationary to the dynamical. As to what this meant for determining the energies of stationary states, we will return to this question soon.

For the time being, Heisenberg set himself to rewriting this equation, using the time evolution operator for an infinitesimal time step $dt$:

\begin{equation}
\varphi (t + dt) = e^{-i \left( \int H_1 dV \right) dt} \varphi(t) = \prod_V e^{-i H_1 dV dt} \varphi(t)
\end{equation}

where in the last step, Heisenberg used the notion of the integral as an infinite (Riemann) sum to factorize the exponential, as well as the fact that the operators $H_1$ at different points in space commute. This could then be generalized to non-infinitesimal time evolution, by letting the product go not only over all of space, but over the entire space-time volume $\Omega$ between initial time $t_0$ and final time $t_1$:

\begin{equation}
\varphi(t_1) = \prod_{\Omega} e^{-i H_1 dx dy dz dt} \varphi(t_0) = \prod_{\Omega} e^{-i H_1 d\omega} \varphi(t_0)
\end{equation}

where $d\omega$ is the space-time volume element. This equation even allowed, as Heisenberg hinted at in his letter, to take $\Omega$ as the volume between two arbitrary space-like surfaces, thereby achieving a fully covariant formulation. This is doubtlessly a very elegant procedure, foreshadowing the introduction of arbitrary space-like surfaces by Tomonaga and Schwinger a decade later. Pauli was enthused (Letter to Heisenberg, 19 January 1937): ``I find your equation [...] very pretty.''

But, of course, it was still simply a reformulation of the regular Schr"odinger equation. Heisenberg had high hopes that it might provide the right starting point for the development of a new quantization procedure, which would also take into account the fundamental length. His program was to modify the time-evolution operator, in order to get a theory which remained relativistically invariant but was also divergence-free. These modifications did not have to involve the fundamental length directly. As we have seen, Heisenberg believed that the fundamental length would already appear in the classical (Fermi) field theory and would then appear in the quantum theory in the correct manner, if one applied the new quantization procedure.  The quantization procedure and the fundamental length were thus not necessarily connected.

Heisenberg's first attempt at a new quantization procedure (January 1937) did still involve the fundamental length, which appeared directly in a modified time evolution operator,  mimicking the well-established but ``horrible'' (\emph{scheu"sliche}) cut-off procedures. These attempts soon floundered (Postcard to Pauli, 27 January 1937). Heisenberg's next attempt (February to April 1937) already no longer directly involved the fundamental length and instead consisted in eliminating certain terms in the perturbation expansion of the time evolution operator, in order to avoid the divergences. Again, Heisenberg soon found the approach to be insufficient: It turned out to be impossible to simultaneously uphold relativistic invariance and remove the divergences (Letter to Pauli, 26 April 1937). 

The details of these failed attempts need not concern us here. More important for us, are the general considerations which Heisenberg connected with these attempts (16 January 1937):

\begin{quote}
[W]hat I like about the discussed proposal is that one gives up the notion of a ``wave function at a certain position'' and only introduces the notion of ``a particle with a certain momentum.''
\end{quote}

This is a bit confusing: Heisenberg appears to be referring to the fact that the wave function $\varphi$ is defined in occupation number space, rather than in (many-particle) configuration space. This was the usual procedure in QFT and made for much more tractable mathematics. However, the equivalence between the two formalisms was well-known and Pauli was consequently not so much confused, but rather, in typical Pauli fashion, enraged (19 January 1937):

\begin{quote}
I do \emph{not at all} agree with the physical viewpoint that the notion ``particle with a certain momentum'' is physically better than the notion ``wave function at a certain position''. For the problem of the former notion is that one demands the momentum of the particle \emph{at a certain, precise time} $t$ (while momentum measurements can never be performed in an arbitrarily short time). Only in the force-free case, where the momenta are time-independent, is this not a problem; as soon as one has interaction terms, one has the same difficulties for $\varphi (N_k, M_k, t)$ [i.e., the wave function in occupation number space] as for $\varphi(x,t)^n$ [i.e., the wave function in $n$-particle configuration space].
 \end{quote}
 
As is often the case in the Heisenberg-Pauli correspondence, it is not entirely clear if Heisenberg had just been imprecise in his wording, or whether he only really thought things through after Pauli's harsh response to his half-baked musings.\footnote{Many years later, Pauli would characterize Heisenberg in the following manner: ``Heisenberg's psychology is that whenever he is refuted, he modifies his assumptions.'' (Letter to K"all\'en, 4 January 1957, \citep{meyenn_2005_wissenschaftlicher})} In any case, he responded on 21 January 1937:

\begin{quote}
I am, by the way, in total agreement with you that one can only speak of a precisely specified momentum for \emph{free} particles, where the measurement can take as long as one wants. My considerations concerning $\prod_{\omega} e^{-i H_1 d \omega}$ [given the notation of his earlier letter, this should of course have been $\prod_{\Omega} e^{-i H_1 d \omega}$]  actually had their starting point in the desire to replace this ``differential'' operator by an integral one, from which one can then only deduce cross sections for the transitions of free particles. But I have not been able to come up with a sensible generalization.
\end{quote}

Since this is basically already the program for the S-Matrix, it is time to briefly take stock and consider again how Heisenberg arrived at this point. A combination of the demand for relativistic invariance (which was the central challenge for a theory of the fundamental length) and an interest in scattering processes, rather than spectroscopy, had led him to a formulation of QFT with the time evolution operator at its center. The next step was now supposed to be the move from the ``differential'' time evolution operator to an integral scattering operator. Such an operator would contain all that was necessary to describe a scattering event. It could of course be calculated in the regular QFT, based on differential equations, but would lead to divergent results. Heisenberg was now looking for a generalization, i.e., a modified calculational procedure, which could not be reduced back to differential equations, which Heisenberg felt were in conflict with a fundamental length. In particular, we see that the instantaneous state, in the form of a Schr"odinger wave function, would be eliminated from such a theory, which would only speak about free asymptotic states.

Heisenberg did not further pursue this approach at the time. In fact, he entirely abandoned his attempts at a reformulation of QFT after the first few had proven unsuccessful. In spring 1937, Heisenberg was increasingly on the defensive concerning the fundamental length and the theory of explosions. New experimental results implied that for low energies the supposed explosions could in fact be understood as electromagnetic cascades  \citep{cassidy_1981_cosmic}. Also, first hints were showing up that the high energy showers could be understood as resulting from the existence of  ``heavy electrons'' and were also electromagnetic in nature (i.e., only involved photons, electrons, and positrons) \citep{galison_1983_the-discovery}. This invalidated Heisenberg's approach in a twofold manner: Nuclear forces (i.e., Fermi's theory of $\beta$ decay) did not seem to be involved in shower production at all, and the showers were no longer viewed as single explosive events.  It also became increasingly doubtful whether the theory of $\beta$ decay was to involve a dimensionful coupling constant at all, when the apparent confirmation of Yukawa's meson theory implied that $\beta$ decay was not actually a four-fermion interaction, but rather mediated by a heavy boson.

How did these developments affect Heisenberg's research program? The fundamental length and its manifestation in the form of cosmic ray explosions (which now, however, had to be taken as a lot less common than originally envisioned) remained cornerstones of Heisenberg's vision for a new theory. This is not true for the other tenet of Heisenberg's work in early 1937, the assertion that the correct semi-classical field theory was already known and one only needed to cook up a new and improved quantization procedure. As already mentioned, the fundamental nature of Fermi's theory of four-fermion interactions was increasingly called into doubt. At the same time, the meson theories, which were supposed to replace it, also ran into ever greater difficulties. After an initial optimism, it became increasingly clear that they were unable to reconcile nuclear and cosmic ray phenomenology. This conundrum was only resolved after the war, with the conceptual separation between the original cosmic ray meson (the muon) and the carrier of the nuclear force (the Yukawa meson or pion).

When Heisenberg thus returned to attempts at constructing a new relativistic quantum field theory in the isolation of wartime, it was no longer with the aim of merely modifying the quantization procedure. Instead, the idea was now to construct a new theory from scratch. This brought his old idea of an integral scattering operator back on the table. 

The motivations (elimination of states and differential equations in order to account for the fundamental length, central role of scattering processes in order to describe explosions) were still valid. And at the same time, his former difficulties were now moot: He had found himself unable to find a suitable generalization of the procedure which led from a correct semi-classical field theory to a quantum time-evolution operator. But there was now anyway no trustworthy semi-classical field theory to start from. So Heisenberg now proposed to take the integral scattering operator (now: S-Matrix) itself as the starting point of the theory.

This was a radical move away from the central ideas of quantum theory, which since Bohr's correspondence principle had always relied on relating classical and quantum theories in some way or another. It was to bring with it major difficulties, which we will discuss in detail. Above all, it needed a new philosophical underpinning. Here again, Heisenberg turned to the glory days of the creation of quantum mechanics. The positivist program of constructing a theory including only observable quantities, which he had laid out in the introduction to his ``Umdeutung'' paper \citep{heisenberg_1925_umdeutung}, was in fact logically distinct from the idea of transferring mathematical structures from a corresponding classical theory. Heisenberg could thus divorce these ideas from the theory of quantum mechanics, to which they had originally led, and re-present them as the basis for his new S-Matrix theory, which was first presented in two papers titled `\emph{Die beobachtbaren Gr"o"sen in der Theorie der Elementarteilchen}'' (``The observable quantities in the theory of elementary particles''), published on the height of World War II \citep{heisenberg_1943_die-beobachtbaren, heisenberg_1943_die-beobachtbarenb}.\footnote{As a curious aside, in the title of the first paper ``observable quantities'' is set in quotation marks, in the title of the second paper it isn't.} It is to these papers that we now turn.

\subsection{Heisenberg's S-Matrix Theory}

The first thing one needs to note concerning Heisenberg's S-Matrix papers is that they do not constitute an actual, full-fledged physical theory. This can be immediately understood from the abandonment of the correspondence idea. In QM, one could write down the abstract Schr"odinger equation

\begin{equation}
H \psi = E \psi
\end{equation}

as a purely quantum equation, but it was the correspondence with a classical theory that provided the input on what sort of an operator $H$ really was. One can, however, learn a lot about the structure of QM without ever invoking a specific Hamiltonian. Heisenberg had something similar in mind for his S-Matrix theory. And because the elements of the S-Matrix should be observables and have an immediate and clear physical interpretation, the general, structural insights into the new theory should be ``at least in principle empirically testable'' \citep[p.514]{heisenberg_1943_die-beobachtbaren}.

No direct physical predictions emerged from Heisenberg's work on the S-Matrix, so this statement should be understood as attempting to reconcile the two almost contradictory aspects of this work: On the one hand, it is a purely formal scheme, on the other hand, it is supposed to be an expression of positivist empiricism, making reference only to observable quantities.

For the time being, Heisenberg had to leave open the question of how the scheme of S-Matrix theory was to be filled with physical content. The more immediate question was what actually were to be the observable quantities on which the theory was based; or rather, which quantities were to be considered \emph{un}observable. Heisenberg had learned his lesson from the debates following the \emph{Umdeutung}. In 1926, he had presented his work in Berlin, followed by a discussion with Einstein, which had impressed him so much that he included it in his autobiography \citep{heisenberg_1969_der-teil}. Einstein had said to him:

\begin{quote}
[I]t may be heuristically useful, to remind oneself, what one is really observing. But from a fundamental standpoint it is entirely wrong, to want to build a theory only on observable quantities. For in reality it is exactly the other way around. It is only the theory that decides, what one can observe.
\end{quote}

This sentiment, which is missing in the \emph{Umdeutung}, is echoed in the introduction to Heisenberg's first S-Matrix paper:

\begin{quote}
[A]lso the future theory should of course primarily contain relations between ``observable quantities.'' Of course only the final theory will determine which quantities are really observable.
\end{quote}

Now, there was no ``final theory.'' But, Heisenberg argued, one knew something about this future theory, namely that it would contain a fundamental length, and this mere ``fact'' already clearly indicated which quantities would no longer be considered observable in the future theory:

\begin{quote}
The existence of a ``smallest length'', i.e., of a universal constant of the dimensions of a length and of the order of $10^{-13}$ cm, makes problematic all of those statements of quantum theory that deal with the precise determination of a position or of a point in time, i.e., with spatio-temporal processes in general. 
\end{quote}

Here is, in a nutshell, the argument for the abolishment of continuous fields and state functions, and the differential equations that govern their behavior. But what were the observable quantities that the new theory would be based on to be instead?  Heisenberg named two. One of them was clearly the (cosmic ray) scattering cross sections, which were derived from the elements of the S-Matrix and which were to describe, in particular, the occurrence of explosions. The other, and these could not be done away with, were the central observable quantities of the old theory, the energies of the stationary states.

Now, there seems to be somewhat of a contradiction here: On the one hand, the elevation of the scattering operator to the central entity of the new theory, with the concomitant abolishment of anything but asymptotic, free states. On the other hand, the wish to keep the central observable and calculable quantity of QM (and of the QFT of the time), the energy of stationary states. This notion was, however, obviously closely tied to the notion of state that was being abandoned. Heisenberg had felt this difficulty already in 1937 - at the time, he was still discussing the (differential) time evolution operator; but already here the discrepancy becomes apparent. On 14 February 1937, he had written to Pauli:

 \begin{quote}
 Of course one still has to find a connection [...] to the magnitude of the rest mass and to the question of the stationary states. I have not been able to obtain such a connection so far; I only have a few tentative attempts, but I do not know, whether they will work out.
 \end{quote}
 
 He voiced similar concerns in the first S-Matrix paper, five years later:
 
 \begin{quote}
 The two [...] kinds of observable quantities at first sight seem to be without inner connection, and one gets the impression that, for spatially enclosed systems with discrete eigenvalues on the one hand, and for non-enclosed collision and scattering processes with continuous energy spectrum on the other, one needs to treat entirely different quantities as ``observable'' .
 \end{quote}
 
But Heisenberg brushed away these concerns in 1943, and made a very vague claim concerning the equivalence of the two types of observables. Christian M\o ller, who began working on S-Matrix theory already during the war, right after the publication of Heisenberg's first papers, soon disproved what he took to be Heisenberg's claim: that ``the discrete energy values in the closed stationary states are at least partly determined by the S-matrix.'' He showed that this was not the case \citep[p. 18]{moller_1945_general}
 
The proof was simple: Start from a scattering potential that allows both scattering and bound states (the classical example of course being a Coulomb potential). In QM this corresponds to a Hamiltonian with both continuous (scattering) and discrete (bound state) eigenvalues. Now, assume we have solved the scattering problem completely, i.e., determined all the eigenfunctions belonging to the continuous eigenvalues. This is of course a lot \emph{more} information than is contained in the S-Matrix, which is only concerned with the asymptotic behavior of these eigenfunctions. Still, all  this allows us to do is determine the eigenspace corresponding to the discrete eigenvalues of the Hamiltonian: It is the subspace of the total Hilbert space that is orthogonal to all the eigenfunctions belonging to continuous eigenvalues. But we can say nothing about the correct eigenbasis in this subspace, nor do we know anything about the corresponding eigenvalues. In other words, we have only partially diagonalized the Hamiltonian, and in order to know the bound state energies we need to fully diagonalize, i.e., explicitly solve the bound state problem. 

This argument is so simple that it is hard to believe that the statement being disproved is really what Heisenberg was claiming. And indeed, looking more closely at Heisenberg's claims of 1943, his claim does look somewhat different:

\begin{quote}
[It] is to be taken as observable the behavior of the in- and outgoing waves at infinity; i.e., in particular the phase difference between the in- and outgoing spherical wave belonging to a certain angular dependence (i.e., to a certain angular momentum of the system). If the system is now enclosed by a spherical shell at a great distance, the energy values of the system become discrete. These energy values, however, only depend on the phase difference between the in- and outgoing wave. If one considers the energy values of the thus modified system to be observable, this is the same as taking the phase difference to be observable [...]
\end{quote}

Heisenberg was thus not, as M\o ller read him, talking about the case where the QM Hamiltonian has both continuous and discrete eigenvalues at the same time. Rather, his starting point was an unenclosed system with continuous eigenvalues; he did not at all touch upon the question whether there is a discrete part of the spectrum corresponding to bound states as well. Instead, he went from this system to an enclosed system, in which the entire energy spectrum is discrete, and then argued that knowing the S-Matrix of the unenclosed system is equivalent to knowing the discrete energy spectrum of the enclosed system.

His argument, therefore, only applied, say, to the problem of a particle in an infinitely high potential well, but not to the more physical cases discussed by M\o ller, where the setup allows both for scattering and for bound states, such as a hydrogen atom. The general problem of bound state energies in a theory which did not deal in states was thus not addressed, even though Heisenberg's wording is very suggestive and certainly initially fooled M\o ller:

\begin{quote}
\emph{A priori} this seems possible, since the asymptotic form of the wave function in great distances, which determines the collision cross sections, depends chiefly on the form of the potential function in small distances, which again is essential for the position of the discrete energy levels. \citep[p.18]{moller_1945_general}
\end{quote}

Maybe Heisenberg even had himself fooled, since he entirely left out the question of bound state energies in his first paper, focusing only on developing the S-Matrix formalism as a relativistically invariant description of scattering processes.

While Heisenberg's theory is known as S-Matrix theory, the central mathematical quantity for Heisenberg was actually the matrix $\eta$, with $S = e^{i \eta}$. $\eta$ is invariant, just like $S$, and since $S$ is unitary, $\eta$ is hermitian. In his first paper, Heisenberg showed how in regular quantum field theory the matrix $\eta$ is related to the Hamiltonian $H$ in the interaction picture. This can easily be established, since the S-Matrix is simply the infinite time limit of the time evolution operator in the interaction picture. In the first order of perturbation theory, the expression for $\eta$ in terms of $H$ is simply

\begin{equation}
\eta = \int_{- \infty}^{\infty} dt H(t)
\end{equation}

For higher orders, $\eta$ is a more complicated functional of the Hamiltonian. What Heisenberg had wanted to demonstrated  was merely ``a certain interrelation between the matrix $\eta$ and the interaction energy'' (p.537). He used this interrelation in the second paper, where he attempted to construct first exemplary S-Matrices. Since the ties between S-Matrix and Hamiltonian established in quantum field theory were to be severed anyway, he argued that one might simply assume that the above perturbative expression was exact and take the resulting $\eta$ as a starting point for the construction of an autonomous S-Matrix theory.

The S-Matrix only describes transitions between states of the same momentum and energy. This meant that if the $\eta$ matrix is finite, so is the S-Matrix, even though it contains arbitrarily high powers of $\eta$. This is in contrast to usual Hamiltonian perturbation theory, where the terms including higher powers of the Hamiltonian contain infinite sums over intermediate states, which need not necessarily have the same energy as the initial and final states, only the same momentum. It was these sums over intermediate states that led to the infinities in quantum field theory. For the higher powers of $\eta$, the sums over intermediate states only went over states with a fixed energy. In modern parlance, the use of $\eta$ instead of the Hamiltonian ensured that all virtual particles are on-shell.

Heisenberg considered the simplest case of uncharged scalar particles interacting via a local interaction, what is nowadays known as $\varphi^4$ theory. The $\eta$ matrix was then:

\begin{equation}
\eta = \epsilon \int (d^4x) \varphi^4
\end{equation}

Now this expression is not finite, as Heisenberg well knew: It contains the infinite self-energy of the scalar particle $\varphi$ due to its self-interaction. But Heisenberg knew how to get rid of this: Normal ordering, i.e., moving all of the annihilation operators included in $\varphi$ to the right and all of the creation operators to the left.

Normal ordering had been used to eliminate divergences already in the earliest days of quantum field theory, when \cite{jordan_1927_zum-mehrkorperproblem} had used it to eliminate the Coulomb self energy of a non-relativistic scalar particle . Heisenberg had always been fond of the ``Klein-Jordan trick,'' but, in constructing QED together with Pauli in 1928/29, found that it did not work in a relativistic field theory.\footnote{See, e.g., a letter from Pauli to Dirac of 17 February 1928, reprinted in \citep{hermann_1979_wissenschaftlicher}.} Even if one enforced normal ordering at the level of the Hamiltonian, one would again have creation operators standing to the right of annihilation operators in higher orders of perturbation theory, leading to diverging integrals over intermediate virtual particles. But now in S-Matrix theory it worked perfectly: If one made the $\eta$ matrix finite by normal ordering, one was guaranteed a finite S-Matrix as well.

This simple ansatz for the $\eta$ matrix could then be gently modified, making sure that both its finiteness and the relativistic invariance were preserved. In this manner, Heisenberg constructed an S-Matrix which would lead to explosions, while at the same time being entirely finite. This S-Matrix was of course not even approximately derivable from a Hamiltonian operator and could thus not be regarded as the result of the quantization of any classical field theory. It was anyway not supposed to be a realistic model of nuclear interactions, for which Heisenberg would have had to go beyond cosmic ray scattering phenomenology and talk about nuclear bound state phenomenology (the deuteron was the classical test case for theories of nuclear interaction). But it furnished a proof of principle that the requirements he had set for a future theory already in 1936 could be met by an S-Matrix theory.

Despite the war, Heisenberg's papers reached physicists all across the world, even his old  pal Pauli in Princeton. Unable to communicate with Heisenberg directly, he instead voiced his usual criticism in a letter to Paul Dirac (21 December 1943),\footnote{All Pauli letters from the 1940s are to be found in \citep{meyenn_1993_wissenschaftlicher}.} pointing out the two main difficulties of the S-Matrix program:

\begin{quote}
Heisenberg did not try to give any theoretical formalism which determines his matrix. Even if such a formalism can be found, other problems would still exist to which the S-matrix is not adapted, as for instance properties of stationary states of compound systems. (For instance the stationary-states of the deuteron as a consequence of the interaction of the protons and the neutrons with the meson-field.)
\end{quote}

I will discuss the attempts to solve these two problems in turn, beginning with the problem of the stationary states. In fact, at about the same time that Pauli was writing his letter, this second problem was being addressed in war-torn Europe by Heisenberg and Hendrik Kramers.

\subsection{The Analytic S-Matrix}

The origin of the study of the analytic properties of scattering amplitudes lies in the study of nuclear collisions in the mid-1930s. In nuclear collisions, in particular in the scattering of neutrons off heavy nuclei, one frequently observed anomalously large cross sections, that is cross sections much larger than the geometric cross section of the scattering nucleus, despite the fact that the nuclear forces were assumed to be short-range, extending not much outside the  nucleus.\footnote{See in particular the excellent contemporary review article by Hans \citet{bethe_1937_nuclear}.} These effects soon came to be understood as resulting from resonance: The total energy of the incident particle $P$ and the scattering nucleus $A$ is approximately equal to an energy level $r$ of the whole system, taken as a many-particle system including both the incident particle and the scattering nucleus. This many-body system was called the ``compound nucleus,'' $C$. The level $r$ is unstable and decays under the emission of some particle $Q$, which typically is not of the same type as the incident particle, leaving behind a residual nucleus $B$. 

This was conceptually very similar to the case of optical dispersion: Here the incoming photon is absorbed, lifting the scatterer into a virtual excited state, which subsequently de-excites, emitting the final, scattered photon. And here, too, one observed anomalous dispersion (and strong absorption) if the frequency of the incident light was equal to a transition frequency of the scattering atoms, or, in particle language, if the energy of the incident photon was approximately equal to the energy difference between the ground state and the excited state of the atom. The equation for the nuclear cross section devised in this picture (first by Breit and Wigner for the case of one resonance level, then generalized by Bethe and Placzek for many resonance levels) consequently carried the name of ``nuclear dispersion equation.'' For the simple case of non-degenerate resonance levels, the equation for the cross section for the absorption of $P$ and the emission of $Q$ was of the form:

\begin{equation}
\sigma = 4 \pi^3 \frac{\hbar^2}{2 M E_P} \left\vert \sum_r \frac{H^{AP}_{Cr} H^{Cr}_{BQ}}{(E_P - E_{Pr}) + \frac{1}{2} i \gamma_r} \right\vert^2
\end{equation}

Here, $E_P$ is the kinetic energy of the incident particle and $E_{Pr}$ is the energy at which resonance with the state $r$ is observed, i.e., where the sum of the energy of the incident particle and of the scattering nucleus $A$ is just equal to the energy of the state $r$ of the compound nucleus. $H^{AP}_{Cr}$ is the matrix element of the Hamiltonian for a transition from the initial state (nucleus $A$, incident particle $P$) to the state $r$ of the compound nucleus. Similarly, $H^{Cr}_{BQ}$ is the matrix element for the transition from the state $r$ of the compound nucleus to the final state (residual nucleus $B$, outgoing particle $Q$). Finally, $\gamma_r$ is the width of the energy level $r$ and $M$ is the reduced mass of the incident particle and the nucleus $A$, considered as a two-body system.

Just as in the calculation of optical dispersion, the nuclear dispersion equation can be, and at first was, derived quite simply using second-order perturbation theory. However, there was a huge difference. For optical dispersion, the relevant interaction between the electromagnetic radiation and an atom can be considered small and the use of perturbation theory is unproblematic (except of course for the divergences appearing at higher order in perturbation theory). For nuclear dispersion, on the other hand, the interaction was assumed to be large. Indeed, this was why one couldn't just describe the problem as the motion of the incident particle in the potential of the nucleus - the interaction was so strong, i.e., there was so much energy exchange going on as soon as the incident particle entered the range, that it had to be described as a many-body process. But this meant that perturbation theory should not actually apply and its use could only be legitimized with mathematical tricks. Alternative derivations of the nuclear dispersion relation were sought that did not rely on perturbation theory. In particular this implied a new interpretation of the resonance energies: In perturbation theory they were given by the stable energy levels of the hypothetical unperturbed compound nucleus. This unperturbed compound nucleus was, however, a mathematical artifact, since the nuclear interactions that made the compound nucleus unstable under emission processes could not be considered small.

A non-perturbative derivation of the nuclear dispersion formula was presented by Arnold \citet{siegert_1939_on-the-derivation}, a German Jew who had actually been a PhD student of Heisenberg back in 1934, but was now working with Felix Bloch in Stanford \citep{dresden_1997_arnold}. Rather than study the full many-body problem of the compound nucleus, he merely treated the toy example of a particle in a short-ranged potential, where a close analog of the dispersion equation could be found for the elastic scattering of the s-wave, i.e., of an incoming particle with zero angular momentum, or, in wave mechanical language, an incoming spherical wave.\footnote{\label{toy} In a sense this was a step backwards, for this is how nuclear scattering had been conceptualized before the work of Bohr, Breit and Wigner had brought home the point that the strong nuclear forces implied that one would always have to take into account the full many-body context. But recently, \cite{kapur_1938_the-dispersion}, in a related attempt to derive the nuclear dispersion equation, had shown how  results from potential scattering could be straightforwardly generalized to the many-body case. They had, thus, first laid out their proof in the simple one-body framework, arguing that ``[a]lthough the complete proof of the dispersion formula is simple in principle, it is obscured by the rather complicated notation which is necessary for the general case.'' This can be viewed as a precursor of developments, which were identified as happening about a decade later by Res Jost (cited in \citep[p. 51]{cushing_1990_theory}), where ``non relativistic wave mechanics serves, not as an analytic tool to calculate some experimental cross section, but rather as an experimental playground for the discovery of general relationships, which might also be useful elsewhere.''}  Siegert realized from the dispersion equation that if one considered the cross section $\sigma$ as a function of a complex $E_P$, the resonance energies would appear as the real part of those energies for which the denominator vanished, i.e., for which the cross section became singular.  He thus wrote down a very general expression for the scattering cross section in his simple toy model, without making any approximations, and studied its singular points, when considered as a function of a complex $E_P$. It turned out that the complex $E_P$ that made the cross section singular were energy eigenvalues of the regular, time-independent Schr\"{o}dinger equation for the particle in the potential, but with special boundary conditions, namely that there were no incoming, only outgoing, waves.\footnote{It briefly needs to be explained how complex eigenvalues can arise as eigenvalues of a regular and supposedly hermitian Hamiltonian operator. The answer is that the Hamiltonian for a particle in a potential is hermitian only with regard to a given space of functions on which it acts. This space is different for different boundary conditions, and for the special case of only outgoing waves the Hamiltonian turns out not to be hermitian and hence can also have complex eigenvalues. Due to this fact the solution of the time-independent Schr\"{o}dinger equation also does not fulfill the continuity equation, which can readily be understood from the fact that the boundary conditions impose outgoing, but no incoming, waves.}

Complex energy eigenvalues had already been introduced more than a decade earlier by \citet{gamow_1928_zur-quantentheorie}, in order to describe the alpha decay of a nucleus, which he had also modeled as the motion of the alpha particle in the nuclear potential (a superposition of the nuclear force and the electromagnetic Coulomb potential).\footnote{Back then, however, this had not been meant as a toy model, but was supposed to give quantitative predictions, cf. footnote \ref{toy}.} He had found that if he imposed the same boundary conditions that Siegert would later obtain, i.e., only outgoing waves, he was forced to introduce complex energy eigenvalues. How were these complex energy eigenvalues now to be interpreted? The real part is the energy of the particle, but the state it is in is unstable, because there is the possibility of radioactive decay and, thus, of the particle escaping to infinity. This is mirrored by the fact that the overall amplitude of the wave function decreases exponentially, with a damping factor equal to the imaginary part of the complex energy.

Siegert could thus offer an interpretation of the denominator of the dispersion relation that did not depend on perturbation theory: The $E_{Pr}$ were the (discrete, as it turned out) energies of the unstable (radioactive) states in the sense of Gamow, and the $\gamma_r$ were the corresponding radioactive decay constants. The important point for the S-Matrix program is now that the whole argument could of course be inverted: Siegert had shown that one could obtain the singular points of the cross section by solving the Schr\"{o}dinger equation for radioactive Gamow states. But of course one could also take the expression for the cross section (taken as a function of complex energy) as primary and obtain the energies and decay constants of the Gamow states as the real and imaginary parts of those complex energy values for which the scattering cross section diverged.

The first one to make this inversion appears to have been Siegfried Adolf Wouthuysen, assistant to Hendrik Kramers in Leiden. Beginning to work on nuclear dispersion at the behest of Kramers in 1940, Wouthuysen studied Siegert's work and was intrigued by the possibility of treating the scattering cross section as a function of a complex energy. So far, only half of the complex plane had been taken into consideration, since (for a particle in a potential) not only the genuine scattering energies, but also the energies of the radioactive states were always positive, ensuring that the particles in question could actually escape to infinity. Wouthuysen now studied the behavior of the cross section (and the simpler scattering amplitudes) as a function of an arbitrary complex energy (also allowing for a negative real part), i.e., as an analytic function defined on the entire complex plane.\footnote{The term ``analytic'' is used here, as in the physics literature of the time, somewhat loosely to simply denote a function of real variables that has been extended to a halfway well-behaved (some singularities are totally fine and indeed necessary) function of complex variables through an unambiguous process of analytic continuation.}

Wouthuysen's notes are not extant,\footnote{On 1 December 1943, Kramers wrote to Heisenberg (Heisenberg Papers, Archive of the Max Planck Society, Berlin, Folder \emph{Correspondence 1933-1948}; I will be referring to this folder simply as ``Heisenberg Papers'' in the following) that he was ``not able to find'' the ``old notes concerning the calculations by Wouthuysen and myself on the analytic properties.'' It seems probable that he did manage to get ahold of them later, though. Wouthuysen recalled (cited in \citep[p. 454]{dresden_1987_h}) that Kramers's oldest daughter Suus transported ``calculations about poles in scattering amplitudes, from Wouthuysen to Kramers,'' when Wouthuysen was in hiding. And indeed, Kramers's next letter to Heisenberg (12 April 1944, Heisenberg Papers) contains a few more details on Wouthuysen's work in the margins, implying that he may have received Wouthuysen's calculations after the letter had already been typed.} nor are there notes on a seminar he gave on the subject in 1942. None of his work was ever published and in Summer of 1942 he went into hiding from the Nazis. We can thus only rely on his own recollections (cited in \citep[p.38-39]{cushing_1990_theory}) and circumstantial evidence to establish what exactly Wouthuysen calculated and how far he got. In particular it is hard to assess his remark that he could ``show the analyticity of the Schr\"{o}dinger scattering amplitude,'' i.e., we cannot say how formal and general a proof he had for the viability of his procedure. However, we can be fairly certain as to what his method was and which examples he applied it to.

Wouthuysen was studying the scattering of a particle in a simple spherically symmetric potential,\footnote{See letter from Kramers to Heisenberg, 1 December 1943 (Heisenberg Papers).} i.e., the same toy model as Siegert. He was using partial wave analysis, i.e., a decomposition of the wave function into terms with a certain angular momentum and thus a definite dependence on the angles in spherical coordinates, this dependence being given by the spherical harmonics.\footnote{In the margins of a letter from Kramers to Heisenberg dated 12 April 1944 (Heisenberg Papers), Kramers remarks: ``4 years ago, together with Wouthuysen, we were thinking of what you now call the `S-Matrix in diagonal form.' '' This is equivalent to a partial wave analysis. The same letter is also in the Kramers Correspondence in the Archive for the History of Quantum Physics, but without these handwritten marginal notes. Kramers apparently did not add these notes to the copy of the letter that he kept.} Wouthuysen's results can most easily be demonstrated by considering only the term with vanishing angular momentum, as Siegert had done. For the scattering problem, the asymptotic form of the wave function is then

\begin{equation}
\label{eq:asymptotic}
\psi{r} = \frac{I}{r} \left(e^{-ikr} + S (k) e^{ikr} \right)
 \end{equation}
 
Where the first summand is the incoming spherical wave (amplitude $I$) and the second summand is the scattered, outgoing wave (also spherical, due to angular momentum conservation, which makes the partial wave analysis possible in the first place) with amplitude $I S(k)$, which depends on the momentum $k$ of the incoming particle. The function $S(k)$ is the scattering amplitude, which Wouthuysen studied as an analytic function. He then discovered that for some complex values of $k$ the scattering amplitude vanishes. Indeed this occurred for negative imaginary values of $k$, which correspond to negative (real) energies, and these negative energies for which $S(k)$ vanished were equal to the energies of the stable bound states of the potential.

It is actually fairly straightforward to see why this is the case: For a negative imaginary $k$, the first term in equation \ref{eq:asymptotic} is an asymptotic exponential decay, while the second summand is an asymptotic exponential increase. The bound state problem is now precisely to solve the Schr\"{o}dinger equation for the boundary condition where there is only asymptotic decay, i.e., where the coefficient of the exponentially increasing term vanishes. And this coefficient is of course the scattering amplitude taken as a function of the purely imaginary $k$. The whole matter is of course not quite as trivial as it might seem: One needs to take into account also non-vanishing angular momentum and also make sure that the extension to complex $k$ is in fact mathematically well-defined. But the fact remains that Wouthuysen's result was really just a mathematical curiosity. While Siegert's results had delivered a new physical interpretation of the resonance energies in terms of radioactive states, there was no need for such a reinterpretation in case of bound state energies. This may be a reason why nothing was ever published on the matter, even though Kramers found it ``interesting'' (according to Wouthuysen).\footnote{Even after Kramers realized the significance of the results for Heisenberg's S-Matrix program, he had no intention to publish them, stating that they merely were to be seen as a ``commentary.'' (Letter to Heisenberg, 1 December 1943, Heisenberg Papers) Of course the reasons for this decision may well have involved political considerations as well \citep[p. 458]{dresden_1987_h}.}

But when Heisenberg came to the occupied Netherlands in October 1943\footnote{For the details of this trip, see \citep{rechenberg_1989_the-early}.} and presented his work on the S-Matrix, Kramers realized that Wouthuysen's results might provide a way to integrate bound states into Heisenberg's theory. For the S-Matrix is really nothing more than a generalization of the one-particle scattering amplitude $S(k)$ to many-particle processes in which also the number of particles need not be conserved. The S-(sub)Matrix for a specific collision process could be viewed as a function of all the in- and outgoing momenta (i.e., of the asymptotic states) and could then also be taken as an analytic function of complex momentum values. Those complex values of the momenta for which the corresponding element of the scattering matrix vanished could consequently be related to the energies of bound states formed by the colliding particles. These appear to have been the general ideas that Kramers suggested to Heisenberg during his visit.

Heisenberg immediately set to work and on 10 January 1944\footnote{Letter and manuscript are in the Heisenberg Papers.} sent Kramers a draft of a new paper, which included a study of the analytic properties of the S-Matrix for systems consisting of two and three particles (without taking into account annihilation or creation processes). This restriction was the result of a major difficulty encountered by Heisenberg: The S-Matrices he had studied in his previous publication, in particular those that included explosions, could not be extended to the complex plane, due to ``discontinuities in the scattering coefficients which do not seem compatible with the analytical properties of S.'' Heisenberg thus cooked up a new expression for S, which had simpler properties when extended to the complex plane. This S-Matrix, however, only described the scattering of two particles and did not describe creation or annihilation processes, and so, of course, no explosions. Kramers later interpreted it as resulting from an interaction between the two particles with infinitely short range (though he could not actually treat this problem relativistically in the usual theory with a Schr\"{o}dinger equation), and indeed Heisenberg's new toy S-Matrix only has non-vanishing elements for the case where the two particles have no relative angular momentum, i.e., when there is a head-on collision. The corresponding scattering matrix element is a function of the momentum of the two particles in the center of mass system, which are of course equal in magnitude and opposite in direction. If the magnitude of the momentum is taken to be purely imaginary, one indeed gets a zero for the S-Matrix. The total energy of the two-particle system for this imaginary momentum turned out to be less than the sum of the rest energies and was consequently interpreted by Heisenberg as the energy of a two-particle bound state. 

This was quite an accomplishment. As Kramers wrote:\footnote{Letter to Heisenberg, 12 April 1944, Heisenberg Papers.}

\begin{quote}
I very much enjoyed the treatment of the simple two-particle model. There one has for the first time a really relativistically invariant stationary state of interacting particles. Very unintuitive [\emph{unanschaulich}], of course, but the whole model is wonderfully consistent.
\end{quote}

Heisenberg had shown that the method of analytic extension worked in principle and that it was possible to build a theory on scattering amplitudes and still be able to describe the energies of bound states. This was indeed a major shift in perspective for quantum theory. Up until then, scattering processes had to be described using the vocabulary of stationary states. The entire scattering event was described by one time-independent wave function with an energy equal to the energy of the incident particle and a specific set of boundary conditions. The scattering coefficients were then determined by identifying different terms in the wave function as incoming and outgoing waves and calculating ratios between the (squared) coefficients of these terms (see, e.g., \citep[chapter IX]{dirac_1935_the-principles}). This conceptual framework had now been inverted by Heisenberg. Certainly, the determination of the bound state energies through an analytic continuation was more \emph{unanschaulich} than regarding the scattering process as a stationary state: What, for example, was to be made of the non-zero values taken by the S-Matrix elements for complex momenta that did not belong to bound state energies? But the proof of principle that scattering could be taken to be the primary concept in a theory of elementary particles, and consequently that a theory which only made reference to asymptotic states could still have something to say about the spectral lines of hydrogen, say, was immensely important for the acceptance of the S-Matrix framework.

Heisenberg's study of the analytic properties of the S-Matrix did not get much further than this proof of principle, though. As already mentioned, the toy models he had constructed before Kramers's suggestion had all turned out not to be compatible with the idea of analytic continuation. And it turned out to be extremely difficult to cook up further examples beyond the very simple two-body theory with contact forces. In the paper he had sent to Kramers, he had tried his hand at a 3-body model, but realized, when the paper was already in print, that his treatment was flawed (see the footnote on p. 98 of the published version \citep{heisenberg_1944_die-beobachtbaren}). He had constructed the phase matrix $\eta$ (recall that $S=e^{i \eta}$) as the sum of the three two-body phase matrices. But, in the unpublished fourth paper on the S-Matrix,\footnote{The manuscript can be found in the Heisenberg Papers and is reprinted in Heisenberg's collected works.} he had to concede that this would only lead to two-body forces, in a much stricter sense than for a potential written as a sum of two-body potentials. For his proposed three-body S-Matrix there was only a two-body bound state, and this bound state was then saturated, i.e., it did not interact with the third particle \emph{at all}. At the end of this manuscript, which he had finished during the war and then left in Switzerland, he had to conclude that the problem of constructing an S-Matrix for a true many-body system was still shrouded in ``impenetrable darkness.'' For the time being, Heisenberg's two-body contact force model (and its trivial, additive extensions to many particles) remained the only known analytic S-Matrix that was not derived from a regular Hamiltonian.

Already here Heisenberg's interest was starting to peter out. Years later, in discussions at the 1961 Solvay conference \citep[p.174]{solvay_1961_la-theorie}, he explained to Geoffrey Chew why he had abandoned S-Matrix theory:

\begin{quote}
I found it very difficult to construct analytical S-matrices. The only simple way of getting (or guessing) the correct analytical behavior seemed to be a deduction from a Hamiltonian in the old-fashioned manner.
\end{quote}

And also M\o ller in his papers was not able to go beyond simple wave-mechanical toy models and the two-body problem. Still, when Heisenberg's third paper and, more importantly, M\o ller's papers, which discussed the question of the analytical continuation in great but abstract detail, became widely available after the war, they caused somewhat of an S-Matrix hype. 1946 was, as Rechenberg has called it, \emph{the} year of S-Matrix theory \citep[p. 564]{rechenberg_1989_the-early}. And indeed, this year saw several proposals for filling the frame given by Heisenberg. Some physicists proposed introducing further restrictions on the S-Matrix, besides the established conditions of unitarity and relativistic invariance, hoping that this might lead in fact to a full determination without recourse to a Hamiltonian. Max Born enthusiastically embraced Heisenberg's general view:

\begin{quote}
I maintain that relativistic quantum mechanics has not the task of describing the world at large, but single collisions. All other physical phenomena are to be regarded as results of one or several collisions. \citep[p. 15]{born_1947_relativistic}
\end{quote}

He then went on to propose that the principle of reciprocity between momentum and coordinate variables, which he had already outlined before the war \citep{born_1938_a-suggestion}, should serve as ``the missing principle [...] which is necessary for narrowing down the theory'' \citep[p. 17]{born_1947_relativistic}. Ralph Kronig proposed to impose relations between the real and the imaginary parts of the S-Matrix elements in analogy to the Kramers-Kronig relations in optics, which could be derived quite generally from a principle of causality \citep{kronig_1946_a-supplementary}. These suggestions of Kronig were, however, not taken up in earnest until about 1950 \citep[p. 57ff]{cushing_1990_theory}.\footnote{Rechenberg mentions some further examples: Pascual Jordan's work on a connection between the S-Matrix and an operator theory in a von Neumann space and model calculations by Harald Wergeland. These attempts are documented by letters sent to Heisenberg in 1946, which I have, however, been unable to locate in Heisenberg's Papers.}

Two attempts went further, and actually led to specific rules for the construction of the S-Matrix. In the following, I will discuss them in turn. The first, Heitler's damping theory, is not essential for the overall story, but it provides necessary contextualization for the second, Stueckelberg's program. Stueckelberg's work shows how, from the S-Matrix, one could then return to regular field theory, but now a field theory that looked quite different. Even though it was not very influential, it does help demonstrate the generality of the historical development that led to Feynman-Dyson QED, which we discuss in the second half of the paper. But first, we turn to Heitler.

\subsection{Filling the Frame}

\subsubsection{Heitler's Damping Theory}

With the question of the bound states at least addressed (if not completely solved) by the analyticity considerations, the problem of constructing an S-Matrix definitely moved to the center of attention. Some simply criticized Heisenberg's scheme for its emptiness, such as (repeatedly) Pauli \citep[p. 564-567]{rechenberg_1989_the-early}, but also, e.g., Peierls (in a letter to Born, 14 June 1946, reprinted in \citep[p. 58]{lee_2009_sir}):

\begin{quote}
I am not at all impressed with Heisenberg's new scheme or M\o ller's work on the same subject. It seems to me quite an empty scheme which is completely indefinite until one formulates the laws to which his matrix S is subjected.
\end{quote}

Others took the openness of Heisenberg's scheme as an opportunity. Since QED and also the quantum field theoretical descriptions of nuclear mesons always led to divergent results in higher order of perturbation theory, one couldn't perform calculations from the Hamiltonian in the usual (quantum mechanical) manner. Rather, one had always had to devise some kind of workaround procedure to extract empirical predictions from the theory. Such ad hoc procedures could now be reinterpreted as methods for calculating the S-Matrix, as Heisenberg later pointed out during a lecture series to England in 1947 \citep[p. 20]{heisenberg_1949_the-present}:

\begin{quote}
[T]he tentative solutions which have been written down in relativistic wave-mechanics so far are actually not defined by a Hamiltonian, but by an S-Matrix. One has, in fact, as a rule started with a Hamiltonian and realized that it has no solutions; one has then used certain matrix elements out of this Hamiltonian to define, arbitrarily, scattering processes for the system concerned; in other words, one has chosen a certain S-Matrix for the system instead of the Hamiltonian, which does not give consistent solutions.
\end{quote}

One such workaround procedure was the damping theory of \citet{heitler_1941_the-influence} (worked out together with his assistant Peng \citep{heitler_1942_the-influence}, and also proposed independently by both \citet{wilson_1941_the-quantum} and \citet{gora_1942_quantentheorie}), which had been developed already before the publication of Heisenberg's S-Matrix papers. It was a theory for calculating scattering amplitudes, not because it assigned any epistemological priority to scattering, but because it was designed to address a very specific problem. At the time, one still identified the meson, discovered in cosmic rays in 1937, as the carrier of the nuclear force proposed by Yukawa in 1935. This, however, led to serious difficulties. In order to be the carrier of the nuclear force, the meson would have to have a substantial interaction with nucleons. But the observed cross sections for the scattering of cosmic ray mesons off nucleons turned out to be much too small.

This statement needs to be qualified: Just like in QED, the quantum field theoretical calculations in meson theory led to divergent results in higher orders of perturbation theory. In calculating the expected value for the meson-nucleon cross sections, one thus only went to the first perturbative order. Damping theory now was a method for calculating scattering cross sections going somewhat beyond this first order, by taking into account the effects of the radiation reaction (damping), without running afoul of the divergences, effectively (and here the relation to Heisenberg's S-Matrix already shows) by keeping all virtual particles close to the mass shell. In this way, results for nucleon-meson scattering were obtained that agreed with experiment a lot better, since, due to the strong coupling, higher-order effects led to significant changes in the results. The damping theory was also quite a general procedure and could equally well be applied to purely electrodynamical scattering processes, such as Compton scattering. When Heitler heard of Heisenberg's works, and in particular of the possibility of obtaining bound state energies through analytic continuation of the scattering amplitudes, he immediately realized that this might turn his limited rough-and-tumble procedure into a new foundational technique, as he stated in a letter to M\o ller (23 February 1946):\footnote{M\o ller papers, Niels Bohr Archives, Copenhagen.}

\begin{quote}
I am very interested in your paper on Heisenberg's theory. In particular the remark [...] concerning the eigenvalues of closed states fascinates me. [If] it is really true that one can determine all eigenvalues from the S-matrix (defined for collisions only) by extending S into the complex plane, then, I think, one could follow up the following preliminary program: One could for instance substitute for S the matrix which one obtains from Peng's and mine theory of damping, which is know to give at least reasonable results for collisions which seems to agree more or less well with the experiments. But then also all eigenvalues are determined, for instance - the self-energie (sic), the magnetic moments of a proton and neutron. (The extension of our damping theory to such problems has occupied Peng and me for years [...])
\end{quote}

Pauli had already praised Heitler's work before hearing of the S-Matrix, but had bemoaned its inapplicability to ``static problems'' (Letter to Rosenfeld, 20 July 1945).\footnote{All letters from and to Pauli in the 1940s are reprinted in \cite{meyenn_1993_wissenschaftlicher}.} When analyticity promised to solve this problem, Pauli immediately endorsed damping theory as the most promising approach for determining the S-Matrix (Letter to M\o ller, 20 May 1946):

\begin{quote}
All depends on how to determine the S-Matrix. I still think that Heitler-Wilson's rules which can be proved to be Lorentz-invariant, are the best what is done in this direction until now.
\end{quote}

It was in this spirit that Heitler presented his damping theory at the 1946 Cambridge Conference on Fundamental Particles (22 to 27 July), where the ``problem of how to determine the S-Matrix'' (M\o ller to Heisenberg, 7 July 1946, Heisenberg Papers) was one of the main points of discussion. Heitler realized that, contrary to Heisenberg's original idea, he \emph{was} actually using the Hamiltonian to calculate the S-Matrix elements, he just wasn't using it in the usual manner:

\begin{quote}
It is quite true then that S is determined, after all, by the Hamiltonian, in fact by the Hamiltonian of the present divergent theories. However, after having omitted he diverging parts [...], $H$ is no longer the actual Hamiltonian of the system in the usual sense. In fact it is doubtful whether in the theory of damping any Hamiltonian exists at all. \citep[p.193]{heitler_1947_the-quantum}
\end{quote}

Heitler continued working on his program through 1947, using the method of analytic continuation to calculate the energies of nucleon-meson bound states \citep{heitler_1947_proton}. While it was the best contender for a determination of the S-Matrix at the time, it was faced with serious difficulties, which led to it quickly being abandoned when renormalized QED was developed in the wake of the 1947 Shelter Island conference.

Some of these difficulties were problems of the S-Matrix approach in general. In 1946, S.T. Ma, Heitler's former collaborator, now with Pauli in Princeton, had found that for non-relativistic wave mechanics the scattering amplitude $S(k)$ for some potentials had zeroes in the complex plane which did not correspond to bound states \citep{ma_1946_redundant}. In full wave mechanics, these could easily be recognized, but if only $S(k)$ was given it was not immediately clear how to identify the false zeroes. Several attempts were made to find further conditions on the zeroes of the S-Matrix in order to distinguish false zeroes from true bound states, which, however, all turned out to be inconclusive. The only failsafe way to exclude them, was to have a potential with a strictly finite range, i.e., which vanished exactly beyond a certain distance \citep{jost_1947_uber-die-falschen}. The importance of the false zeroes for the abandonment of the S-Matrix program was stated quite strongly by \citet{grythe_1982_some}, but this has, rightly, been identified as an overstatement by \citet[p. 319]{cushing_1990_theory} and Rechenberg.\footnote{The main case mentioned by Grythe, Christian M\o ller, also had other reasons for turning to different topics at the time. In a letter to Pauli (13 November 1946), he wrote:  ``During the last months, I have been exclusively occupied by writing a book on oldfashioned relativity for the Clarendon Press; thus I have left the S-Matrix for the moment.'' The book was not published until several years later \citep{moller_1952_the-theory}.} The question of the redundant zeroes was rather a mathematical curiosity,\footnote{Res Jost, the man who had published the (for the time being) definitive last word on the subject, himself later remembered how his work on potential scattering, which developed out of his study of the redundant zeroes, made him somewhat of an outsider in the physics community of the time \citep{jost_1984_erinnerungen}.} a problem to be tackled by S-Matrix theory, but certainly not an unsolvable one. Also Heisenberg conceded, in a letter to Pauli (20 June 1947)

\begin{quote}
[M]aybe we just need to accept that the existence of a singularity\footnote{Singularities on the positive imaginary axis correspond to zeroes on the negative imaginary axis, due to the unitarity of $S$. Some physicists preferred to focus on the zeroes, others on the singularities.} of $S$ is a necessary, but not a sufficient condition for the existence of a stationary state.
\end{quote}

To be sure, the appearance of the redundant zeroes made it doubtful, whether analyticity could be imposed as a requirement in constructing the S-Matrix. In another conceding move, Heisenberg remarked (letter to Res Jost, 10 January 1947, Heisenberg Papers):

\begin{quote}
All in all, you are probably right that the notion of analytic continuation should not be included into the foundations of a theory. In the final theory there will therefore have to be some mathematical scheme with which the S-Matrix can be determined from some simpler basic functions (which need not, however, be Hamiltonian functions). This calculational method then ensures by itself the correct analytic behavior of the S-Matrix.
\end{quote}

But still in 1948, Heitler's former collaborator Ning \citet{hu_1948_on-the-application} used the analytic properties of the S-Matrix in a very general manner (i.e., not with respect to a specific model of meson physics, as for the case of the nucleon isobars), without feeling the need to reference the question of the redundant zeroes, in order to derive the nuclear dispersion formula, thereby incidentally bringing the theory of the analytic S-Matrix back full circle to its beginnings in the work of Siegert.\footnote{Hu cited Siegert's work, but was most probably not aware of the roundabout way on which it had led to the theory of the analytic S-Matrix that he was now using.} This work was praised by M\o ller (Letter from M\o ller to Bohr, 3 April 1948, AHQP, Bohr Scientific Correspondence, M/f No. 30).

As opposed to this general difficulty of S-Matrix theory, there were other more pressing difficulties specific to damping theory. By discarding the infinite terms, damping theory also discarded some effects which were essential for a full theory. On the one hand, there was the difficulty that the elimination of higher orders in perturbation theory led to a reintroduction of the infrared divergence.\footnote{See \citep{blum_2015_qed}.} And, on the other hand, damping theory also discarded all terms which might be responsible for the observed anomalous magnetic moments of the nucleons (see, e.g., \citep[p. 194]{heitler_1947_the-quantum}). These decisive shortcomings were much more important for the demise of damping theory.

While Heitler's damping theory was the prime contender for a theory of the S-Matrix in the immediate postwar period, there was another approach, by Stueckelberg, which actually turned out to be more successful. However, it took some time to mature, and by the time it was ready it had been scooped by the renormalized, diagrammatic QED of Feynman and Dyson, which we shall discuss in the second part of the paper. It is, however, important for our overall argument, because it provides another example of how, through the catalyst of the S-Matrix theory, a new field theory emerged in which the notion of the state was marginalized. In addition, the Stueckelberg case also further underlines the point that, in the period immediately before the advent of renormalized QED, rules for determining the S-Matrix were the immediate concern of many physicists working on quantum field theory, at least in Europe.

\subsubsection{Stueckelberg and the S-Matrix}

Ernst Carl Gerlach Stueckelberg de Breidenbach was the one to carry Heisenberg's S-Matrix program the farthest in the 1940s, independently of the analytic properties of the S-Matrix, which never played a central role for him.\footnote{Besides brief references to Heisenberg's work on the matter, he never published anything on the question of determining the energy values of bound states in the S-Matrix approach. There is one comment in a letter to Arnold Sommerfeld, dated 26 December 1944, where he mentions that he independently arrived at Heisenberg's results for bound states, though not through analytic continuation, but rather through ``nicht-diagonales Erg"anzen'' of the matrix. It is unclear what is meant by this method, which apparently did not need to make use of the analyticity of the S-Matrix. I would like to thank Michael Eckert for making the Stueckelberg-Sommerfeld correspondence, which is in the Sommerfeld Papers in Munich, available to me.} 

Stueckelberg, professor at the University of Geneva, was interested in Heisenberg's S-Matrix theory from the very beginning, but missed a colloquium that Heisenberg held on the subject in November 1942 for health reasons \citep[p. 555-556]{rechenberg_1989_the-early}. But in spring of 1943, Heisenberg's papers were discussed in the theoretical physics seminar in Zurich, which Stueckelberg regularly attended.\footnote{Personal recollections of Fritz Coester, PhD student of Gregor Wentzel in Zurich at the time. Prof. Coester communicated these recollections to me by e-Mail.}

\paragraph{Stueckelberg and Classical Electron Models}

Stueckelberg's interest in S-Matrix theory came from a direction quite different from Heisenberg's original motivation. As we have seen, Heisenberg's S-Matrix program emerged first from attempts to modify the quantization procedure and then from attempts to construct an autonomous quantum theory, independent of a classical field theory. It can thus be placed in a school of thought that viewed the difficulties of QFT as stemming from the quantum theory proper. All through the 1930s and 40s there was another popular viewpoint, which viewed the divergence difficulties as an inheritance from classical theory. The proposed solution was thus a modification of the classical theory. Such attempts tended to focus less on the theory of nuclear interactions and more on electrodynamics, where the classical theory and its difficulties were well-established. These attempts can again be divided into two distinct approaches:\footnote{This is not merely a historian's categorization: The distinction was made explicitly, both by \citet[p. 149]{dirac_1938_classical} and by \citet[p. 52]{stueckelberg_1941_un-nouveau}.} One attempted to modify the classical Maxwellian field theory - we will later see a prime example of this in the work of Feynman. The other approach instead attempted to construct a new classical model of charged particles that avoided the difficulties of classical Lorentz electron theory (instability for extended electrons, divergent field energy for point-like electrons).

This last line of research was initiated by \citet{dirac_1938_classical}. He indeed managed to arrive at a classical theory of the electron that avoided all divergences. Dirac's model, however, problematized the notion of state, already at the classical level. The equation of motion for Dirac's electron included, besides the usual Lorentz force, a radiation reaction term, proportional to the time derivative of the acceleration (the jerk). A similar equation of motion had already been obtained by \citet[p. 49]{lorentz_1916_the-theory}, using a classical model of an extended electron. For Lorentz this had only been an approximate expression for the radiation reaction: There were other terms involving time derivatives of even higher order, which explicitly depended on the structure (the charge distribution) of the electron. In Dirac's theory, the equation of motion was exact and could also be given in a fully Lorentz-covariant manner.

This equation of motion, depending as it did on the time derivative of the acceleration, did not fit into the usual scheme of Hamiltonian mechanics. One might, at first glance, think that all that was needed was an extension of the usual Hamiltonian state space, i.e., phase space. In order to solve the equations of motion, one would need to define not just the initial position and velocity of the electron, but also its initial acceleration. However, as Dirac had already shown, if one defined the initial state by values for position, velocity and acceleration, one ended up with runaway solutions, where the velocity of the electron increased exponentially, even in the absence of an external field. In order to obtain physically meaningful solutions, one had to replace the initial condition on the acceleration by a final one, i.e., impose that for $t \rightarrow \infty$ the acceleration went to zero. Dirac called this ``a striking departure from the usual ideas of mechanics.'' To top it off, the resulting motions, even though they no longer showed the obviously problematic runaway behavior, did have the surprising property that the electron would in general start accelerating somewhat before it encountered an external field. They thus showed a certain degree of acausality, albeit only on a microscopic scale, with noticeable acceleration setting in when the distance of the electron to the external field was of the order of the classical electron radius.

As we shall later see again with Feynman, the loss of the classical notion of the state seemed hardly troubling. And even the microscopic acausality was not viewed as a problem of principle. The main problem was that the loss of the Hamiltonian structure prevented a straightforward quantization of the classical theory. Stueckelberg's first forays into electron theory aimed at finding a model that would allow for quantization along the usual lines by introducing an additional scalar field interacting with the electron, which compensated the divergences from the electromagnetic field      \citep{stueckelberg_1939_a-new, stueckelberg_1941_un-nouveau}. Similar ideas had been put forth by Bopp, and were later advanced by Pais and Sakata.\footnote{See, e.g., \citep{blum_2015_qed} and references therein.} But when Stueckelberg encountered Heisenberg's S-Matrix approach, he soon realized that this could provide the right framework for quantizing electron theories of the Dirac type. He hinted at this idea in a brief Nature note \citep{stueckelberg_1944_an-unambiguous} and then expanded on it in a longer paper in the Helvetica Physica Acta \citep{stueckelberg_1944_un-modele}.

Stueckelberg's main idea was that in the Dirac electron theory, while it was not possible to provide the usual Hamiltonian structure as a basis for quantization, one was able to define a classical analog of the S-Matrix, which could provide the basis for transforming it into a (quantum) S-Matrix theory. It looked like the perfect match: A classical theory that resisted quantization joined with a quantum theory that had lost contact with classical theory, due to the abandonment of the correspondence principle. From the wording of the Nature note, it appears that Stueckelberg originally believed that one would obtain a Heisenbergian S-Matrix directly from a quantization of the Dirac electron.\footnote{This reading is also supported by a letter from Stueckelberg to Sommerfeld from 21 July 1943 (Sommerfeld Papers, Deutsches Museum, Munich), where he states that the application of Heisenberg's S-Matrix to electrodynamics corresponds exactly to Dirac's theory.} It was indeed quite typical of Stueckelberg to send out somewhat premature notes and then to work out difficulties only in later extended papers. In his long Helvetica Physica Acta paper, Stueckelberg found that he needed to generalize Dirac's theory by allowing for a more general equation of motion of the form:\footnote{Dirac had also considered the possibility of a more general equation of motion, but had argued that ``they are all much more complicated [...], so that one would hardly expect them to apply to a simple thing like an electron.''}

\begin{equation}
\label{eq:models}
m \xi \left( - d^2/dt^2 \right) \ddot{\mathbf{x}} - \frac{2}{3} \frac{e^2}{c^3} \dddot{\mathbf{x}} = \mathbf{F}_{(\mathrm{inc})}
\end{equation}

Here $\mathbf{F}_{(\mathrm{inc})}$ is the force due to the external (incoming) electromagnetic field, while $\xi$ is some function of one variable,\footnote{Stueckelberg used $\eta$ instead of $\xi$, but I have decided to modify his notation in order to avoid confusion with Heisenberg's phase matrix $\eta$.} so that $\xi \left( - d^2/dt^2 \right)$ is some (time) differential operator. The constant term in a series expansion of $\xi$ is always 1, giving the usual mass times acceleration term of the Newtonian law of motion. The above expression thus contains, in addition to the usual equation of motion of a charged particle in an external field, terms proportional to time derivatives of the position of an order greater than two. The term proportional to $\dddot{\mathbf{x}}$, which is given explicitly, is the radiation reaction term also appearing in the Dirac equation of motion. For the simplest case of $\xi = 1$, Stueckelberg's equation thus reduced to that of Dirac. More complicated functions $\xi$ introduced higher-order time derivatives, as in Lorentz's original theory of an extended electron.

Lorentz had not derived the terms containing higher time derivatives of the position, as these would anyhow have depended explicitly on the exact model of the electron. Inverting Lorentz's approach, Stueckelberg now took a classical model of the electron to be \emph{defined} by the exact form of the higher order terms, i.e., by the function $\xi$ rather than by an extended charge distribution, which would have led to the old problem of electronic stability. Stueckelberg stil wanted to think of electrons as pointlike (for all possible $\xi$), although this is by no means an unambiguous interpretation of the equation of motion: Dirac had argued that his equation of motion could be ``interpreted in a natural way, [...] if we suppose the electron to have a finite size.''

The next step was to define the classical analog of the S-Matrix. Here, Stueckelberg had to restrict the scope of his approach: He only studied a single scattering process, namely the scattering of electromagnetic radiation by an electron described by one of the electron models.\footnote{Stueckelberg then also extended this somewhat by giving his point particle an additional, internal degree of freedom, in order to model an atom with excited states, but I will not be discussing this slight extension much further.} He also assumed for simplicity that both the electromagnetic field and the electron were scalars (the latter assumption was necessary because his classical equations of motion made no reference to the electron spin).

For this scattering process, he defined the classical S-Matrix as an operator that transformed the Fourier coefficients $c_{\mu} (-T)$ of an incoming electromagnetic radiation field into the Fourier coefficients $c_{\mu} (T)$ of the outgoing field, including the radiation emitted by the electron. This was something that could be calculated readily using Stueckelberg's electron models: One first solved the equation of motion (approximately) for a given incoming field. The charge density as a function of space and time could then directly be calculated from the electron's world line and inserted into the inhomogeneous Maxwell equations, which could then again be solved approximately to give the Fourier coefficients of the outgoing field in terms of the Fourier coefficients of the incoming field. The resulting expression was rewritten as a classical S-Matrix operator acting on the Fourier coefficients of the incoming field:

\begin{equation}
c_{\mu'} (T) = S_{\mu' \mu}(T) c_{\mu}(-T)
\end{equation}

The classical S-Matrix clearly depended on the choice of the electron structure function $\xi$, but quite generally it could be written as a series expansion in terms of an elementary operator $\alpha$ that did not depend on $\xi$:

\begin{equation}
\alpha_{\mu \mu'} = \frac{e^2 \pi}{m V}  \cos{\theta} \frac{\delta(\omega'-\omega)}{\omega}
\end{equation}

where $V$ is the spatial volume used to discretize the modes of the electromagnetic field (which drops out when calculating scattering cross sections), $\omega$ and $\omega'$ are the frequencies belonging to the modes $\mu$ and $\mu'$, respectively, and $\theta$ is the angle between the propagation directions of the two modes. The structure function $\xi$ then only defined the exact functional form of $S(\alpha)$. For the simple Dirac model, it was:

\begin{equation}
\label{eq:heitler}
S(\alpha) = \frac{1- \frac{1}{2} i \alpha}{1+ \frac{1}{2} i \alpha}
\end{equation}

Stueckelberg could show that for any arbitrary function $\xi$, $S$ is always a unitary matrix (i.e., classically, the scattering of electromagnetic radiation by an electron is always elastic).

One can here see the difficulty that had led Stueckelberg to move from the simple Dirac model to his more general electron models: The classical S-Matrix is now supposed to be related to the quantum S-Matrix and $\alpha$ to the phase matrix $\eta$. But for the Dirac model, $S(\eta)$ then has the wrong functional form. Stueckelberg's generalization of the Dirac model addressed this difficulty in a twofold way: On the one hand, he could show that there \emph{was} a specific structure function $\xi$ (which he called the ``quantum'' model of the electron, even though it defined a classical theory) which led to the expected functional form $S(\alpha) = e^{i \alpha}$. On the other hand, he could show that for an arbitrary $\xi$ the resulting quantum S-Matrix $S(\eta)$ with a hermitian $\eta$ would be a Lorentz-invariant, unitary operator, and thus that there were many different possibilities to construct the quantum S-Matrix for a given $\eta$, which could be regarded as corresponding to different classical models of the electron.

All of this did not yet address the question of how to obtain $\eta$, which had been Heisenberg's main concern. Rather, Stueckelberg had only added an additional degree of arbitrariness to Heisenberg's theory, by also generalizing the possible functions $S(\eta)$. While these different possibilities could be related to different classical models, this only delegated the question to the classical theory, where there was also no grounds to prefer any particular structure function $\xi$, besides the argument from simplicity that Dirac had brought forth. As for the construction of $\eta$, he thought that one might simply take over Heisenberg's prescription of setting $\eta$ equal to the (normal ordered) expression obtained in first order of perturbation theory in regular quantum electrodynamics, or rather equal to the expression obtained in the lowest order of perturbation theory that gives a non-vanishing contribution to the process in question (second order perturbation theory for Stueckelberg's central example of electron-photon, i.e., Compton, scattering). One sees here the close relation to Heitler's damping theory, and indeed Stueckelberg could show that, for this choice of $\eta$, Heitler's theory was a special case of his, corresponding to the choice $\xi = 1$, i.e., to the Dirac model.\footnote{Consequently, Stueckelberg's general ``unambiguous method'' had the same serious flaws as damping theory: Along with the divergences, it also eliminated physically meaningful terms, as was realized early on by Pauli (letter to Wentzel, 24 November 1944). Incidentally, the editorial comment in the edited Pauli correspondence references the wrong Stueckelberg paper.} He could also show that $\eta$ was equal to the classical operator $\alpha$ for Compton scattering in the limit where $h \rightarrow 0$, which could be interpreted physically in this case as the limit where the photon momentum can be neglected. This correspondence between $\eta$ and $\alpha$ and between $S(\eta)$ and $S(\alpha)$ was supposed to furnish a new correspondence principle between a classical theory (described by a structure function $\xi$) and a quantum S-Matrix theory. The principle appeared to be generalizable, and Stueckelberg also applied it to his somewhat more complicated example of the inelastic scattering of radiation by a two-level atom.

In his following papers, Stueckelberg then went all-out: He cast his electron models, which did not fit into the regular scheme of Hamiltonian mechanics, as representative of a more general ``functional mechanics,'' which was to be a generalization of Hamiltonian mechanics \citep{stueckelberg_1945_mecanique}. Such a general functional mechanics would then be related to a functional quantum mechanics (an S-Matrix theory), using a new correspondence principle, a generalization of the ``quantization'' technique outlined above.

So far, this was only a generalization of what he had already done one year earlier, but Stueckelberg now sought to eliminate all the ambiguities in constructing the S-Matrix by introducing a second correspondence principle, which would relate the functional mechanics to the usual Hamiltonian mechanics. The limit in this case was of course not $h \rightarrow 0$, but rather the non-relativistic limit $c \rightarrow \infty$. As one can see from equation $\ref{eq:models}$, this gets rid of the radiation reaction term proportional to $\dddot{\mathbf{x}}$ and of all the non-trivial terms in $\xi$, since the terms involving higher order time derivatives will contain dimensionful coefficients, which are also taken to be proportional to inverse powers of $c$. In the non-relativistic limit, thus, Stueckelberg's general equation of motion for the electron turns into the regular equation of motion for the electron, containing only the Lorentz force due to an external field, which conforms to the usual structure of Hamiltonian mechanics. This second correspondence principle could also be applied on the level of the quantum theory. It then read: In the non-relativistic limit, the S-Matrix should be the same as that obtained from regular quantum theory.

Now, we have seen earlier that in regular quantum theory the matrix $\eta$ will in general contain higher-order terms in the perturbation expansion. And Stueckelberg indeed found that in order to satisfy his second correspondence principle it was necessary to go beyond leading order and add higher order terms (all of which needed to be normal ordered, in order to avoid the divergences). He then showed that, starting from the S-Matrix for (scalar) electron-electron scattering\footnote{Although Stueckelberg does not state it explicitly, it is easy to see why he switched his paradigmatic example from Compton to electron-electron scattering: In the strict non-relativistic limit there is no Compton scattering, since the electromagnetic field is eliminated and replaced by an instantaneous interaction between electrons. In his  new functional mechanics, Stueckelberg was able to write down more general classical S-Matrices, which could give the relations between any quantity in the initial and final states, not just the Fourier coefficients of the electromagnetic field.} with a suitable choice of higher-order terms for $\eta$, one got in the non-relativistic limit the S-Matrix that one obtained in regular quantum theory for the scattering due to an instantaneous Coulomb (or, for a massive scalar photon, Yukawa) interaction between two particles.\footnote{Stueckelberg somewhat confusingly calls this ``Rutherford scattering.'' This term is nowadays mainly reserved for the scattering process, where one of the particles is infinitely heavy, so that the light particle is simply scattered in a static Coulomb field. This latter problem can be solved exactly and does not necessitate the use of perturbation theory.}

This by no means fixed the arbitrariness in the construction of the S-Matrix. In order to show that the second correspondence principle was fulfilled by his new choice of $\eta$, Stueckelberg still needed to assume a functional form $S(\eta)$. And even then, all one could show was that the resulting S-Matrix fulfilled the second correspondence principle. There was no constructive principle to obtain the S-Matrix, much less a prescription for constructing the S-Matrix for processes other than electron-electron scattering.

It was at this point that Stueckelberg shifted focus, with the question of causality becoming prominent. Before we address this point, it is time to briefly take stock of what Stueckelberg had really achieved so far, leaving aside all the elaborate conceptual framework of classical electron models and functional mechanics: He had obtained a generalization of Heitler's damping theory, defined by the function $S(\eta)$. And, for the construction of $\eta$, he had gone beyond the usual prescription of taking the lowest order term from the expression obtained in usual quantum mechanics (normal ordered), by including higher order terms, which ensured (if chosen accordingly) that his second correspondence principle was fulfilled. All of this did not address the difficulties of damping theory discussed above, and from a modern viewpoint Stueckelberg's simple-minded trick of normal ordering the operators in $\eta$ seems a ``wild proposal'' \citep[p. 96]{wanders_2009_stueckelberg}. Stueckelberg himself, in a letter to Arnold Sommerfeld from 21 July 1943 (Sommerfeld papers), writes that

\begin{quote}
What seems the most astonishing to us, to Mr. Weigle [Stueckelberg's colleague at the theoretical physics department in Geneva], Mr. Wentzel and to myself, is that before no-one arrived at this change in quantum electrodynamics, which I need to make in order to make it conform with Heisenberg's ideas and to evade the divergences.
\end{quote}

In contrast to Heitler, no comments by Stueckelberg on the problems of this approach, published or unpublished, are extant. If it had stayed at this, Stueckelberg's work on the S-Matrix would rightly be forgotten. But then Stueckelberg hit upon the idea of replacing his second correspondence principle with the demand for causality.

\paragraph{Stueckelberg and Causality}

Already in 1943, Stueckelberg pointed out a strange feature of Heisenberg's simplest toy S-Matrix, the normal ordered $\varphi^4$ theory mentioned above.\footnote{Personal recollections of Fritz Coester.} While Stueckelberg never gave much of a derivation of this strange feature in his published work, his point can be understood quite clearly from the following calculation. The calculation, which involves going to the second order of perturbation theory, is entirely mine, but involves no elements that were unavailable to Stueckelberg in 1943/44 (and uses whenever possible elements that he actually used in another context) and thus hopefully at least parallels Stueckelberg's own reasoning.

As we saw earlier, Heisenberg's method consisted of setting the $\eta$ matrix equal to the time integral of the interaction Hamiltonian, which is what one gets in first order perturbation theory for regular quantum field theory: 

\begin{equation}
\label{eq:pert}
\eta_{(1)}  =  \int_{-\infty}^{\infty} dt H(t) 
\end{equation}

where the subscript $(1)$ is supposed to indicate the first order in perturbation theory. Going to the second order of perturbation theory in regular quantum field theory, the next term in the perturbative expansion of $\eta$ is (this expression was given explicitly for the first time in print in \citep[p.20]{stueckelberg_1944_un-modele}):

\begin{equation}
\eta_{(2)}   =  - \frac{1}{2} i \int_{-\infty}^{\infty} dt \int_{-\infty}^{t} dt' \left[H(t), H(t') \right]
\end{equation}

Consequently, the perturbative expansion of $S$ itself is given as:

\begin{equation}
S = 1 - i  \eta_{(1)} -  \left(i \eta_{(2)} + \frac{1}{2} \eta_{(1)}^2 \right) + \ldots 
\end{equation}

One can write the term in brackets, i.e., the second order term in a perturbative expansion of the S-Matrix, as

\begin{equation}
\label{eq:stueck}
i \eta_{(2)} + \frac{1}{2} \eta_{(1)}^2 =  \frac{1}{2} \int_{-\infty}^{\infty} dt \int_{-\infty}^{t} dt' \left[H(t), H(t') \right] + \frac{1}{2} \int_{\infty}^{\infty} dt \int_{-\infty}^{\infty} dt' H(t)  H(t')
\end{equation}

The second summand can be split in two, one term for $t>t'$, another for $t<t'$, i.e.:

\begin{equation}
\int_{-\infty}^{\infty} dt \int_{-\infty}^{\infty} dt' H(t)  H(t') = \int_{-\infty}^{\infty} dt \int_{-\infty}^{t} dt' H(t)  H(t') + \int_{-\infty}^{\infty} dt \int_{-\infty}^{t} dt' H(t')  H(t)
\end{equation}

where in the second summand the names of the two variables have been exchanged to give a unified notation. Similarly, the first summand can be split in two, by writing out the commutator:

\begin{equation}
\int_{-\infty}^{\infty} dt \int_{-\infty}^{t} dt' \left[H(t), H(t') \right] = \int_{-\infty}^{\infty} dt \int_{-\infty}^{t} dt' H(t)  H(t') - \int_{-\infty}^{\infty} dt \int_{-\infty}^{t} dt' H(t')  H(t)
\end{equation}

Reinserting these two expressions into equation \ref{eq:stueck}, we see that two terms cancel while the other two add up, giving:

\begin{equation}
\label{eq:pertend}
i \eta_{(2)} + \frac{1}{2} \eta_{(1)}^2 =  \int_{-\infty}^{\infty} dt \int_{-\infty}^{t} dt' H(t)  H(t')
\end{equation}

This expression (which is of course valid for an arbitrary Hamiltonian) has the feature that the operator belonging to the earlier time $t'$ stands to the right of the operator belonging to the later time $t$ (it is time-ordered). But now Heisenberg had suggested to take equation \ref{eq:pert} as an exact expression for $\eta$, i.e., to set $\eta_{(2)} = 0$ (and all higher-order terms as well). The resulting S-Matrix, where $\eta$ contains no $\eta_{(2)}$ term, but only $\eta_{(1)}$ is perfectly acceptable on the grounds of unitarity and relativistic invariance alone: It is unitary because $\eta_{(1)}$ is hermitian, and it is relativistically invariant, because all terms in the perturbation expansion of $\eta$ transform in the same way. And, of course, it avoids all possible difficulties with divergences that may show up in higher orders. But the field operators in the resulting S-Matrix are then no longer time-ordered. One might have a particle being created at time $t$ and then annihilated at an earlier time $t'$.\footnote{This is of course a very loose manner of speaking, used also by Stueckelberg, since the photon will not be created at a sharp instant - otherwise it would have a totally undefined energy. But a certain degree of localization of the emission and absorption events is of course possible, and these will be centered around the times $t'$ and $t$, respectively. This point was clarified by \citet{fierz_1950_uber-die-bedeutung}. Also note that the inverse time ordering also occurs if $H$ is normal ordered, since this does not imply that $H(t) H(t')$ is also normal ordered. It also does not play a role, whether this virtual particle is on-shell or not.}

What is the significance of this? In a short note that Stueckelberg submitted to Nature in early 1944, his first paper on S-Matrix theory, he described this behavior as ``acausal'' \citep{stueckelberg_1944_an-unambiguous}. This is a tricky point and before elaborating on what Stueckelberg meant by acausality, we need to say a few words about Stueckelberg's treatment of the infinite-time limit in S-Matrix theory. Stueckelberg was not dogmatic about the S-Matrix only describing asymptotic states at infinite times. He viewed Heisenberg's S-Matrix as the infinite-time limit of the interaction-picture time evolution operator, which had of course been Heisenberg's starting point in 1937. Stueckelberg was ambiguous about whether the techniques he was now developing would also apply to finite time evolutions: In his papers, he often avoided taking an infinite-time limit and in the Nature note even stated that ``for any given $t$, we can describe the quantum mechanical state of the system.'' But he consistently used Heisenberg's S-Matrix terminology and focussed exclusively on scattering processes. Also, several results clearly only hold in the infinite-time limit.\footnote{A striking example is the calculation in equation (6.21) in \citep{stueckelberg_1944_un-modele}.} This remained true as Stueckelberg moved ever closer back to field theory and away from Dirac-type electon models, which had been the original motivation for looking at infinite times only. For the moment, it is merely important that the reader should not be irritated (or only mildly) by finite times showing up in what effectively is a pure S-Matrix theory.

What now did Stueckelberg mean by acausality? Even if one takes the time evolution for a finite time interval instead of the S-Matrix, both the times $t'$ and $t$ lie within this time interval, i.e., if we look at the wave function at the final time, both the cause (particle emission) and the effect (particle absorption) have occurred, and it is unclear how one could unambiguously determine the order in which they occurred from the final state wave function. This is of course true \emph{a fortiori} when going over to the $t \rightarrow \infty$ limit, i.e., to the S-Matrix. One sees here a certain tension between a theory which deals only with in- and outgoing states and the fact that at the macroscopic level there is such a thing as infinitesimal (and causal) time evolution. In a nutshell, this criticism of the S-Matrix approach had, again, been voiced early on by Pauli (Letter to M\o ller, 18 April 1946):

\begin{quote}
I am still not convinced; not only the whole frame of concepts is empty (no theory is given which determines S), but the rather complicated formalism does not contain the classical mechanics as a limiting case.
\end{quote}

How, for example, were the observables of classical theory, which are continuous functions of space and time, related to those of the S-Matrix theory? The emergence of the classical theory, which is of course already tricky in regular quantum theory, became even more of a mystery in S-Matrix theory. And Stueckelberg gave no indication how the acausal behavior he identified in the S-Matrix theory would be detectable or how it would manifest itself at the macroscopic level. Still, he claimed that our knowledge of the causal behavior observed in electrodynamic phenomena prevents us from constructing the S-Matrix for QED by, e.g., dropping the higher order terms in the expansion for $\eta$. At this point, however, Stueckelberg did not yet consider causal behavior as an intrinsic requirement for constructing the S-Matrix. Rather, he pointed out that for nuclear interactions empirical evidence did not enforce a causal description even for distances on the order of $10^{-15}$ m, i.e., on the order of the size of nuclei. One thus had the freedom to adopt non-causal S-Matrices for the description of the nuclear force, a possibility he seems to have considered legitimate at the time.\footnote{In Stueckelberg's Nature note there appears to be a misprint: ``...if we go back to a causal description at small distances'' should probably be ``large distances.'' I think with this replacement, the argument is no longer enigmatic, cf. \citep[p.96]{wanders_2009_stueckelberg}. This of course does not change anything concerning the difficulties with the emergence of classical physics and macroscopic acausality, which Stueckelberg simply did not touch upon.}

Stueckelberg did not pursue these thoughts on causality much further at the time. It first reappeared  in two talks, at the meeting of the Swiss Physical Society on 4 May 1946 \citep{stueckelberg_1946_une-propriete} and at the Cambridge conference mentioned earlier (22-27 July 1946) \citep{stueckelberg_1947_the-present}, after he had completed his grand scheme of functional mechanics. Both conference papers are even more elliptic than usual for Stueckelberg. In the paper presented to the Swiss Physical Society, Stueckelberg presented the general problem of acausality in an S-Matrix theory, which he had already sketched in his 1944 Nature note. The interpretation of the problem was, however, quite different from the one he had given in 1944: Instead of speaking of the virtual particle being emitted at a time $t'$ and then being absorbed at an earlier time $t$, he now spoke of a particle of negative energy being emitted at time $t$ and then being absorbed at time $t'$. He thus offered a re-reading of the process, where the causal ordering of events is presupposed and instead the positive-energy particle traveling backwards in time is interpreted as a negative-energy particle traveling forwards in time. In other words, the absorption of the positive energy particle coming from the future is instead viewed as the emission of a negative energy particle into the future. Given that the S-Matrix theory only makes definite statements about the in and the out states, both interpretations are equally possible, and it should be noted that this talk presents an exception in that Stueckelberg explicitly only considers an infinite final time.

Stueckelberg now argued that one should get rid of these negative energy states that appear if one ultimately wants to give a spatiotemporal description of the scattering process. He claimed that there was ``one and only one way'' to do this, and that was to add specific higher order terms to $\eta$, with $S(\eta)$ given by equation \ref{eq:heitler}, and that these were the same terms one would get by applying the second correspondence principle described above. He thus claimed that, given a leading order term $\eta_{(1)}$, the demand for the absence of intermediate negative-energy particles uniquely determined the S-Matrix, functional form $S(\eta)$, higher order terms in $\eta$, and all. No derivation of any sort was given; it seems to have been another case of Stueckelberg's intellectual ejaculatio praecox, since in the paper he submitted to the proceedings of the Cambridge conference, he was again a lot more careful:

\begin{quote}
The condition of unitarity and relativistic invariance is by no means sufficient to determine $S$. The theory is too general and it contains numerous effects contradicting macroscopical causality. We therefore conclude that additional conditions must be looked for limiting the choice of $S$ or [$\eta$]. They must be of a form so as to admit a causality [sic] (=premonition) only for microscopical time intervals...
\end{quote}

Reverting from talking of negative energy states back to talking about causality, there was now also no more talk of the demand for causality uniquely determining the S-Matrix. Also, he had apparently reverted back to demanding only the absence of macroscopic acausality, as opposed his talk at the Swiss Physical Society, where he attempted to remove the intermediate negative energy states entirely. It seems that while Stueckelberg certainly felt that the causality condition would provide fruitful in constructing the S-Matrix, he was still very unsure how this was to be done and how it would go beyond the limited success of his second correspondence principle. A pause in publications ensued, which went along with a personal crisis, as Stueckelberg divorced his first wife in 1946 \citep[p. 143]{crease_1986_the-second}. The reception of Stueckelberg's ideas on new correspondence principles and causality was meagre. They are, however, mentioned rather prominently in a review article by Gregor Wentzel, then still in Zurich \citep{wentzel_1947_recent}. Wentzel, who got the inside scoop on Stueckelberg's work from his former PhD student Coester, who was now a postdoc in Geneva, was, however, quite skeptical concerning the prospects of Stueckelberg's program. Concerning the second correspondence principle he stated that it was ``certainly not sufficient to remove \emph{all} ambiguities.'' As for the arguments from causality, he wrote:

\begin{quote}
It is not quite clear yet to what extent the conclusions drawn from this argument coincide with those obtained from the ``second correspondence principle.'' But, generally speaking, it is to be expected that closest correspondence with the wave-mechanical theory will automatically do away with all ``non-causal'' predictions, since in a rigorous wave-mechanical theory based on the Schr\"{o}dinger equation the succession of events in time is necessarily ``causal.''
\end{quote}

We see here that there were two ways to look at Stueckelberg's approach: A more foundational one, where causality is used as a constructive principle for writing down the S-Matrix, in the spirit of Heisenberg's work of the 1940s. And a more modest one, where a covariant formulation of the usual field theory (or, in any case, the S-Matrix obtained in this formalism) is gently modified to get rid of the divergences, with causality acting as a guiding principle that ensures that these modifications do not destroy those properties of the usual theory that one wants to preserve. This is more akin to Heisenberg's original program of 1937, with the major difference that Heisenberg's restriction of preserving relativistic covariance is now supplemented with the demand of preserving causality. 

The more foundational viewpoint appears to have been expounded in a paper that Stueckelberg submitted to the Physical Review, which was, however rejected. The manuscript had been written by Stueckelberg in the spring of 1947 in Copenhagen. He had met M\o ller at the 1946 Cambridge conference and inquired whether it might be possible for him to visit M\o ller (and Bohr) in Copenhagen.  An invitation from M\o ller followed\footnote{Letter from 29 November 1946, M\o ller Papers, Niels Bohr Archive, Copenhagen.} and Stueckelberg ended up spending a month at Niels Bohr's institute in Copenhagen (April 9 to May 4 1947, Institute for Theoretical Physics register book for foreign guests, Archive for the History of Quantum Physics). During this time he finished the manuscript ``Theory of the S-Matrix for Nuclear Interaction'', which is no longer extant and has acquired somewhat of a legendary status. On 3 June 1947, Stueckelberg wrote to Bohr (Bohr Scientific Correspondence (BSC), Niels Bohr Archive, Copenhagen) thanking him for the useful discussions in Copenhagen. This is followed by a shocked letter of 25 September 1947 (BSC): ``To my great surprise, the article that I wrote in your institute has not been accepted by the Physical Review.'' The letter of rejection is appended to this letter (which is how we know the title of the paper). Wentzel refers to the manuscript in an article on causality in non-relativistic wave mechanics, submitted in December 1947. \citet{wanders_2009_stueckelberg} mentions the manuscript, but fails to point out that it is not extant, at least not in any of the archives I checked.\footnote{Stueckelberg, in an interview with \citet[p.143]{crease_1986_the-second}, mentions a manuscript dealing with renormalization techniques that was rejected by the Physical Review. He gives the year as 1942 or 1943, but it seems highly unlikely that Stueckelberg had \emph{two} major papers rejected by the Physical Review (given that he hardly ever wrote in English), with both manuscripts being lost. Then again, Stueckelberg stated that this rejected manuscript was about QED, while the paper rejected in 1947 was (inferring from its title) about nuclear interactions. Stueckelberg went on to say that he later found out that the referee who had rejected his paper was Gregor Wentzel. Since Wentzel was in Switzerland during the war, it seems highly unlikely that he would have acted as a referee for the Physical Review in 1942/43. Then again, Wentzel in his 1947 paper states that Stueckelberg had kindly allowed him to see the still unpublished manuscript, which would be quite a bold statement if he had anonymously refused publication just a few months earlier. With the current archival evidence it is impossible to decide whether the two papers are identical.} Stueckelberg's paper was not rewritten and (re-)submitted, as Stueckelberg stated to Bohr:

\begin{quote}
I continue with my work, but I do not have the time, for the moment, to rewrite it [the paper]. Nonetheless, the results hold and Mister Rivier, one of my students, is currently calculating the magnetic moment of the proton and the neutron.
\end{quote}

Stueckelberg's new approach was thus first presented in a short note in the Physical Review \citep{rivier_1948_a-convergent} and then in Dominique Rivier's thesis \citep{rivier_1949_une-methode}. In these papers, Stueckelberg and Rivier presented the work from the more modest viewpoint of modifying regular field theory. I will focus on these papers in the remainder of this section. Stueckelberg did return to the foundational approach soon after, beginning with \citep{stueckelberg_1950_causalite}. But this later work was even less received. In any case, the main feature I want to work out is how, despite the return to usual quantum field theory, the S-Matrix heritage shows in the final formulation, in order to compare this with the parallel development in Feynman's work. This feature can be worked out nicely in the work with (and of) Rivier. Stueckelberg's later work certainly deserves closer historical study, both for a detailed comparison with orthodox quantum field theory and for premonitions of the renormalization group, which are to be found therein.\footnote{Some work has been done on the matter by \citep{wanders_2009_stueckelberg} and \citep{peters_2004_schonheit}.} But this would lead us too far away from the main focus of this paper.

How then did the modified field theory proposed by Stueckelberg and Rivier look? The main feature was a new renormalization procedure that went beyond Stueckelberg's original normal ordering idea. It was based on an insight into the general structure of S-Matrix elements. They can be written in terms of the Hamiltonian, as in equations \ref{eq:pert} to \ref{eq:pertend}. But they can also be written out explicitly in terms spatial integrals over the Hamiltonian density, which can in turn be expanded in terms of annihilation and creation operators. One then has for a given process the operators annihilating the incoming particles and the operators creating the outgoing particles. All the other annihilation and creating operators are contracted to give the creation and subsequent annihilation of intermediate virtual particles. Using completeness relations, the infinite sums over possible intermediate particles can be turned into singular functions of two space-time points, corresponding to the locus of creation and annihilation of the intermediate particle (of course these coordinates are integrated over, so there is no notion of a single point where the intermediate particle is created). Now, Stueckelberg was an expert on such singular two-point functions appearing in the completeness (and commutation) relations for relativistic field theories and had written two mathematical papers on the matter in the early 1940s, before becoming involved with the S-Matrix program \citep{stueckelberg_1942_solutions,stueckelberg_1942_solutionsb}. And he realized that this was the point in the theory where causality can be ensured or destroyed: Stueckelberg's hypothesis was that a theory is causal if and only if it contains only one type of singular two-point function in the space-time integrals appearing in the S-Matrix elements, namely the causal function $D_c (\mathbf{x}, t; \mathbf{x}', t')$ (nowadays known as a Feynman propagator; we shall later see why).\footnote{The equivalence of Feynman's propagator with Stueckelberg's causal function was pointed out to Dyson by Pauli: ``[T]he function called there $D_c$ in honour of `Causality' is identical with the function which you call $D_F$ (in honour of Feynman).'' \citep[p. 595]{meyenn_1993_wissenschaftlicher}}

What the causal function does is precisely to ensure that there are only causal processes in Stueckelberg's sense discussed above. $D_c$ could be written out explicitly as (cf. the very useful appendix of \citep{rivier_1949_une-methode})

\begin{equation}
\label{eq:causal}
D_c (\mathbf{x}, t; \mathbf{x}', t') = \frac{i}{16 \pi^3} \int \frac{d^3 \mathbf{k}}{\omega (\mathbf{k})} e^{-i \mathbf{k} \cdot (\mathbf{x} - \mathbf{x}')} \left[ \theta (t-t') e^{i \omega(\mathbf{k}) (t-t')} + \theta (t'-t) e^{-i \omega(\mathbf{k}) (t-t')} \right]
\end{equation}

Here $\omega (\mathbf{k})$ is the positive energy belonging to the momentum vector $\mathbf{k}$ and $\theta$ is the Heaviside step function. If the space-time integrand in the expression for the S-Matrix elements is thus interpreted as describing actual physical processes and not just as a numerical expression arising in perturbation theory (for all reference to space-time points of course vanishes in the final S-Matrix element, where the integrations have been performed), the function $D_c$ describes the emission of plane waves of positive energy moving forwards in time and plane waves of negative energy moving backwards in time (that is the absorption of positive energy plane waves coming from the past), i.e., causal processes in Stueckelberg's sense.\footnote{It is interesting to note that, as opposed to Feynman, Stueckelberg did not make use here of the notion of anti-particles as particles going backward in time, which he himself had introduced \citep{stueckelberg_1941_la-signification}, but rather treated particles and antiparticles on the same footing, as positive energy particles moving forward in time.} The question of whether terms in a perturbation expansion should be taken as in some way representing physical processes has been discussed in particular in the context of Feynman's work, where they are represented by the suggestive Feynman diagrams. W"uthrich has argued that such a conception certainly played an important role in the genesis of Feynman diagrams \citep{wuethrich_2010_the-genesis}. We see here that Feynman was not alone in this: Also for Stueckelberg it was essential to treat the intermediate particles of perturbation theory as really propagating in space and time, in order to restrict and determine these processes by imposing his causality condition.\footnote{\label{fn:fierz} It should perhaps be pointed out that already at the time the unsoundness of this heuristic was explicitly pointed out, e.g., by Markus Fierz in a letter to Pauli from 17 June 1949: ``In the `causal interpretation' of the formulae one [Feynman] now pretends as if configuration space were actual space; I find this highly confusing. Stueckelberg does the same thing. But if one takes these interpretations not too seriously, only as a heuristic point of view for the private use of people who like that sort of thing (\emph{f"ur den Privatgebrauch gewisser Naturen}), one can instead look at it in the following manner...''}

The causal function was of course most essential for the constructive aspect of the causal theory, as it was the main building block for perturbatively constructing the S-Matrix without referring to field-theoretic Hamiltonians. It was, however, also important for the more pedestrian approach of modifying regular quantum field theory (in particular making it finite) that we are discussing here. As already stated in the above quote from Wentzel, it was of course to be expected that an S-Matrix calculated from the regular quantum theory would obey Stueckelberg causality and would thus only contain singular two-point functions of the $D_c$ type. This was explicitly proven in 1949 by Stueckelberg's collaborators in Geneva \citep{houriet_1949_classification}. And, Stueckelberg realized, it was the singularities of this function (for the integration over an infinity of plane waves results in $\delta$-function type singularities), or rather of higher powers of this function, which were responsible for the infinities appearing in usual QFT perturbation theory.

The program outlined in the note with Rivier and in Rivier's thesis was thus to modify the singular two-point functions in such a manner that they become finite, without destroying relativistic invariance, unitarity and Stueckelberg causality.\footnote{In Stueckelberg's full constructive causality program, he argued, the demand for causality does not determine the value of the function $D_c$ for $(\mathbf{x}, t) = (\mathbf{x}', t')$, since causality can only make statements about what happens between two distinct space-time points. Thus, in this approach, the regularization method is really a specification of an underdetermined function, rather than the modification of a singular function. He viewed this as being preferable to having a theory in which infinities first necessarily appear and then have to be removed \citep{stueckelberg_1950_causalite}.} This was done by subtracting two-point functions corresponding to auxiliary, very massive particles (the mass enters $D_c$ through the dispersion relation $\omega (\mathbf{k})$). This was a throwback to the auxiliary massive particles that Stueckelberg had first proposed in 1939. The method still had some defects (for example uniqueness, as it did not yet make contact with the physical ideas underlying the renormalization theory being developed at the time) and was further developed both by Stueckelberg and Rivier \citep{stueckelberg_1950_a-propos}, as well as in the well-known Pauli-Villars renormalization scheme \citep{pauli_1949_on-the-invariant}.\footnote{On the Stueckelberg-Rivier-Pauli-Villars connection, see \citep[pp. 582ff]{schweber_1994_qed}.}

If the development of renormalized, covariant QED had not been well underway by this point, the work of Rivier and Stueckelberg, which was clearly quite independent of these developments and resulted from a decade-long research program, might well have been the starting point of the revitalization of QFT and the end of the age of infinities.\footnote{Stueckelberg's lack of impact on the wider physics community, despite a great number of innovative ideas, has often been discussed and lamented (most notably in a recent edited volume studying his many-faceted work \citep{lacki_2009_ecg}), and we need not go into this matter here. I would, however, like to add a short personal comment: Stueckelberg's work for all its prescience, really is very hard to read - elliptic and overly short at times, overly general with complicated notation at others. On this last point, it should be noted that part of the reason that his 1947 Physical Review paper was rejected (according to the letter of rejection that Stueckelberg forwarded to Bohr) was that ``the manuscript at present presents quite a formidable problem for both the compositor and the printer.''} For it was quite clear by now that what they had constructed was not a novel S-Matrix theory along Heisenberg's lines, but rather a slight modification (through regularization) of the established quantum field theory of the 1930s. But the historical path dependence is quite clear: The time evolution operator is still called $S$, in general it is only defined for infinite time intervals, and the new renormalization techniques could only directly be applied to this operator and not to the (until then) usual differential equation form of quantum field theory. As Rivier wrote in his thesis:

\begin{quote}
[T]he proposed integral form of the theory is equivalent to the usual Hamiltonian form in its results, with regards to the phenomena of scattering and collision, the only ones studied here.
But its point of view is quite different: The characteristic operator is the integral operator $\mathbf{S}$, instead of $\mathbf{h}$ [the Hamiltonian]. It is on this operator that the attempts to improve the theory are built, in particular the attempts to suppress the infinities. (p. 290)
\end{quote}

The focus on scattering is particularly striking, as it was not determined by the physical problems at hand: The physical quantity to be calculated in Rivier's thesis was the anomalous magnetic moment of the neutron. Rivier thus had to calculate the S-Matrix (to a given order of perturbation theory) for the scattering of a particle with a given magnetic moment in an external field from the usual Hamiltonian formalism. He then calculated the S-Matrix for the scattering of a neutron (interacting with a pseudoscalar, charged nuclear meson field) in an external field, and extracted the expression for the (time average of the) magnetic moment by comparing it with the general expression. The formulation of the field theory in terms of an S-Matrix theory was thus not determined directly by the physical problems at hand, but rather, as we have reconstructed, by its origins in the S-Matrix program, which aimed to get rid of instantaneous states entirely. That is not to say that it would have been easy or even practicable to construct the theory otherwise at the time (though we will take a brief look at the Schwinger's more traditional approach later on). All I want to argue is that the reason these methods were so readily available and also so readily accepted as a new formulation of quantum field theory was that they had been prepared in the much more radical S-Matrix theory.

It is also not the case that the deficiencies of a pure S-Matrix formulation of QFT were not clear to physicists of the time. Stueckelberg almost immediately set to work and tried to make renormalized, covariant QFT mathematically consistent also for finite times \citep{stueckelberg_1951_relativistic}.\footnote{See also \citep{cianfrani_2008_stueckelberg}.} But this was already done within a framework of quantum field theory in which the description of scattering processes between asymptotic states at infinite times had become the standard formulation, while the extension to finite times was just that: an extension of the theory. Of course, Stueckelberg's work had little to do with the establishment of this new framework. This was mainly due to the work of Richard Feynman, which had its origin in a different attempt to get rid of the instantaneous quantum mechanical state and then only merged with the S-Matrix program through the work of Freeman Dyson. This development will be discussed in the second half of this paper.

\section{Feynman's QED}

Heisenberg's S-Matrix was largely a European affair. During wartime, its influence was restricted to some countries neighboring Germany, occupied and not, which Heisenberg visited personally (Kramers in the Netherlands, M\o ller in Denmark, Stueckelberg in Switzerland), and after the war it spread to the United Kingdom and Ireland, but it never became a big thing in the United States, where Pauli had given it a bad name early on, before moving back to Switzerland, where he became somewhat more involved, but never enthusiastic. Modern, renormalized QED, however, as is well-known, was to a large extent developed in the US. How then did Heisenberg's attempt to get rid of the quantum mechanical state come to have such a large impact on the newly adopted formulation of quantum field theory? Here, two factors came together. First, there was an entirely independent attempt to solve the difficulties of QED by getting rid of states, by Richard Feynman. And, second, Feynman's work was then brought together with Heisenberg's S-Matrix by Freeman Dyson, from Britain, who was well-versed in the work done on the S-Matrix in Europe, leading to the modern formulation of QED viewed mainly as a theory for determining S-Matrix elements. I will discuss these developments in the second part of this paper, beginning with the work of Feynman. Just as for Stueckelberg, it has its origins in an attempt to solve the divergence difficulties at the classical level, but not by modifying the theory of point charges, but by modifying Maxwell's theory itself, by eliminating the concept of the electromagnetic field altogether.

\subsection{Absorber Theory}

This idea to get rid of the electromagnetic field came to Richard Feynman, according to his own recollections, as an undergraduate student at MIT (1935-39), after studying QED in the standard textbooks of the time by Dirac and Heitler.\footnote{The following narrative is heavily indebted to \citep{mehra_1994_the-beat} and \citep{schweber_1994_qed}.} Interestingly, these two textbooks are good representatives of the two schools of thought on the origin of the divergence difficulties of quantum field theory discussed earlier. Heitler clearly argued for the quantum nature of the divergence difficulties \citep[p. 183]{heitler_1936_the-quantum}, while Dirac explicitly stated:

\begin{quote}
The limitations in the applicability of quantum electrodynamics [...] correspond precisely to those of classical electrodynamics. The amendments required in classical theory in order to make it applicable accurately to the elementary charged particles are thus not provided by the passage to the quantum theory... \citep[p. 296-297]{dirac_1935_the-principles}
\end{quote}

This view motivated Dirac's attempts at finding a new ``classical'' theory of the electron, discussed above. While Feynman initially tended towards Dirac's view on this matter, it appears to have been Heitler's detailed elaboration of the origin of the divergences that directly influenced his search for  a solution. Heitler identified two sources for the central divergent quantity, the self-energy of the electron. There is first the electrostatic self-energy, which is due to  the self-force, ``the force which the field produced by the charge exerts on the charge itself'' (p. 29). And then there is the transverse self-energy, which is due to virtual photons in the intermediate states of higher-order perturbation calculations. It is divergent because ``the theory gives in general non-converging results in cases where \emph{the number of intermediate states is infinite}'' (p. 184), i.e., because the electromagnetic field has an infinite number of degrees of freedom.\footnote{Feynman recalled that, at the time, he mistakenly believed that the infinite number of degrees of freedom only showed up in the infinite zero-point energy.  Both \citep[p.379-380]{schweber_1994_qed} and \citep[p.89]{mehra_1994_the-beat} read this as Feynman being mistaken on the whole difficulty of the infinite degrees of freedom, not taking into account that in the literature of the day the infinite degrees of freedom were made responsible not just for the zero-point but also for the transverse self-energy.}

To Feynman the immediate solution to both of these problems seemed to be the abolishment of the electromagnetic field. This would eliminate the infinite number of degrees of freedom. The theory could then be recast as a theory of action-at-a-distance between electrons, where the action would have to be taken as retarded and not as instantaneous. This would allow the self-consistent elimination of the interaction of an electron with itself, which was not possible if the interaction was described through a universal field.

Although this was probably not the young Feynman's immediate priority, it is clear that already this general proposal problematizes the notion of an instantaneous state: For a retarded action-at-a-distance it is not sufficient to know the initial conditions and let them evolve. Rather, one needs to know a lot about the past evolution of the system. 

When Feynman went to Princeton as a graduate student in 1939, his new advisor, John Wheeler, pointed him to a much more immediate problem in this approach. If one eliminated the action of an electron on itself, one not only got rid of the troublesome infinite electrostatic self-energy, but also of radiative reaction, i.e., the recoil experienced by an electron emitting electromagnetic radiation.

Together, Wheeler and Feynman modified Feynman's original proposal. They could show that if one replaced the retarded action-at-a-distance by a combination of retarded and advanced interactions (in field theory this would correspond to taking the average of the advanced and retarded electromagnetic potentials) one could include the radiative reaction: The radiative reaction emerged as the reaction of the emitting electron to the advanced back-reaction of all the other electrons in the universe, if one assumed that there were sufficiently many of these to ensure that all emitted radiation would eventually be absorbed (hence, Wheeler-Feynman electrodynamics is also known as absorber theory).

There remained the central difficulty of the mathematical formulation: Even more than in a theory of retarded interactions, a theory of advanced and retarded interactions could not be formulated as a Hamiltonian theory of the time evolution of instantaneous states. But Wheeler happened upon a paper by the Dutch physicist Adriaan Fokker in which this difficulty was solved.

In 1929, Fokker had in fact constructed a formulation of the electrodynamics of point particles as a combination of advanced and retarded actions-at-a-distance. He had not been concerned with the question of the radiation reaction, but he had arrived at the same combination of advanced and retarded actions-at-a-distance merely through the demand of a relativistically invariant action functional \citep{fokker_1929_wederkeerigheid}. The formulation using the principle of least action avoided the difficulties of the Hamiltonian formalism: The entire world-lines (space-time trajectories) could be obtained from the minimization of the action functional instead of constructing them from inifinitesimal time evolution. In Feynman's notation, the Fokker action reads (for a set of electrons, labeled by an index $a$, with coordinates $x_a$, proper times $\tau_a$, charges $e_a$, and masses $m_a$):

\begin{equation}
\label{eq:fokker}
S = - \sum_a m_a c \int_{-\infty}^{\infty} \sqrt{- \dot{x}_a^{\mu} \dot{x}_{a \mu}} d \tau_a + \sum_{a < b} \frac{e_a e_b}{c} \int_{-\infty}^{\infty} \int_{-\infty}^{\infty} \delta \left( \left( x_a^{\mu} - x_b^{\mu} \right)^2 \right) \dot{x}_a^{\nu} \dot{x}_{b,\nu} d \tau_a d \tau_b
\end{equation}

Here, the first term is simply the kinetic energies of the single electrons, while the second term gives the pairwise interactions between the electron's currents, with the $\delta$ function ensuring that only those points on the world lines interact that are on each others light-cones, both forward and backward.

When supplemented with the absorber initial conditions, this action provided an elegant reformulation of classical electron theory without involving the notion of a field. But this was of course only the starting point for the actual program: The construction of a new quantum electrodynamics. Wheeler was constantly announcing that the quantization of the absorber theory was almost ready, while not communicating any of the details, so Feynman had to ponder the question by himself, taking the Fokker action as his starting point.

Fokker himself had already presented his action as an alternative starting point for the construction of a quantum theory of electrodynamics, in a German republication of his work \citep{fokker_1929_ein-invarianter}, which had appeared shortly after the publication of the original Heisenberg-Pauli QED. The full extent of the divergence difficulties had not yet been realized at the time, so Fokker's main argument for the superiority of his least action formulation was its manifest relativistic invariance. But Fokker did not pursue the quantization of his theory any further. An obvious reason is that, while Heisenberg and Pauli could rely on the established quantization procedures of QM, it was unclear how to construct a quantum theory from a classical theory which was only formulated using a least action principle. The emphasis here is on \emph{only}. A general classical theory can of course also be written using the principle of least action, with an action functional of the form

\begin{equation}
\label{eq:action}
\mathcal{A} = \int_{t_0}^{t_1} L (q,\dot{q} ) dt
\end{equation}

with the Lagrangian $L$ some function of the dynamical variables $q$ and their time derivatives $\dot{q}$. From the Lagrangian one can then construct the Hamiltonian (leaving aside questions of gauge invariance, which forms an entirely unrelated difficulty) in the usual manner, which forms the starting point for canonical quantization. But the Fokker action differs from this usual form in two decisive aspects. The first is that the integrations are over the proper times of all of the individual particles instead of over some universal time coordinate. The second is that the integrations are carried out from $-\infty$ to $\infty$, instead of from an initial time $t_0$ to a final time $t_1$.

It is this latter difficulty which is related to the central focus of this paper: For an action formulated in terms of a Lagrangian, the time integration range can be taken infinitesimally small. The action is then minimized in each infinitesimal step and thus contact is made with the differential time evolution of instantaneous states. This is not possible for the Fokker action: Since the interaction is retarded (and advanced), one always needs to take into account the entire trajectories. This is another way of stating that the notion of state is lost in such an action-at-a-distance theory. 

Over a decade later, Feynman was faced with the same difficulty. But, different from Fokker, he could draw on the work of Paul Dirac, who had made some first steps towards the quantization of a classical theory formulated using the least action principle. It is to this work of Dirac's that we now turn.

\subsection{Dirac's Lagrangian Quantum Theory}

The relation between quantum theory and relativity is a \emph{leitmotif} in the work of Paul Dirac. Early on, he attempted to reconcile matrix mechanics and relativity by elevating time to the status of q number \citep{dirac_1926_relativity}. In 1928, he presented the relativistic generalization of the Schr"odinger equation \citep{dirac_1928_the-quantum1}, certainly his best-known contribution to this problem context. In 1933, he presented another step towards a more relativistic quantum theory \citep{dirac_1933_the-lagrangian}, his paper ``The Lagrangian in Quantum Mechanics.'' Parts of this paper are then taken up in the second edition of Dirac's textbook \citep{dirac_1935_the-principles}. This seems to have been Feynman's source: In his thesis (reprinted in \citep{brown_2005_feynmans}), only Dirac's book is cited, the reference to the 1933 article (which was published in the \emph{Physikalische Zeitschrift der Sowjetunion}) only appears as a ``see also'' in the later article based on the thesis \citep{feynman_1948_space-time}. We thus will only concern ourselves with the material reprinted in 1935 and not with the (even) more speculative parts of Dirac's paper, which concern the generalized transformation function and which were also very influential, especially for the further development of quantum field theory in Japan.

The starting point of this work is Dirac's dissatisfaction with the usual Hamiltonian formalism, but for a reason unrelated to Feynman's difficulty: its lack of manifest covariance. Indeed, the Hamiltonian of a system is not a relativistic invariant, it is rather the time component of a four vector. Dirac was thus looking for a formulation of QM which was closer to the classical Lagrangian formalism, since at least the time integral of the Lagrangian function \emph{is} a relativistic invariant. This was, however, far from straightforward, as he outlined in the introduction to the 1933 paper:

\begin{quote}
A little consideration shows, however, that one cannot expect to be able to take over the classical Lagrangian equations in any very direct way. These equations involve partial derivatives of the Lagrangian with respect to the coordinates and velocities and no meaning can be given to such derivatives in quantum mechanics. The only differentiation process that can be carried out with respect to the dynamical variables of quantum mechanics is that of forming Poisson brackets and this process leads to the Hamiltonian theory.\\
We must therefore seek our quantum Lagrangian theory in an indirect way.
\end{quote}

This indirect way relied on another role played by the Lagrangian in classical mechanics, besides its appearance in the Euler-Lagrange equations of motion, which result from the extremization of the action. The Lagrangian also appears in the expression for Hamilton's principal function $S$ (not to be confused, of course, with Heisenberg's S-Matrix), that is the action functional $\mathcal{A}$ (Equation \ref{eq:action}) evaluated on the actual, physical path. Since this path is obtained by extremizing the action functional, the principal function is the extremum value of the action functional. In the classical theory, this principal function, taken as a function of the beginning and end points of the trajectory, $q_T$ and $q_t$, can be viewed as the generator of a canonical transformation that connects the canonical variables at times $T$ with those at time $t$, i.e.

\begin{eqnarray}\label{eq:canonical}
p_t & = & - \frac{\partial S}{\partial q_t} \nonumber\\
p_T & = & \frac{\partial S}{\partial q_T}
\end{eqnarray}

Instead of introducing the Lagrangian into quantum mechanics via the equations of motion, as it was done for the Hamiltonian (be those equations of motion the Heisenberg picture equations of motion, which include the commutator of the Hamiltonian with a given observable, or the Schr"odinger equation), Dirac could now introduce the Lagrangian through the relation between classical canonical transformations and quantum unitary transformations. 

This relation had been an important issue in the early days of matrix mechanics \citep{lacki_2004_the-puzzle}. It was made obsolete by the establishment of transformation theory and modern quantum mechanics. Some questions were, however, left open, including the one at stake here: Given a classical canonical transformation, what is the corresponding unitary transformation in quantum mechanics? Dirac attempted to give an answer to this question in 1933. His argument is, however, quite opaque. There is one thing that he unambiguously shows: Just as for classical canonical transformations, unitary transformations in quantum mechanics can also be expressed by a generating function (which is a function of the old and the new canonical coordinates, as was the case above). This is done in the following way: Let the scalar product between an eigenstate of the old canonical coordinates $q$ with the eigenvalue $q'$ and an eigenstate of the new canonical coordinates $Q$ with the eigenvalue $Q'$ be given by:

\begin{equation}
\Braket{q' | Q' } = e^{i F(q', Q') / \hbar}
\end{equation}

where $F$ is some generating function. Dirac could then show that the matrix elements of the old and new canonical momenta (and, imposing some restrictions concerning the ordering of operators, also the momentum operators) obeyed relations analogous to the classical transformation equations (Equations \ref{eq:canonical}):

\begin{eqnarray}
\Braket{q' | p | Q' } & = & \frac{\partial F(q',Q')}{\partial q'}  \Braket{q' | Q' }\nonumber\\
\Braket{q' |P | Q' } & = & - \frac{\partial F(q',Q')}{\partial Q'} \Braket{q' | Q' }
\end{eqnarray}

The decisive question is now: What is the relation between the classical and the quantum generating function? From Dirac's elaborations, both in the 1933 paper and the 1935 book, it appears that he did not really consider this question. He introduces the generating function in the quantum theory as a distinct mathematical object, without giving any relation between the quantum and classical generating functions. They are merely treated as analogous. \footnote{Indeed no general relation between quantum and classical generating functions was established by Feynman, who only deals with infinitesimal time translations. If equation 68 on page 113 of Dirac's book were intended to imply the general numerical equality of quantum and classical generating functions, it would be plain wrong - it would not even hold for finite time evolutions. But Dirac clearly identifies the $S$ in this equation as merely denoting the quantum generating function. I therefore believe that W"uthrich's argument that Dirac here anticipated Feynman's work \citep[p.53]{wuethrich_2010_the-genesis} is fallacious. Julian Schwinger's take on the whole question was the following: ``Now, we know, and Dirac surely knew, that to within a constant factor the ``correspondence'' for infinitesimal $dt$ is an equality when we deal with a system of nonrelativistic particles possessing a coordinate-dependent potential energy $V$. [...] Why, then, did Dirac not make a more precise, if less general, statement? Because he was interested only in a general question: What, in quantum mechanics, corresponds to the classical principle of stationary action ?'' \citep{schwinger_1989_a-path}}\footnote{Dirac's tantalizing analogy had also been noticed by M\o ller, who suggested that it might provide a starting point for constructing the S-Matrix directly from the classical theory, without going through the Hamiltonian and canonical quantization (Lectures delivered at the H.H. Wills Physical Laboratory, University of Bristol, Spring 1946, lecture notes by I.N. Sneddon). However, he did not pursue this idea any further, nor did he offer any ideas on how to read Dirac's analogy. I would like to thank Thiago Hartz for pointing me to M\o ller's lecture.}

Applying this to the case of the classical time-evolution canonical transformation discussed above, this implied that also in the quantum case, there would be a generating function related to unitary time evolution, which is (and this certainly adds to the confusion) also denoted by $S$ and is given by

\begin{equation}
\Braket{q_t' | q_T' } = e^{i S(q_t', q_T') / \hbar}
\end{equation}

where $\ket{q_T'}$ and $\ket{q_t'}$ are eigenkets of the canonical coordinate operators (in the Heisenberg representation) at the initial and final times. $S$ can then be considered the quantum analogue of the classical principal function. Since the precise nature of this analogy is not defined, the only upshot of Dirac's study was that a quantum theory modeled more closely on the classical Lagrangian formulation would be based on the (transformation) matrix elements $\Braket{q_t' | q_T' }$, rather than on a quantum state evolving in time. Such a formulation could thus be a better starting point for a relativistic generalization of QM.

Dirac's program (which he himself did not pursue much further) was thus very similar to that of Heisenberg in early 1937: rewrite quantum theory in a more explicitly covariant manner, in order to have a better starting point for obtaining the correct relativistic theory. And, indeed, they ended up at very similar formulations: Dirac's transformation matrix elements are simply the matrix elements of the Schr"odinger (or interaction) picture time evolution operator, expressed in the Heisenberg picture. So, just like Heisenberg in 1937, Dirac's attempts at re-writing quantum theory in a more relativistic fashion, had led him to an approach in which the procedural took precedence over the study of stationary states.

What Dirac did not supply was any novel way of calculating these transformation matrix elements. The question of what the quantum generating function would be was left entirely open. It was Feynman who was to fill this gap, at the same time taking the step that Heisenberg had also taken, namely to move from a theory which still made explicit reference to arbitrary initial and final states of a process, to a theory in which only free, asymptotic states appeared and the notion of the differential evolution of an instantaneous state was gone. And in both cases, it was the attempt to solve the divergence difficulties of QFT which motivated this next step: for Heisenberg it had been the fundamental length, for Feynman it was action-at-a-distance electrodynamics.

\subsection{Feynman's Path Integrals}

Feynman's realization that there is a very close relation (indeed proportionality) between the classical and quantum generating functions for the special case of infinitesimal time evolution occurred in the Spring of 1941. The amusing anecdote of how he hit upon this fact, with the help of German physicist Herbert Jehle, is recounted by himself in his Nobel lecture. Using this fact, and iterating the infinitesimal time evolutions to obtain an expression for finite time evolutions, he arrived at his famous path integral formulation of (non-relativistic) quantum mechanics. The details of this discovery need not concern us here.\footnote{They are discussed, e.g., in \citep{wuethrich_2010_the-genesis}. Also, the gradual shift from a more formalistic view to the more visualizable sum-over-paths view is discussed in \citep{schweber_1986_feynman}.} What Feynman ended up with was an expression for arbitrary matrix elements between an initial state $\psi$ at time $t_0$ and a final state $\chi$ at time $t_1$. As a simplest case, this gives an expression for Dirac's transformation matrix elements, generalized to an arbitrary initial and final state:

\begin{equation}
\label{eq:amplitude}
\Braket{\chi (t_1) | \psi (t_0)} =   \int  \chi^{\ast} (x_k, t_1) \; \mathrm{exp}\left[ \frac{i}{\hbar} \sum_{i=0}^{k-1} L \left( x_{i+1}, \frac{x_{i+1}-x_i}{\epsilon} \right) \epsilon \right] \psi (x_0, t_0) \frac{dx_0}{A} \frac{dx_1}{A} \cdots \frac{dx_{k-1}}{A} dx_k
\end{equation}

where

\begin{equation}
\epsilon = \frac{t_1 - t_0}{k}
\end{equation}

is the infinitesimal time interval introduced through the iteration of the infinitesimal time evolution and $A$ is a normalization factor. In the limit where $\epsilon$ is taken to zero, the familiar path integral arises, but Feynman eschewed the highly problematic notion of functional integration, both in his thesis and his first paper on path integrals in 1948. Still the limit of continuous time was conceptually important, because in this limit the sum in the exponential would be replaced by an integral (and the difference quotient in the argument of the Lagrangian by a time derivative), giving, and this was Feynman's second great insight, the classical action functional of equation \ref{eq:action}.

This realization allowed Feynman to go beyond a mere reformulation (albeit a very elegant and hugely influential one) of regular non-relativistic QM to the quantization of theories which had no Hamiltonian, only an action. This had been, after all, his goal: to quantize the absorber theory starting from the Fokker action.

Here now, in discussing in general the quantization of a classical theory with only an action, Feynman really began to wrestle with the difficulty of the absence of states. For after all, even in the action method (assuming a regular, Lagrangian-based action) the matrix elements still explicitly depend on the initial and final quantum mechanical states at some finite times $t_0$ and $t_1$. Feynman's solution was to assume that a non-Lagrangian nature of the system would only exist for a finite time:

\begin{quote}
This difficulty may be circumvented by altering our mechanical problem. We may assume that at a certain very large positive time $T_2$, and at a large negative time $T_1$, all of the interactions (e.g., the charges) have gone to zero and the particles are just a set of free particles (or at least their motion is describable by a Lagrangian). We may then put wave functions, $\chi$ and $\psi$, for these times, when the particles are free, into [the equation for the matrix elements]. (We might then suppose that the motion in the actual problem may be a limit of the motion as these times $T_1$ and $T_2$ move out to infinity).
\end{quote}

Feynman thus, just like Heisenberg, arrived at the point where the theory could only speak about transitions between asymptotic free states, while the evolution of the state in some intermediate area of interaction was inaccessible to the theory. This establishment of this view may be viewed as stemming from philosophical preferences. Indeed, Feynman himself in his Nobel lecture stated that at this time he developed a preference for an ``overall space-time point of view,'' where one always only considered the full evolution of systems, and a ``disrespect for the Hamiltonian method of describing physics,'' where ``things are discussed as a function of time in very great detail.''  \citet[p. 393]{schweber_1994_qed} also claims that Feynman's recasting of QM was ``clearly influenced by the S-matrix viewpoint Wheeler had expounded to him.''

What is this ``S-Matrix viewpoint'' of Wheeler's? For one, \citet{wheeler_1937_on-the-mathematical} was the first one to use the S-Matrix --- years before Heisenberg, though Heisenberg claimed he developed the idea independently \citep{rechenberg_1989_the-early}, which is plausible, given that he had already arrived at the idea of an integral scattering operator in 1937. In any case, for Wheeler the S-Matrix was merely a calculational tool in non-relativistic nuclear physics and not the central quantity of the current theory or of a future one. Rather, Schweber is referring to a view of Wheeler's ``that all quantum-mechanical descriptions of physical phenomena could be construed as scattering processes'' \citep[p. 379]{schweber_1994_qed}. Wheeler jokingly referred to these ideas of his as ``everything as scattering,''  and they most certainly influenced Feynman's diagrammatic approach to perturbation theory. But these developments were still far off in 1942, and Wheeler himself placed his work on Wheeler-Feynman electrodynamics in the context of a different, albeit similarly ambitious, program of ``everything as electrons.'' \citep{wheeler_1989_the-young}\footnote{Indeed, Wheeler's recollections appear to imply that the ``everything as electrons'' program was the actual starting point of Wheeler-Feynman electrodynamics, not Feynman's ambition to eliminate field and self-energy. It is impossible to resolve this priority debate. In all their published work, Wheeler and Feynman only made reference to Feynman's motivation, as Wheeler was at this time still very shy about his grander schemes for physics. As he remarked in an interview with Charles Weiner and Gloria Lubkin on 5 April 1967 (https://www.aip.org/history-programs/niels-bohr-library/oral-histories/4958): ``And this I think illustrates a weakness of my approach at that time, to have this secret hope nursed internally and talk about it occasionally with close friends but not feeling particularly at ease about bringing it out on a public platform except insofar as one could talk of some specific thing that was a clean cut result that one could write up and publish, but publish it without making clear what the longer term goal is...'' } Also, the idea of ``everything as scattering'' is clearly distinct both from Heisenberg's philosophical starting point of reducing everything to observables (which played no apparent role for Feynman) and from Feynman's own space-time point of view. So, while such general philosophical considerations certainly played a role for all the involved actors, they do not explain the surprising convergence of the programs of Heisenberg and Feynman.

The common origin of their theories of free, asymptotic states is rather the belief that the failure of QED should be understood from the misguided ambition of giving a detailed picture of infinitesimal time development. Their reading of why such a picture is impossible is very different (fundamental length v. action-at-a-distance), but the conclusions are strikingly similar. It is indeed probable that Feynman was driven to the adoption of his viewpoint by the difficulties of QED, rather than by some overarching philosophical program. Schweber himself cites another interview with Feynman that Schweber himself conducted \citep[p. 396]{schweber_1994_qed}:

\begin{quote}
The reason in my philosophy not to descend to the Schr"odinger equation and to do as much physics as I could without doing that, is that I really believed at that time, in 1941-42, that this back action, this Wheeler-Feynman thing was really a forward step. That's why I was doing everything. I wanted to get the quantum mechanics of that, and that was in the form of a path integral; it had no Hamiltonian.
\end{quote}

And also in his thesis, we see Feynman wrestling with these issues, as in the following long quote, where he argues that one should at least keep the notion of an asymptotic, free state in order to interpret the theory:

\begin{quote}
It is not unreasonable that it should be impossible to find a quantity like a wave function, which has the property of describing the state of the system at one moment, and from which the state at other moments may be derived. In the more complicated mechanical systems [of action-at-a-distance] the state of motion of a system at a particular time is not enough to determine in a simple manner the way the system will change in time. It is also necessary to know the behavior of the system at other times; information which a wave function is not designed to furnish. An interesting, and at present unresolved, question is whether there exists a quantity analogous to a wave function for these more general systems, and which reduces to the ordinary wave function in the case that the action is the integral of a Lagrangian. That such exists is, of course, not at all necessary. Quantum mechanics can be worked entirely without a wave function, by speaking of matrices and expectation values only. In practice, however, the wave function is a great convenience, and dominates most of our thought in quantum mechanics. For this reason we shall find it especially convenient, in interpreting the physical meaning of the theory, to assume our mechanical system is such that, no matter how complex between the time $T_1$ and $T_2$, outside of this range the action is the integral of a Lagrangian. In this way, we may speak of the state of a system at time $T_1$ and $T_2$, at least, and represent it by a wave function.
\end{quote}

We can thus see, how the two war year programs that involved an elimination of the notion of state from quantum mechanics can both be viewed as directly resulting from the perceived inability of a quantum field theory to address questions of incremental time evolution. But like the S-Matrix formalism, the path integral was only a general frame for a quantum theory, yet to be filled with specific particles and interactions. How now did Feynman's least action formalism fare as the foundation of a new QED? What became of the attempt at using it to quantize Wheeler-Feynman electrodynamics? It is to this question that we now turn.

\subsection{The Quantization of Action-at-a-Distance Theories}

The short answer is: Feynman did not quantize Wheeler-Feynman electrodynamics in his thesis. The first major difficulty was in fact not the action at a distance, but rather the relativistic kinematics of the particles. This was also the problem Wheeler was struggling with \citep[130]{mehra_1994_the-beat}. Feynman would return to these questions only after the war, when his struggle with the Dirac equation began, which is recounted in great detail in \citep{wuethrich_2010_the-genesis}. We will also return to this point. For his thesis, Feynman restricted himself to non-relativistic theories.

Apart from the difficulties of relativistic kinematics, there was a further major, and more general, problem in Feynman's approach to quantizing non-Lagrangian theories. He was, in fact, very unsatisfied with circumventing the problem of the lack of a state described by a wave function by talking only of free, asymptotic states. In the section following the long quotation above, he went on to write (p.49):

\begin{quote}
The physical interpretation which is given in the above section, although the only consistent one available, is rather unsatisfactory. This is because the interpretation requires the concept of states representable by a wave function, while we have pointed out that such a representation is in general impossible. We are therefore forced to alter our mechanical problem so that the action has a simple form at large future and past times, so that we may speak of a wave function at these times, at least. [...] We have not defined precisely what is to be done when the action does not become simple at times far from the present.\\
One possibility that suggests itself is to devise some sort of limiting process so that the interpretation of the last section could be used, and the limit taken as $T_1 \rightarrow - \infty$ and $T_2 \rightarrow + \infty$. The author has made several attempts in this direction but they all appear artificial, having mathematical, rather than physical, content.
\end{quote}

Feynman thus attempted to construct a formulation of the theory in which there was no more talk of initial and final states at all:

\begin{quote}
An alternative possibility is to avoid the mention of wave function altogether, and use, as the fundamental physical concept, the expectation value of a quantity, rather than a transition probability.
\end{quote}

This attempt was plagued by difficulties: Feynman could not devise a general method to get only real expectation values for physical quantities. This was especially problematic concerning the expectation value of the energy, where a complex expectation value implied the loss of unitarity (i.e., probabilities did not add up to one). In his Nobel lecture, Feynman stated that he wrote up his thesis in ``in one of the short periods during which I imagined I had laid [the difficulty] to rest.'' But even in his thesis he characterized his approach as ``very incomplete and the results tentative.'' And indeed, in the following years he reached the conclusion that it would not work:

\begin{quote}
During the war, I didn't have time to work on these things very extensively, but wandered about on buses and so forth, with little pieces of paper, and struggled to work on it and discovered indeed that there was something wrong, something terribly wrong.
\end{quote}

Returning to these questions after the war in earnest, he somewhat grudgingly accepted working with asymptotic free states, as witnessed, e.g., by the following quote from a later paper \citep{feynman_1950_mathematical}:

 \begin{quote}
 [W]e can imagine the charges to be turned on after [$T_1$] adiabatically and turned off slowly before [$T_2$] [...]. Hereafter we shall [...] consider the range of integration of $t$ to be from $- \infty$ to $+ \infty$, imagining, if one needs a definition, that the charges vary with time and vanish in the direction of either limit.
 \end{quote}
 
 While it is not necessary for our story to study in detail Feynman's attempts in his thesis at eliminating even the asymptotic states, it is important to understand how this affected his approach to the quantization of non-Lagrangian theories. The incompleteness of his quantization program implied that he could not simply take an action for a non-relativistic theory of action-at-a-distance and plug it into a path integral. Instead, he took a more conservative approach.
 
Recall that Wheeler-Feynman electrodynamics is removed from the classical field theory in two steps, which are related to the two difficulties of QED that Feynman had learned as an undergrad. First, there is the elimination of the field degrees of freedom and their replacement by a half-retarded, half-advanced direct interaction. The resulting theory of particles interacting at a distance is equivalent to the field theory - the only modification is that a certain solution of the field equations has been singled out. The second step is the removal of the electron's self-interaction, which makes the resultant theory inequivalent to the field theory.

Instead of now directly quantizing a non-Lagrangian theory, Feynman only attempted to find a quantum mechanical generalization of the elimination of the field. The resultant quantum theory would then of course not have a Lagrangian and was supposed to be describable only in terms of Feynman's ``expectation value'' formulation of the path integral for non-Lagrangian theories. But at the same time, he was able to keep a clear relation to a theory formulated in the well-established language of Hamiltonian QM. Expressed positively, as Feynman did in the abstract of his thesis, this meant that ``the results serve as a confirmation of the proposed generalization.'' One can of course also read this, as saying that Feynman did not dare venture too far into unknown territory with the tentative (and soon to be discarded) tools he had at hand.

The non-relativistic toy model for the elimination of the field was the interaction of two particles (coordinates $y$ and $z$, respectively) interacting via a harmonic oscillator (coordinate $x$, mass $m$, frequency $\omega$), described classically by the Lagrangian (in which the oscillator is not yet eliminated) :

\begin{equation}
L = L_y + L_z + \left(\frac{m \dot{x}^2}{2} - \frac{m \omega^2 x^2}{2} \right) + (I_y + I_z) x
\end{equation}

where $I_y$ and $I_z$ are arbitrary functionals of $y(t)$ and $z(t)$, respectively, describing the interaction of the particles with the oscillator. This was a natural starting point, since QED could be formulated, and this is how Feynman learned it from Fermi's influential review article \citep{fermi_1932_quantum}, as a set of particles interacting via a continuum of harmonic oscillators, one for each mode of the electromagnetic field.

In the classical theory, one can now eliminate the oscillator degree of freedom by solving the Euler-Lagrange equation of motion for the oscillator and by then plugging the chosen solution (which will be a functional of $y(t)$ and $z(t)$ and a function of the boundary conditions imposed on the oscillator) into the equations of motion for the particle. The important fact is that the oscillator equation of motion can be solved independently because the interaction term is linear in $x$, making the inhomogeneous term in the oscillator equation of motion merely a function of $y$ and $z$.

The important question is then whether the new equations of motion for the particles, which are no longer of the Euler-Lagrange form, can still be obtained from the minimization of some new action, which is then of course no longer simply the time integral of a Lagrangian. Feynman could show that this depended on what type of boundary conditions one used to specify the solution of the harmonic oscillator equation of motion, and what kind of parameters would consequently show up in the action-at-a-distance particle equations of motion.

If one specified the initial conditions (position and velocity, i.e., the usual Cauchy boundary conditions) of the oscillator, one ended up with equations of motions for the particles that could not be derived from an action. If one instead specified the position at some initial and at another final time (Dirichlet boundary conditions), one did end up with a theory that could be cast into an action. Feynman finally considered imposing conditions on two specific linear combinations $R_0$ and $R_T$ of initial ($t=0$) and final ($t=T$) positions and velocities (a special case of Robin boundary conditions):

\begin{eqnarray}
R_0 & = & \frac{1}{2} \left[ x(0) + x(T) \cos{\omega T} - \dot{x} (T) \frac{\sin{\omega T}}{\omega} \right] \nonumber\\
R_T & = & \frac{1}{2} \left[ x(T) + x(0) \cos{\omega T} - \dot{x} (0) \frac{\sin{\omega T}}{\omega} \right]
\end{eqnarray} 

These boundary conditions could be interpreted in the following way:

\begin{quote}
$R_T$ is the mean of the coordinate of the oscillator at time $T$ and what the coordinate would have been at this time if the oscillator had been free and started with its actual initial conditions. Similarly, $R_0$ is the mean of the initial coordinate and what that coordinate would have had to be, were the oscillator free, to produce the actual final conditions at time $T$.
\end{quote}

They also led to an action-at-a-distance particle action, which depended on the parameters $R_0$ and $R_T$. Of particular interest to Feynman was the special case where both of these parameters are set equal to zero. The resulting action for this parameter choice had the special feature that the initial and final times could unproblematically be taken to negative and positive infinity, respectively. The action in this limit 

\begin{equation}
\label{eq:fokkertoy}
\mathcal{A} = \int_{-\infty}^{\infty} \left[ L_y + L_z \right] dt + \frac{1}{2 m \omega} \int_{-\infty}^{\infty} \int_{-\infty}^{t} \sin{\omega (t-s)} \gamma(t) \gamma (s) ds dt
\end{equation}

where $\gamma$ is short for $I_y + I_z$, was then also  time-translation invariant and thus allowed for the definition of a conserved energy, which now of course was a function of the full particle trajectories. 

The special significance of this action for Feynman is not immediately apparent upon a cursory reading of his thesis. \citep[p. 133]{mehra_1994_the-beat} claims that Feynman somehow proved the uniqueness of this choice of boundary conditions, given the constraints of obtaining a time translation invariant action. This is, however, certainly an exaggeration (e.g., Feynman considered neither Neumann boundary conditions, nor more general Robin boundary conditions) and Feynman never makes such a claim. Mehra further identifies the choice of boundary conditions as chosing ``a definitely determined solution of the oscillator equation, a symmetric one which included one-half advanced and one-half retarded interaction between the atoms.'' But this physical interpretation is very unclear: If there is only one oscillator, there is no dispersion relation and consequently neither a group or a phase velocity which might give a meaning to the notion of advanced or retarded interactions; and again, Feynman makes no such claim. What he did remark, off-handedly, was that

\begin{quote}
In electrodynamics it [the action of equation \ref{eq:fokkertoy}] leads to the half advanced plus half retarded interaction used in the action at a distance theory.
\end{quote}

Indeed, as Feynman showed in his thesis, one can generalize the procedure outlined above to the case of a large number of oscillators. In particular (and this Feynman did not do explicitly in his thesis), one can generalize to the case of electrodynamics, with one oscillator for each radiation mode of the electromagnetic field. Further generalizing to the case of velocity-dependent interactions (which Feynman had refrained from doing in his thesis (p. 18) ``for simplicity only''), one can take the interaction of a particle (described by its position vector $\mathbf{y}$ and possessing a charge $q$) with the oscillator corresponding to the radiation mode of wave vector $\mathbf{K}$ (frequency $c \vert \mathbf{K} \vert$) and polarization vector $\mathbf{e}$ to be given by the usual minimal coupling of electron theory\footnote{Note that for each such oscillator there is another one with the cosine replaced by a sine in the interaction term.}

\begin{equation}
I_y = \sqrt{8 \pi} q_y \left( \mathbf{e} \cdot \dot{y} \right) \cos{\left( \mathbf{K} \cdot  \mathbf{y} \right)}
\end{equation}

After eliminating all the oscillators, with boundary conditions $R_0= R_T=0$ for each one of them, the resultant action indeed describes half-advanced, half-retarded electromagnetic interactions, i.e., it is the Fokker action, but with an additional self-energy term. Feynman's above comment and the central role that the action of equation \ref{eq:fokkertoy} plays in his thesis clearly indicate that he was aware of this.

Then why didn't Feynman include such an analysis in his thesis, making the connection between his toy model and Wheeler-Feynman electrodynamics more explicit? There is an additional difficulty in obtaining half-advanced, half-retarded action at a distance through the elimination of the electromagnetic mode oscillators: In addition to the Fokker action (plus self-energy), there are additional boundary terms, which can only be made to vanish by assuming that the interaction is adiabatically turned off in the distant past and future. And as we have seen, Feynman was very hesitant to make such assumptions at the time. This may well explain why he only hinted in passing at the connection between his toy model and the Fokker action.

Restricting himself to the one-oscillator toy model, the next question still was: How was the elimination of the oscillator to be carried over into the quantum theory? The general idea was clear: Given a path integral involving both the particles and the oscillator, one would perform the integration only over the coordinates of the oscillator and arrive at a new expression for the matrix element in question that would now only involve (functional) integration over the coordinates of the particles. Feynman showed how this could be done in general, effectively developing techniques for the functional integration of Gaussian integrals. But what now was the analogue of the classical choice of boundary conditions for the oscillator?

The obvious answer would seem to be setting the initial and final quantum mechanical states of the oscillator. But this obvious answer was not available to Feynman at the time, since he was attempting to eliminate the notion of state altogether. The alternative he presented in his thesis is quite involved, but basically boils down to simply choosing some boundary conditions for the oscillator coordinate and its time derivative, in full analogy to the classical theory. The (rather unspectacular) result was then that a given choice of boundary conditions always led to the same action-at-a-distance theory, no matter whether the oscillator was eliminated classically or quantum mechanically. This is where Feynman left things when heading off to Los Alamos. 

\subsection{Applying Path Integrals to QED proper}

His work immediately after the war was, as already mentioned, mainly devoted to the Dirac equation. Tentative, but unsatisfactory results of this work was presented at the Shelter Island conference in June 1947 (Conference Notes, Gregory Breit papers, Yale University Archives). Feynman was not much involved in the debates on the Lamb Shift at the conference, but upon returning to Cornell after the summer he attended a lecture by Hans Bethe, who had just completed his non-relativistic calculation of the Lamb Shift. In his talk, Bethe discussed the prospects of a fully relativistic calculation, stating that this would need some covariant technique for making the infinite integrals --- appearing in the calculation of the electromagnetic self-energy of the electron --- finite, in order to perform unambiguous subtractions.

To this task Feynman now set himself. The general idea of how to make the divergent integrals finite (replacing $\delta$ functions by something more smeared out) appears to have been clear to him from the start; the method was published in two papers in the summer of 1948, which make no reference at all to his path integral methods \citep{feynman_1948_a-relativistic, feynman_1948_relativistic}. But Feynman also wanted to apply his path integral techniques to the problem. Now, the original motivation was gone: The self-energy problem of QED could apparently be brought under control through the newly developing renormalization techniques (to which Feynman was already contributing), yes, it was in fact necessary to not eliminate the self-interaction of the electron entirely, since it manifested itself physically in measurable effects such as the Lamb shift. But Feynman was coming to realize that there were reasons for him to stick to his path integral, independently of the quantization of action-only theories, independently even of the new physical picture it offered: It was actually a great calculational tool. As he later remarked to Jagdish Mehra (p. 229):

\begin{quote}
[O]f course, the final answer for a physical problem like the scattering of two electrons [...] is simple, but it was the result of a rather complicated bunch of terms  [...] whereas when I started with my path integrals [...] I knew which terms went together, how they went together, and how to generalize to four dimensions from the two transverse dimensions. It was obvious; it would work; that was the fun of it. \emph{It would always work.} [...] I began to realize that I already had a powerful instrument; that I was sort of flying over the ground in an airplane instead of having so many terms.
\end{quote}

For Feynman's path integrals to fulfill this role, several challenges had to be overcome:

\begin{itemize}
\item First, since he was now convinced that the entirely state-free method for eliminating field oscillators that he had devised for his thesis did not work, he needed to come up with a new one. 
\item Second, the resulting action of direct particle interaction would not be of the Lagrangian form, and Feynman still had no real recipe for dealing with that case. 
\item And finally, Feynman needed to address how to treat a problem such as the Lamb Shift, which required the calculation of the energy levels of stationary states, in a theory such as his which only dealt with transitions between asymptotic states.
\end{itemize}

I will discuss these three challenges in turn.

The first problem, the elimination of the field oscillators, had probably already been met before Bethe's talk. Feynman's solution consisted of imposing genuinely quantum boundary conditions, by determining the quantum states of the field oscillators for asymptotic times $t=\pm \infty$. He thus re-instated at least the asymptotic states for the field oscillators, which he had attempted to get rid of in his thesis work. When exactly he made this shift is not entirely clear, but it shows up (for just one oscillator) for the first time in print in \citep{feynman_1948_space-time}, which was penned in Pittsburgh in the weeks immediately following the Shelter Island conference, in the summer of 1947. The elimination of all the field oscillators is found for the first time in handwritten notes, also from the summer of 1947,\footnote{\citep[p. 479]{schweber_1986_feynman} understates the novelty of these calculations. The calculations done for the thesis did not use quantum boundary conditions and the calculations in the published paper only dealt with the case of one oscillator.} but does not appear in print until \citep{feynman_1950_mathematical}. Interestingly, with quantum boundary conditions the resulting action showing up in the exponential in the path integral is \emph{not} simply the Fokker action. Rather, e.g., for the simplest quantum boundary conditions, i.e., for all field oscillators being in their quantum ground state initially and finally, the action reads:

\begin{equation}
\label{eq:fokker2}
S = - \sum_a \int_{-\infty}^{\infty}  \frac{1}{2} m_a \dot{\mathbf{x}}_a^2  d t + \frac{1}{2} \sum_{a, b} \frac{e_a e_b}{c} \int_{-\infty}^{\infty} \int_{-\infty}^{\infty} \delta_{+} \left( c^2 \left( t-s \right)^2 - \left( \mathbf{x}_a (t) - \mathbf{x}_b (s) \right)^2 \right) \left( c^2 - \dot{\mathbf{x}}_a (t) \cdot \dot{\mathbf{x}}_{b} (s) \right) dt ds
\end{equation}

This differs from the Fokker action in several obvious ways: The action that Feynman derived in 1947 assumed non-relativistic particle kinematics, as he had still not been able to incorporate the relativistic kinematics of the Dirac electron in his path integral formalism. Consequently, the kinetic energy for the particles (first summand) takes the non-relativistic form and the integrals are over (absolute) time coordinates, rather than over proper times. For consistency, also all the other expressions have been written in non-covariant form, even though they are equivalent to their analog in the Fokker action, e.g., $c^2 \left( t-s \right)^2 - \left( \mathbf{x}_a (t) - \mathbf{x}_b (s) \right)^2$ is equal to $\left( x_a^{\mu} - x_b^{\mu} \right)^2$. Also, the self-interaction of the particles is included in this action, and consequently the sum in the interaction term (the second summand) goes over all pairs of particles, and double-counting is remedied by an additional factor of $1/2$. But all of this was to be expected from the start.

The essentially novel thing is that the Delta Function of the Fokker action is replaced by the function $\delta_+$, which is obtained by dropping the negative frequency terms in the usual Fourier expansion of the Delta Function. It is complex, and thus the above action cannot be given a straightforward classical interpretation. But the restriction to negative frequencies can be understood, and was understood by Feynman \citep[p.772]{feynman_1949_space-time}, as only positive-frequency intermediate photons mediating the direct particle interactions. This is of course closely related to Stueckelberg's positive-frequency restrictions, which led him to his causal function. In particular, it implies that blatant acausalities as they appear in Wheeler-Feynman electrodynamics, removed only by the absorber (spatial) boundary conditions, did not occur in Feynman's path integral QED with quantum (temporal) boundary conditions. It is not clear whether this was a deliberate step by Feynman in response to his discussion with Bethe, since as already pointed out, they first appear in the Reviews of Modern Physics article which was drafted before that discussion.\footnote{\label{fn:bethe} \citep[p. 231]{mehra_1994_the-beat} is very vague on this question. \citep[p. 479]{schweber_1986_feynman}, more forthrightly, states that the above action results from the elimination of the oscillators in the retarded formalism. This is certainly misleading, as there is nothing in the theory that distinguishes the retarded and the Wheeler-Feynman case \emph{before} the imposition of boundary conditions. Still, it is quite possible that the quantum boundary conditions were added to the draft of the 1948 RMP paper only after the discussion with Bethe. There is no evidence for such a hypothesis; the only hint is that Feynman does explicitly acknowledge Bethe's input in the last paragraph of that paper.} In any case, even when attempting to remove the hindsight goggles, quantum boundary conditions grow quite naturally out of Feynman's path integral idea and in turn lead directly to the above action, with its more standard causal properties.\footnote{It displays these properties when it shows up as the action in the path integral. As already stressed, it is not clear how to interpret it as a regular classical action.} Indeed, when he finally published the full elimination of all the radiation oscillators \citep{feynman_1950_mathematical}, Feynman still felt the need to express his surprise that quantum boundary conditions so naturally led to retarded, instead of Wheeler-Feynman interactions:

\begin{quote}
[O]ne might have anticipated that $R$ [the action of equation \ref{eq:fokker2}] would have been simply $R'$ [Fokker action with self-interaction]. This corresponds, however, to boundary conditions other than no quanta present in past and future. It is harder to interpret physically. For a system enclosed in a light tight box [i.e., absorber boundary conditions], however, it appears likely that both $R$ and $R'$ lead to the same results.\footnote{On this last point, Feynman had claimed only a year earlier \citep[fn 7]{feynman_1949_space-time} that the equivalence between half-advanced, half-retarded interaction with absorber boundary conditions and regular retarded interactions, which Wheeler and he had demonstrated in the classical theory, could be shown by ``[a]nalogous theorems [...] in quantum mechanics,'' stating that their discussion would ``lead us too far astray.'' A year later, he adopted the more cautions wording above.} (Footnote 10)
\end{quote}

Instead of ending up with a new quantum theory of electrodynamics, as originally envisioned, Feynman instead had a path-integral formulation of regular QED. This new formulation turned out to be immensely powerful for the problems at hand, which required an explicit, manifest covariance. Such covariance had, incidentally, been Dirac's original motivation for his attempts at a Lagrangian formulation of quantum theory. They had now been realized in Feynman's formulation of QED.

What about the states? The oscillators had been eliminated leaving a direct interaction between particles. Since this interaction was not instantaneous, Feynman again faced the difficulty of defining instantaneous states for the particle, i.e., he was again faced with the difficulty of treating an action-only theory, the second difficulty mentioned at the beginning of this section. Recall that Feynman's original suggestion in his thesis had been to assume that the system was ``non-Lagrangian'' only for a finite period of time, i.e., that the interaction would cease at some point in the past and in the future, and that the particles could be assumed to be in asymptotic free states before and after. He had abandoned this suggestion in favor of his ill-fated attempts at an entirely state-free theory. In 1947, Feynman again returned to free, asymptotic states. This was a natural and consistent step, as the elimination of the field oscillators using quantum boundary conditions already necessitated the assumption that the electromagnetic interaction vanish for asymptotic times. It was even more natural in the context of the path-integral perturbation theory that Feynman devised\footnote{Again, it is impossible to say whether he did this before the discussion with Bethe, since it appears in the 1948 RMP paper, or whether those passages were later added to the original draft. See footnote \ref{fn:bethe}.} and which was of prime importance for the application to electrodynamics, where also in the usual formulation most calculations could only be performed in perturbation theory. Here, the non-Lagrangian interaction term (the second summand of the action of Equation \ref{eq:fokker2}) is considered to be non-vanishing only for a finite number of pairs of time points $(t,s)$; at all other points in time, and in particular for asymptotic times, the system is perfectly Lagrangian, as the first summand of equation \ref{eq:fokker2} is of the form $\int L_0 dt$.

But what about the applicability of this formalism to the specific problems at hand, the third challenge outlined at the beginning of this section? The extension to scattering problems involving external photons (an unthinkable process in Wheeler-Feynman electrodynamics, which had no notion of external photons) was quite straightforward: One simply replaced the asymptotic ground state for some of the radiation oscillators with an excited state, corresponding to the initial or final number of photons in that mode. But these were not the problems of immediate concern in the late 1940s. Rather, the rebirth of QED was related to classical spectroscopic problems, in particular the Lamb Shift, which Bethe had asked Feynman to calculate. How now did Feynman tackle this problem in his framework, which did not include the concept of a stationary state, much less of the energy of such a state?

The solution was very simple: The Path Integral method could also be used to treat the trivial case of no scattering, where the system is simply in some initial state and then stays in that state forever. The transition probability is then trivially 1. The transition amplitude is also easily calculable in regular, time-dependent quantum mechanics as $\exp{(i E t/\hbar)}$, where $E$ is the energy of the stationary state and $t$ is the time difference between the initial and final states. Thus, the energy of the stationary state enters directly into the transition amplitude for staying in a stationary state as the phase difference, similar to how Heisenberg had envisioned the calculation of energy levels in S-Matrix theory before hitting upon analyticity. There is of course still the problem that Feynman's path integral formalism was generally equipped to handle only infinite times: Otherwise, as we have seen, the elimination of the radiation oscillators would also lead to boundary terms, which only vanished as one took the time difference to infinity. But as soon as one knew the structure of such terms, one could simply drop them also for finite time intervals, assuming that the coupling between electron and radiation vanish not just for infinite, but already for finite times. This did not change the content of the theory: One was still dealing only with transitions between asymptotic free states. However, by introducing the artifice of a finite time $t$, Feynman was able to read the energy of a stationary state out of the transition amplitude from that state to itself.

This worked not just for free particles, where the no-scattering transition amplitude has the intuitively clear interpretation of the particle just moving in a straight line with constant energy, but also for bound states of a particle in a potential, in particular for the case of a single electron bound in the Coulomb potential of a hydrogen nucleus. For an unstable state (i.e., one from which real and not just virtual radiation would be possible), one would get imaginary contributions to the energy, because the transition probability is not 1 but rather $\exp{(- \lambda t)}$, with $\lambda$ the state's decay constant. But the energy can still be read off the real part of the phase. There was only one problem: While the hydrogen atom is an easily solvable problem in regular wave mechanics, Feynman had not yet solved it in his path integral formalism.\footnote{Indeed, this was not done until the 1980s \citep{duru_1982_quantum}.} In fact, the only problem he had fully solved using path integrals was the free, non-relativistic particle. This made the evaluation of the intervals of free, Lagrangian time evolution problematic, independent of all perturbation theory. Feynman here had to calculate the relevant expressions (i.e., the matrix elements and energies for the unperturbed problem) in regular wave mechanics and plug them into his path integrals. One sees here that, rather than providing a full calculational framework, Feynman's path integral technique now provided a guideline for how to fit together diverse inputs into a coherent and covariant final expression for a transition amplitude. Parallel to the genesis of Feynman Diagrams one sees here the genesis of Feynman Rules.

In this way, Feynman was able to obtain a (still divergent) perturbative expression for the self-energy of the electron $\Delta$ in a given state $\psi_0$ of unperturbed energy $E_0$ as (Feynman Papers, Caltech Archives, Box 12, Folder 9):\\

\begin{equation}
\Delta = \frac{e^2}{4 c} \sum_n \int \frac{d^3 \mathbf{k}}{\left\vert \mathbf{k} \right\vert} \int d^3 \mathbf{x} \int d^3 \mathbf{x}' \frac{\psi^{\ast}_0 (\mathbf{x}') \dot{x}'_{\mu} \psi_n (\mathbf{x}') \psi^{\ast}_n (\mathbf{x}) \dot{x}_{\mu} \psi_0 (\mathbf{x})}{c \left\vert \mathbf{k} \right\vert + E_n - E_0} \cos{\mathbf{k} \cdot (\mathbf{x}-\mathbf{x}')}
\end{equation}

where the sum goes over all the unperturbed states $\psi_n$ with energies $E_n$, expressions that were not derivable in the path integral formalism but had to be taken from regular wave mechanics. Note also that $\dot{x}'_{\mu} \dot{x}_{\mu}$ is Feynman's convenient short-hand notation for $\left( c^2 - \dot{\mathbf{x}}' \cdot \dot{\mathbf{x}} \right)$; the equation is not otherwise Lorentz covariant.

This was basically equivalent to the result obtained in the usual perturbation theory, which had been the starting point for Bethe's lamb shift calculation.\footnote{Bethe had additionally imposed the dipole approximation, where the photon momentum $\mathbf{k}$ is neglected. The relation to Bethe's expression can be most easily seen in Schweber's rewriting of Feynman's expression using exponentials instead of the cosine, Equation 8.8.37 of \citep{schweber_1994_qed}. Note that there appears to be an extraneous factor of $1/\hbar$ in Schweber's equation. Also the exact numerical coefficients do not match up; given that I need not reproduce numerical results, I have given up hunting down those factors of 2, $\pi$, and $(-1)$.} But Feynman's calculation had easily led him to an expression that contained both the Coulomb self-energy and the transverse self-energy, since in his formalism all degrees of freedom of the electromagnetic field were eliminated and not just, as in the usual treatment, those degrees belonging to the electrostatic field. Bethe's expression thus only contained two components of the velocity four-vector $\dot{x}_{\mu}$ (those space-like components that are orthogonal to the polarization of the virtual photon), instead of all four.\footnote{The expression he obtained is thus not identical to that of Bethe, as stated in \citep[p.422]{schweber_1994_qed}, nor is it the relativistic generalization, as claimed in \citep[p.  233]{mehra_1994_the-beat}: The electrons are still treated non-relativistically, the only relativistic effect is the inclusion of retardation, which Bethe neglected (dipole approximation).} This was not of much relevance for the non-relativistic case (the Coulomb self-energy is the same for each state and does not play a role for the Lamb Shift), but it would be an immense advantage, when calculating (and then regularizing and renormalizing with the methods Feynman was devising at Bethe's behest) the self-energy in a relativistic framework:

\begin{quote}
I was very surprised to discover that it was not known at that time that every one of the formulas that had been worked out so patiently by separating longitudinal and transverse waves could be obtained from the formula for the transverse waves alone, if instead of summing over only the two perpendicular polarization directions you would sum over all four possible directions of polarization. [...] I thought it was general knowledge and would do it all the time. I would get into arguments with people, because I didn't realize they didn't know that; but it turned out that all their patient work with the longitudinal waves was always equivalent to just extending the sum on the two transverse directions of polarization over all four directions. This was one of the amusing advantages of the method. (Nobel Lecture)
\end{quote}

By the fall of 1947, Feynman had worked out how to treat (regular retarded) electrodynamic scattering processes (and, even more importantly, the energies of one-electron states conceptualized as non-scattering) using his path integral method and how to apply regularization and renormalization techniques, and all of that in a potentially fully covariant manner, which greatly simplified calculations. Only \emph{potentially} covariant, because what was still lacking was a way to treat relativistic kinematics. The old bugbear of Wheeler-Feynman electrodynamics, the Dirac electron, was still with him. As he wrote to the Corbens, who had been his summer hosts in Pittsburgh, on 6 November 1947 (cited after \citep[p. 423]{schweber_1994_qed}):

\begin{quote}
It [...] seems that I have guessed right, that the difficulties of electrodynamics and the difficulties of the hole theory of Dirac are independent and one can be solved before the other. I am now working on hole theory...
\end{quote}

\subsection{Path Integrals and the Dirac Equation}

It was clear to Feynman that his path integral was poorly equipped to handle the Dirac equation. In a sense, the Dirac electron offers the opposite problem to Wheeler-Feynman electrodynamics. The latter had an action, but no Hamiltonian. The former had a Hamiltonian, but no action. This statement requires some clarification. In the modern understanding of the Dirac equation as a field equation for quantized, fermionic fields, there is of course a Lagrangian density (and consequently a regular action) that gives the Dirac equation as its (field) equations of motion. This Lagrangian density was known and used already in the 1930s. But Feynman's path integrals could not yet deal with fermionic fields: The modern day formulation with Grassmann-valued variables was still far off. Feynman's starting point could thus only be the one-particle Dirac equation. This of course brought with it the difficulty of dealing with the negative energy states, to which we will turn in a second. But even if straightforwardly interpreted as a relativistic generalization of the Schr"odinger equation, the Dirac equation offered severe difficulties to the path integral approach. The Dirac equation can be written (and was originally written by Dirac) in a not manifestly covariant manner, so as to closely resemble the non-relativistic Schr"odinger equation:

\begin{equation}
i \hbar \frac{\partial \psi}{\partial t} = \gamma^0 c \left( mc - i \hbar \left( \mathbf{\gamma} \cdot \mathbf{\nabla} \right)\right) \psi
\end{equation}

The operator acting on the wave function on the right-hand side can then be identified as the Dirac Hamiltonian. However, it remains an operator, even when replacing the differential operator $-i\hbar \mathbf{\nabla}$ with its classical correspondent, the momentum $\mathbf{p}$, for the Dirac matrix structure remains. In quantum theory this is easily understood, as the wave function is not just a function of space and time, but also a 4-spinor, on which the quantum Hamiltonian acts with its Dirac matrix structure. Classically, the Hamiltonian doesn't act on anything, it's simply a number, and it is not clear how to interpret a classical Hamiltonian that is a 4-by-4 matrix. Still, formally a classical matrix Hamiltonian can be extracted. But that's as far as it goes: Already the construction of a classical Lagrangian fails, because the velocities, as determined from the Hamilton equations of motion, turn out to be matrix-valued, too. There is thus also no straightforward way to write down a classical action for a Dirac particle, which would be the necessary starting point for a Feynman path integral. Feynman made a somewhat misleading comment on this matter in the last section of his RMP paper, where he states that he had found a Lagrangian from which ``[t]he Dirac equation results.'' However, this is qualified in the next sentence: ``What results directly is the square of the usual Dirac operator.'' This can indeed be confirmed by direct calculation, and while the second-order differential equation thus obtained has some interesting properties,\footnote{It is not simply the Klein-Gordon equation, but contains an additional spin-oribit coupling term. Feynman would return to this equation years later in the context of the weak interactions \citep{feynman_1958_theory}.} the fact remains that it is \emph{not} the Dirac equation and nothing in the path integral formalism can help draw the square root.

Feynman tried long and hard to find some sort of action/path-integral description of the Dirac equation.\footnote{\label{fn:misled} See the excellent reconstruction in chapter 4 of \citep{wuethrich_2010_the-genesis}. Wuethrich, however, seems to have been somewhat misled by Feynman's comments mentioned above, and appears to claim (e.g., on p. 77 and p. 113) that Feynman had an action that led to the Dirac equation, even though, as I have stressed above, all that action provided was its square.} But eventually, he gave up, and shifted gears. It is here that the notion of a propagator takes center stage.

To understand this, we take a step back and recall the double nature of the S-Matrix, which we have encountered several times in the first half of the paper: The S-Matrix can be regarded as simply giving transition amplitudes between initial and final states, or it can be regarded as a time evolution operator in the infinite time limit, acting on an initial state wave function and turning it into a final state wave function. Similarly, one can rewrite equation \ref{eq:amplitude} as describing the time evolution of the initial state wave function to the final state function:

\begin{equation}
\label{eq:amplitude2}
\psi (t_1, x_k)  =   \int  \; \mathrm{exp}\left[ \frac{i}{\hbar} \sum_{i=0}^{k-1} L \left( x_{i+1}, \frac{x_{i+1}-x_i}{\epsilon} \right) \epsilon \right] \psi (x_0, t_0) \frac{dx_0}{A} \frac{dx_1}{A} \cdots \frac{dx_{k-1}}{A} 
\end{equation}

One sees here how the path integral can be understood as a formal solution of the Schr"odinger equation with the initial conditions set by $\psi (x_0, t_0)$. The path integral can be understood as the Green Function of the Schr"odinger equation, i.e.,

\begin{equation}
G(x_k, t_1; x_0, t_0) = \int  \; \mathrm{exp}\left[ \frac{i}{\hbar} \sum_{i=0}^{k-1} L \left( x_{i+1}, \frac{x_{i+1}-x_i}{\epsilon} \right) \epsilon \right] \frac{1}{A} \frac{dx_1}{A} \cdots \frac{dx_{k-1}}{A}
\end{equation}

But for a free particle (or any Schr"odinger equation where a complete set of time-independent solutions $\phi_n$ with energy eigenvalues $E_n$ can be obtained in a non-perturbative manner) the Green Function can be calculated in a much easier manner than by evaluating the path integral as \citep[eq. 3]{feynman_1949_theory}

\begin{equation}
\label{eq:green}
G(x_k, t_1; x_0, t_0) = \sum_n \phi_n (x_k) \phi_n^{\ast} (x_1) e^{-i E_n (t_1 - t_0)} \Theta (t_1-t_0)
\end{equation}

Feynman had already had to use this method for the hydrogen atom, as discussed above. This method also worked just fine for the free one-particle Dirac equation.\footnote{Actually, Feynman first derived the Green Function for the two-dimensional (1+1) Dirac equation in a much more complicated fashion already in early 1947, i.e., before Bethe set him on the Lamb Shift track, cf. \citep[sec. 4.1.3]{wuethrich_2010_the-genesis}. At the time, he used heuristic path-counting techniques, where specific numerical values are assigned to the possible turns the electron might take in a discretized, two-dimensional space-time. These cannot, however, be considered actual path integral techniques, as the assignments of numbers to the turns of the electron cannot be reduced to an overall action or Lagrangian. W"uthrich seems to claim otherwise on p. 77, but I believe that he is here being misled by Feynman's comments in \cite{feynman_1948_space-time}. See also Footnote \ref{fn:misled}.} When used, however, as the basis for a perturbative calculation of the scattering of a Dirac electron in a potential, it led to the possibility of scattering the electron into negative energy states. This could of course easily be avoided for the final states - one simply only considered the scattering from positive energy states into another positive energy state. But for the intermediate states that appeared in the sum of Equation \ref{eq:green} the negative energy states could not simply be dropped. Here now, Feynman brought in one of his most famous inventions, the notion of positrons as electrons going backward in time.

The notes on this idea date from late 1947, after the time when Feynman had figured out how to deal with the difficulties of electrodynamic interactions. His new method was to modify the Green Function obtained from the Dirac equation taken as a one-particle equation, Equation \ref{eq:green}, so that the intermediate negative energy states would propagate backwards in time, i.e.

\begin{equation}
\label{eq:green2}
G(x_k, t_1; x_0, t_0) = \sum_{n^{+}} \phi_{n^{+}} (x_k) \phi_{n^{+}}^{\ast} (x_1) e^{-i E_n (t_1 - t_0)} \theta (t_1-t_0) - \sum_{n^{-}} \phi_{n^{-}} (x_k) \phi_{n^{-}}^{\ast} (x_1) e^{- i E_n (t_1 - t_0)} \theta (t_0-t_1)
\end{equation}

where the first sum is over all the positive energy states, the second sum over all the negative energy states, so that $E_n$ is positive in the first sum and negative in the second. It should first be noted that this is basically equivalent to the singular invariant function of Equation \ref{eq:causal}, which Stueckelberg obtained around the same time, using the requirement of causality rather than Feynman's heuristic of positrons as negative energy electrons going backwards in time. The central difference is the minus sign between the two terms, which arises when using the Dirac rather than the Klein Gordon equation.\footnote{It should be noted that, even though this certainly was not clear to Feynman at the time, in the resulting theory the electron and the photon are basically treated on the same footing, as the $\delta_+$ function appearing in Equation \ref{eq:fokker2}, which describes electromagnetic interaction between electrons, i.e., photon propagation, is also closely related to the (massless vector case of the) Stueckelberg-Feynman propagator: one is the four-dimensional Fourier Transform of the other.}  It had, as far as I can see, implicitly been derived correctly by Stueckelberg and Rivier in the spinorial generalization of the scalar causal function (Appendix III.d of Rivier's thesis), but caused Feynman great headaches. It did not arise naturally from his heuristic, and Feynman's only guideline here was the comparison with the results obtained ``by the more tedious and old-fashioned methods'' (Interview with Mehra, p. 240), based on second quantization and hole theory. This further cemented the modular and rule-based character of Feynman's formulation of QED:

\begin{quote}
I would get to a point and say, ``with the sign plus or minus,'' and later ``the sign is equal to the number of something or other.'' Then it wouldn't work. Then I would try again. So essentially I was discovering the rules by a kind of cut and try scheme, which I have used ever since.
\end{quote}

This unique style of doing physics that Feynman developed here - Feynman himself refers to it as the ``intuitive method'' in his Nobel lecture - has been much discussed and I have here only touched upon those aspects that are most relevant to my overall argument. Galison convincingly argues for the origin of this method in Feynman's war work \citep{galison_1998_feynmans}; in particular, the general method of modifying a theory by first rewriting it as an integral equation and then tweaking the kernel clearly began in Los Alamos. But while the techniques and approaches Feynman was now applying to QED can be traced to his applied work in the Manhattan project, one also needs to see that this did not necessarily mean that he preferred this approach to a more axiomatic approach starting from fundamental equation. As Wuethrich has argued,\footnote{Unpublished talk by Adrian W"uthrich at the 2014 Seven Pines Symposium.} and I think my reconstruction backs this view, it was very specific theoretical problems that led him to adapt this approach. Feynman would certainly not have been unhappy if QED scattering amplitudes could have been directly calculated using his path integral. But the path integral simply was not powerful enough at the time to deal with more complex problems, such as the hydrogen atom or positron theory, and so Feynman was forced to import results and expressions derived in the usual formulation of QED, leading to the patchwork style we now associate with him. Galison cites Feynman's aversion to this usual formulation (p. 430), but this aversion was based on the complicatedness of the calculations involved, not on the fact that they were derived from first principles \emph{per se}.

This latter point is underscored by the fact that Feynman actually worked hard to base his calculational techniques on a solid theoretical ground. One way to do this would have been to prove generally (and not just for specific results and processes) that the Feynman rules followed from the usual formulation of QED. This was the path later taken by Dyson, but it appears that Feynman, who was only just learning that regular formulation, was not yet able to do this (see, e.g., \citep[p. 144]{wuethrich_2010_the-genesis}).  Instead, Feynman worked hard to put his path integral on firmer mathematical ground and at the same time enhance its intrinsic calculational capabilities, a work that resulted in Feynman's ordered operator calculus. This calculus was only published in \citep{feynman_1951_an-operator}, but was ready enough by the spring of 1948 that Feynman dared to use it to introduce and derive his methods at the Pocono conference (March 30 to April 1), the follow-up conference to Shelter Island. As for his methods, he now had almost everything in place to derive general scattering amplitudes in QED, with the sole exception of the problem of vacuum polarization, which was to keep him occupied for another year. Still, his Pocono presentation famously flopped. Neither Feynman's operator calculus, nor his more heuristic derivations, which he then tried to present, were greeted with much enthusiasm. And even years later, Feynman felt that both of these approaches were still ill understood. Concerning the operator calculus, he remarked to Norbert Weiner in 1966:\footnote{Interview on 27 June 1966, https://www.aip.org/history-programs/niels-bohr-library/oral-histories/5020-3.}

\begin{quote}
I had invented a new mathematical method of dealing with operators, with ordering the operators according to a parameter which I to this day feel is a great invention, and which nobody uses for anything, and which nobody pays any attention to, but I just take this opportunity to say that I think that that thing is someday going to be - I mean, maybe in history you'll find out that the guy knew it was good or thought it was good and it never was good, or whatever, but I still think it's an important invention, a very important invention.
\end{quote}

And as for his more heuristic approach, in his Nobel Lecture:

\begin{quote}
Often, even in a physicist's sense, I did not have a demonstration of how to get all of these rules and equations from conventional electrodynamics. [...] As a result, the work was criticized, I don't know whether favorably or unfavorably,\footnote{A certainly unfavorable criticism can again be found in the letter from Fierz to Pauli, cited in Footnote \ref{fn:fierz}, where Fierz writes that Feynman's ``heuristics do not hold up in the face of a serious critique.''} and the ``method'' was called the ``intuitive method.'' [...] In the face of the lack of direct mathematical demonstration, one must be careful and thorough to make sure of the point, and one should make a perpetual attempt to demonstrate as much of the formula as possible. Nevertheless, a very great deal more truth can become known than can be proven.
\end{quote}

It would take the translation efforts of Freeman Dyson to turn Feynman's approach into the standard way of doing quantum electrodynamics.

\subsection{Dyson and the S-Matrix}

Freeman Dyson, a trained mathematician, had learned modern physics, and in particular quantum field theory, from Nicolas Kemmer at Cambridge in 1946/47.\footnote{The following draws heavily on \citep[ch. 9]{schweber_1994_qed}.}  When he joined Bethe (and Feynman) at Cornell in September 1947, he was already well-versed in the established framework of quantum field theory and immediately started working on a relativistic calculation of the Lamb Shift for a scalar (Klein-Gordon) electron, published in \citep{dyson_1948_the-electromagnetic}, using the combination of old-fashioned, not manifestly covariant techniques and the new ideas and techniques of renormalization, pioneered by Bethe. In the first half of 1948, he got further acquainted with the new manifestly covariant formulations of QED of Tomonaga and Schwinger. He also, through his personal contact at Cornell (recall that Feynman had as yet hardly published any of his work), got to know Feynman's methods. And although he was as deeply mystified by Feynman's approach as the audience at the Pocono conference, he did appreciate the incredible speed with which Feynman was able to perform calculations once his rules had been obtained in the idiosyncratic manner described in the previous sections.

The essential revelation appears to have come to Dyson upon moving to Princeton after a year at Cornell, in September 1948. He was puzzling over Feynman's set of rules and diagrams, this black box for deriving scattering amplitudes (and energy eigenvalues of stationary states using the no-scattering amplitude approach discussed above), garnished with nice heuristics, which gave an overall idea of the actual processes occurring but did not amount to an actual detailed space-time description of an evolving physical state between the asymptotic in and out states. Dyson had set himself the goal of relating this, apparently immensely useful, approach to the more understandable, yet computationally cumbersome, formalism of Schwinger and Tomonaga. Then it hit him: Feynman was presenting an algorithm for calculating S-Matrix elements, and Feynman's theory was essentially a fulfilment of Heisenberg's S-Matrix program.

Coming from England, Dyson was much more aware of Heisenberg's S-Matrix work than most American physicists. His physics mentor, Nicolas Kemmer, had attended the 1946 Cambridge conference, discussed earlier, and had imparted his knowledge of S-Matrix theory on Dyson. Up until September 1948, Dyson had simply thought of the S-Matrix as ``new physics,'' an attempt to solve the problems of QED by radically breaking with the established quantum field theory. Now suddenly it dawned on him that Feynman's method, with all its visualizable trajectories and inputs from traditional field theory, was in fact an S-Matrix theory. And what was more, this realization could help Dyson to bring Feynman's work in contact with the more traditional work of Tomonaga and Schwinger. To better understand this latter point, we need to briefly discuss the Tomonaga-Schwinger reformulation of QED, its relation to the concept of a quantum state and to the S-Matrix.

In the following, I will not distinguish between the contributions of Schwinger and Tomonaga. Suffice it to say that while Tomonaga worked out the foundational equations more clearly than Schwinger (at least in the opinion of Freeman Dyson and myself), Schwinger worked out in much greater detail their applications and the implementation of renormalization techniques. The Tomonaga-Schwinger theory is based on the (aptly named) Tomonaga-Schwinger equation, which is a covariant generalization of the (time-dependent) Schr"odinger equation in the interaction picture:

\begin{equation}
i \hbar c \frac{\partial \Psi [\sigma]}{\partial \sigma (x)} = H (x) \Psi [\sigma]
\end{equation}

The wave function $\Psi$ describing the state is now no longer a function of time, but rather a functional of some space-like hypersurface $\sigma$. Instead of time evolution, the Tomonaga-Schwinger equation describes the change in $\Psi$ under an infinitesimal deformation of the space-like hypersurface near the space-time point $x$ (on the hypersurface). This change is determined by the Hamiltonian operator $H$, which is simply the field-theoretic Hamiltonian in the interaction representation (for QED it is the invariant expression $\left( - j_{\mu} A^{\mu}/c \right)$, with $j_{\mu}$ the electromagnetic four-current of matter). 

This is a straightforward generalization of the Schr"odinger equation to relativistic space-times, with infinitesimal time evolution replaced by infinitesimal shifts of a space-like hypersurface into the future. And consequently, one can write down a ``time-evolution operator,'' i.e., an operator that transforms the wave function on an initial hypersurface into that on a final hypersurface, which involves an integration over the space-time volume between those two hypersurfaces, as already anticipated by Heisenberg in 1937. And, just as in regular quantum mechanics, one could take the infinite-time limit of this operator to get an S-Matrix. This was a simplification, for in the infinite-time limit one did not have to worry about the foliation of space-time, one simply integrated over all of space and time. This operator had in fact been defined by Schwinger, who had called it the ``invariant collision operator'' \citep[sec. 4]{schwinger_1948_quantum}. It had, however, not been the central quantity of Schwinger's theory. In particular, the Lamb Shift was not conceptualized in terms of scattering, but rather as a matrix element of a suitably transformed (and renormalized) Hamiltonian.

What Dyson now showed in \citep{dyson_1949_the-radiation} was that the S-Matrix as calculated from Feynman's rules was the same as that calculated from the more traditional Tomonaga-Schwinger formulation. In effect, Feynman's rules could thus have been derived from the Tomonaga-Schwinger formulation. However, the systematization of the calculations through Feynman's modular approach was hardly a natural outgrowth of the Tomonaga-Schwinger formalism. When applied straightforwardly, that formalism led to a ``long and rather difficult analysis'' \citep[p. 491]{dyson_1949_the-radiation}.\footnote{Mehra recounts that Schwinger's response to this was the following: ``Well, it wasn't so long and it wasn't so difficult, but nevertheless it was not the most ecomomical way of going on to higher order effects... He [Dyson] did recognize that, as I think Feynman probably didn't, that the Feynman theory does operate with a statement about initial and final states, which is a concentration on the overall evolution of the system. And that was a useful thing. [...] And as soon as I understood that, I immediately incorporated it into my own next version as well.'' \citep[p. 288n]{mehra_2000_climbing} So, the Heisenberg-Feynman approach of bypassing instantaneous states spread and influenced Schwinger's action formalism of quantum field theory in the Heisenberg picture.}  When presenting Feynman's approach at the Basel ``International Congress on Nuclear Physics and Quantum Electrodynamics'' (\emph{Internationaler Kongress "uber Kernphysik und Quantenelektrodynamik}, 5 to 9 September 1949), Dyson characterized it in the following manner:

\begin{quote}
[O]ne may consider it [the Feynman theory] as a supplement to the Tomonaga-Schwinger theory, not differing from the latter in any of its basic assumptiions but only in its method of handling problems. [...] From this viewpoint, the contribution of the Feynman theory lies only in the following; by concentrating attention on the overall behaviour of a physical system as Heisenberg has done in his S-matrix program, and ceasing to ask questions about the instantaneous state of the system at intermediate times, one finds quick and general ways to derive results which are otherwise obtained only by more laborious and special considerations.\citep[p. 240]{dyson_1950_the-radiation}
\end{quote}

It were these ``quick and general ways to derive results'' that turned Feynman's method into the core of the new quantum electrodynamics and led to the rapid spread of Feynman's rules and diagrams into the many fields where quantum field theory was applied or was now found to be applicable, and it was around this core that extensions of the theory were performed for treating non-scattering processes using Feynman-type techniques.\footnote{On this rapid spread, see \citep{kaiser}. A prominent example of the extension of Feynman's S-Matrix method to non-scattering problems is the work of \cite{bethe_1951_a-relativistic}, which deals with extending Feynman's method to actual two-body bound state problems (i.e., not just one particle in an external potential, as in the Lamb Shift).} The calculation of S-Matrix scattering amplitudes became the new paradigmatic problem of quantum field theory, replacing the stationary state energies of quantum mechanics. So, even though the theoretical foundations of the theory were hardly changed in the re-invention of QED in the late 1940s,\footnote{The conservative nature of renormalized QED was stressed by Dyson time and again, see, e.g., \citep{dyson_1965_tomonaga}.} we observe a different kind of paradigm shift, where the formulation of the theory is so modified that different physical problems are more easily calculated and thus become the paradigmatic problems of the theory.

Consequently, Dyson's famous proof of the renormalizability of quantum electrodynamics to all orders of perturbation theory in his follow-up paper \citep{dyson_1949_the-s}, was really just a proof of the renormalizability of the S-Matrix, i.e., of the finiteness of the renormalized S-Matrix, calculated using Feynman's rules, in an arbitrarily exact approximation, as Dyson himself pointed out:

\begin{quote}
It is still an unanswered question, whether the finiteness of the S-Matrix automatically implies the finiteness of all observable quantities, such as the bound-state energy levels, optical transition probabilities, etc., occurring in electrodynamics. An affirmative answer to the question is in no way essential to the arguments of this paper. Even if a finite S-Matrix does not of itself imply finiteness of other observable quantities, it is probable that all such quantities will be finite... \citep[p. 1736]{dyson_1949_the-s}
\end{quote}

And similarly in a letter to Peierls (16 January 1949, reprinted in \citep{lee_2009_sir}):

\begin{quote}
In the paper I have dealt with scattering problems exclusively, and shown that for them at least the theory always gives finite and sensible results. [...] It seems to me now not at all likely that the bound-state problems will give worse troubles than the scattering problems, though of course the calculations will always be tougher.
\end{quote}

The centrality of the S-Matrix grew as the traditional spectroscopic experiments that had led to the formulations of renormalized QED were more and more supplanted by high-energy scattering experiments, not just the cosmic ray experiments that had occupied Heisenberg in the 1930s, but more importantly the experiments made possible by the development of particle accelerators.\footnote{It would be an interesting research question to see how much the theoretical primacy of scattering, established around 1949, actually drove the experimental developments. One might think of Einstein's remark to Heisenberg, quoted earlier, that the theory determines what is measurable.} As, in turn, also the mathematical treatment of the S-Matrix matured, by the late 1950s time was once again ripe for an attempt to get rid of states altogether, even if only as an element of the underlying theory. But that, as they say, is another story \citep{cushing_1990_theory}.

\section{Conclusions}

I hope to have convinced the reader that in the course of the 1930s and 40s a shift occurred in the paradigmatic problem to be treated in a quantum theory of fields, from determining the properties of instantaneous states to the determination of scattering amplitudes between free, asymptotic states. I have identified several important factors for this shift. The full implementation of relativistic covariance is one of these, and the recurrence of the many-faceted work of Paul Dirac on this matter underlines his importance for the conceptual development of QFT well beyond his pioneering studies on second quantization and relativistic wave equations in 1927/28.\footnote{I hope that this will finally convince Suman Seth, with whom I long ago had a memorable, if historiographically questionable, debate over who was the greater physicist, an argument in which I championed Dirac and he championed, of all people, the inventor of the stationary quantum state, Bohr. Dirac's towering role in the development of QFT in the 30s and 40s is also emphasized by \cite[p. 573]{schweber_1994_qed}.} But my main focus has been on the attempts to save QFT (before it was then saved by renormalization) by getting rid of instantaneous states altogether. These attempts paved the way for the paradigm shift: They provided the calculational techniques and more generally prepared the physics community for the shift, so that when it actually occurred within renormalized QFT in the late 1940s, it no longer needed to be announced and justified, appearing rather as the natural next step. This is all the more true as, in renormalized QFT, physicists could actually take a step back, re-introducing instantaneous states, but assigning them a far more marginal place in the theory. Heisenberg and Feynman had abolished the state entirely; instead it merely withered away. I have further discussed how experimental developments served both as a motor for this development (Heisenberg's attempts at finding a theory that could describe cosmic ray explosions), but also as a constant challenge, as physicists were forced to think about how to calculate the energies of stationary states in such a framework, when such spectroscopic questions briefly moved to the center of attention again in the late 1940s.

I believe, and even hope, that this paper is not the last word on this paradigm shift. In particular, I have hardly touched upon the ramifications that this shift has, both for the philosophy of quantum field theory and for the further development of physics. I will allow myself here merely a few brief comments on these matters. Especially in the last sections, I have emphasized how important the question of calculational ease was for this shift to occur: The methods of Feynman and Dyson won out, because they were a lot easier to use. Philosophers of Science tend to give short shrift to the question of calculational ease: Any formulation of a theory is as good as another to tease out its ontological, epistemological and metaphysical implications. But the striking fact that (relativistic) QFT is so readily formulated as a theory of scattering, and that the discovery of this reformulation was so important for the development of the theory, even for its acceptance as a consistent physical theory, calls for an explanation. In what sense \emph{is} QFT really a theory of scattering processes, and in what sense is the prevalence of scattering problems merely a historical contingency and an effect of good ol' American pragmatism? This, I believe, calls for further philosophical investigation, and I hope that the present study can contribute to such an investigation.

Philosophical considerations of currently established QFT aside, it is a well-recognized historical fact that different formulations of a theory imply different  ways for going beyond it. The natural historical follow-up question to the analysis presented in this paper is: How did (and still does) the focus on scattering first established in the late 1940s determine the further development of theoretical high-energy and particle physics and the construction of quantum field theories for the nuclear interactions and condensed matter physics? Most interesting, maybe, is the question how it still determines the currently ongoing quest for a quantum theory of gravity. The QFT focus on scattering certainly plays an important role here. Approaches to quantum gravity coming from a modern QFT background (such as string theory) have tended to focus on the calculation of gravitational scattering amplitudes  and celebrated the derivation of covariant expressions for such amplitudes as central constituents of the emerging theory. On the other hand, approaches to quantum gravity that draw more on traditions in general relativity and QM have tended to focus more on other observables, or the concept of observables in general. One might even argue that what started out merely as a shift in the paradigmatic problem of QFT led, in the confrontation of general relativity and quantum theory, to a separation of world-views that begins to look more like a ``full'' conceptual paradigm change. I hope this paper also provides a first step towards addressing these questions.

\section{Acknowledgments}
I would like to thank several colleagues at the Max Planck Institute for the History of Science for helpful comments on the manuscript at various stages of completion: Roberto Lalli, Christoph Lehner, and J\"{u}rgen Renn. I had the opportunity of presenting an earlier version of this work at the 20134 Seven Pines Symposium and would like to thank the organizers for inviting me, as well as all participants, in particular Sam Schweber and Thiago Hartz, for helpful discussion. I would also like to thank the two anonymous referees for their helpful (and kind) comments.

\bibliography{habil}
\bibliographystyle{apalike}
\end{document}